\RequirePackage{etex}
\PassOptionsToPackage{table,dvipsnames}{xcolor}
\documentclass[a4paper,fleqn]{cas-sc}

\usepackage[numbers]{natbib}
\usepackage[T1]{fontenc} 
\usepackage{graphbox} 
\PassOptionsToPackage{hyphens}{url} 
\usepackage{hyperref,bookmark}
\hypersetup{
    colorlinks,
    citecolor=blue,
    filecolor=black,
    linkcolor=black,
    urlcolor=blue,
    breaklinks
}
\usepackage{graphicx}
\usepackage{multirow}
\usepackage{booktabs}
\usepackage{caption}
\usepackage{xcolor}
\usepackage{array}
\usepackage{amsmath}
\usepackage{amssymb}
\usepackage{subcaption}
\usepackage[utf8]{inputenc}
\usepackage{adjustbox}
\usepackage{xspace}
\usepackage{paralist}
\usepackage{import} 
\usepackage{bibentry} 
\usepackage{siunitx}
\usepackage{soul}
\usepackage{tabularx}

\newcommand{\mul}{{\texttt{MULTI}}\xspace}
\newcommand{\unfl}{{\texttt{UNFL}}\xspace}
\newcommand{\infl}{{\texttt{INTFL}}\xspace}
\newcommand{\ind}{{\texttt{INDE}}\xspace}
\newcommand{\mono}{{\texttt{MONO}}\xspace}

\newcommand{\rtw}{{\texttt{RTW}}\xspace}
\newcommand{\rpl}{{\texttt{RPL}}\xspace}
\newcommand{\men}{{\texttt{MEN}}\xspace}
\newcommand{\urld}{{\texttt{URL}}\xspace}

\newcommand{\hash}{{\texttt{HST}}\xspace}

\newcommand{\russia}{{\texttt{IORussia}}\xspace}
\newcommand{\uk}{{\texttt{UK}}\xspace}

\newcommand{\louv}{{\textsc{Louvain}}\xspace}
\newcommand{\info}{{\textsc{Infomap}}\xspace}
\newcommand{\glouv}{{\textsc{Generalized Louvain}}\xspace}
\newcommand{\ginfo}{{\textsc{Generalized Infomap}}\xspace}

\makeatother

\begin{document}

% \begin{frontmatter}

\let\WriteBookmarks\relax
\def\floatpagepagefraction{1}
\def\textpagefraction{.001}

% Short title
\shorttitle{Multimodal Coordinated Online Behavior: Trade-offs and Strategies}

% Short author
\shortauthors{Mannocci et~al.}

\title[mode=title]{Multimodal Coordinated Online Behavior: Trade-offs and Strategies}\tnotemark[1]
\tnotetext[1]{Post-print of the article published in Information Sciences 123125, 2026. Refer to the published version: \href{https://doi.org/10.1016/j.ins.2026.123125}{10.1016/j.ins.2026.123125}}

\author[1,2]{Lorenzo Mannocci\corref{cor1}}
\cortext[cor1]{Corresponding author}
\ead{lorenzo.mannocci@di.unipi.it}

\author[2]{Stefano Cresci}
\ead{stefano.cresci@iit.cnr.it}

\author[3]{Matteo Magnani}
\ead{matteo.magnani@it.uu.se}

\author[1]{Anna Monreale}
\ead{anna.monreale@unipi.it}

\author[2]{Maurizio Tesconi}
\ead{maurizio.tesconi@iit.cnr.it}

\affiliation[1]{organization={University of Pisa},
    city={Pisa},
    country={Italy}}

\affiliation[2]{organization={Institute for Informatics and Telematics, National Research Council (IIT-CNR)},
    city={Pisa},
    country={Italy}}

\affiliation[3]{organization={InfoLab, Department of Information Technology, Uppsala University},
    city={Uppsala},
    country={Sweden}}

\begin{abstract}
Coordinated online behavior, which spans from beneficial collective actions to harmful manipulation such as disinformation campaigns, has become a key focus in digital ecosystem analysis. 
Traditional methods often rely on monomodal approaches, focusing on single types of interactions like co-retweets or co-hashtags, or consider multiple modalities independently of each other. However, these approaches may overlook the complex dynamics inherent in multimodal coordination. 
This study compares different ways of operationalizing multimodal coordinated behavior, examining the trade-off between weakly and strongly integrated models and their ability to capture broad versus tightly aligned coordination patterns. By contrasting monomodal, flattened, and multimodal methods, we evaluate the distinct contributions of each modality and the impact of different integration strategies. Our findings show that while not all modalities provide unique insights, multimodal analysis consistently offers a more informative representation of coordinated behavior, preserving structures that monomodal and flattened approaches often lose. This work enhances the ability to detect and analyze coordinated online behavior, offering new perspectives for safeguarding the integrity of digital platforms. 
\end{abstract}

%%Research highlights
\begin{highlights}
\item We introduce an operationalization to detect multimodal coordinated behavior, which fully exploits the potential of the interdependence among different modalities (e.g., co-retweets, co-hashtags, co-URLs), by exploiting a multiplex coordination network.

\item We analyze the contributions of different monomodal approaches in detecting coordinated online behavior, revealing whether various data modalities provide complementary or overlapping insights.

\item We compare different levels of multimodal integration, highlighting the trade-offs between broadly capturing coordination patterns and strictly identifying tightly coordinated users, providing guidance for robust multimodal analyses.
\end{highlights}


\maketitle

\makeatletter
\let\oldthefootnote\thefootnote  \renewcommand\thefootnote{}      

\footnotetext{\tolerance=1000
\href{https://orcid.org/0000-0002-5556-3746}{0000-0002-5556-3746} (Lorenzo Mannocci), \href{https://orcid.org/0000-0003-0170-2445}{0000-0003-0170-2445} (Stefano Cresci), \href{https://orcid.org/0000-0002-3437-9018}{0000-0002-3437-9018} (Matteo Magnani), \href{https://orcid.org/0000-0001-8541-0284}{0000-0001-8541-0284} (Anna Monreale), \href{https://orcid.org/0000-0001-8228-7807}{0000-0001-8228-7807} (Maurizio Tesconi)
}

\let\thefootnote\oldthefootnote  \makeatother

\makeatletter
\def\@makefnmark{\hbox{\@textsuperscript{\normalfont\@thefnmark}}}
\makeatother
\setcounter{footnote}{0}

\section{Introduction}

The digital age has transformed the way individuals and organizations interact online, giving rise to novel forms of coordination in social and political arenas. Coordinated behavior, a phenomenon where multiple actors engage in synergic actions in pursuit of an intent~\cite{mannocci2024detection}, has become increasingly relevant in the study of online ecosystems.

Coordination in online behavior can manifest across a wide spectrum of intentions and outcomes. It may emerge organically or be deliberately orchestrated, and its perceived value often depends on the observer’s perspective. For instance, coordinated efforts are found in activism~\cite{ng2022combined,nizzoli2021coordinated}, boycott~\cite{lucchini2022reddit}, and protests~\cite{magelinski2022synchronized,steinert2015online}, which some may view as socially beneficial, while others may see them as disruptive or even harmful. Conversely, coordination can also be employed for goals widely recognized as problematic, such as disinformation campaigns~\cite{vargas2020detection}, election interference~\cite{nizzoli2021coordinated}, or astroturfing~\cite{keller2020political}. Even in cases like coordinated botnets~\cite{mannocci2022mulbot}, which are typically associated with manipulation, there may be applications aligned with legitimate objectives. 
Ultimately, the evaluation of coordinated behavior -- whether deemed harmful or beneficial -- is often subjective and context-dependent.
Hence, since coordination in social media exhibits a dual nature, the scope behind coordination can vary widely, ranging from benign collective action to malicious intent aimed at manipulation or deception.

\begin{figure}[!htbp]
    \centering
    \includegraphics[width=0.5\textwidth]{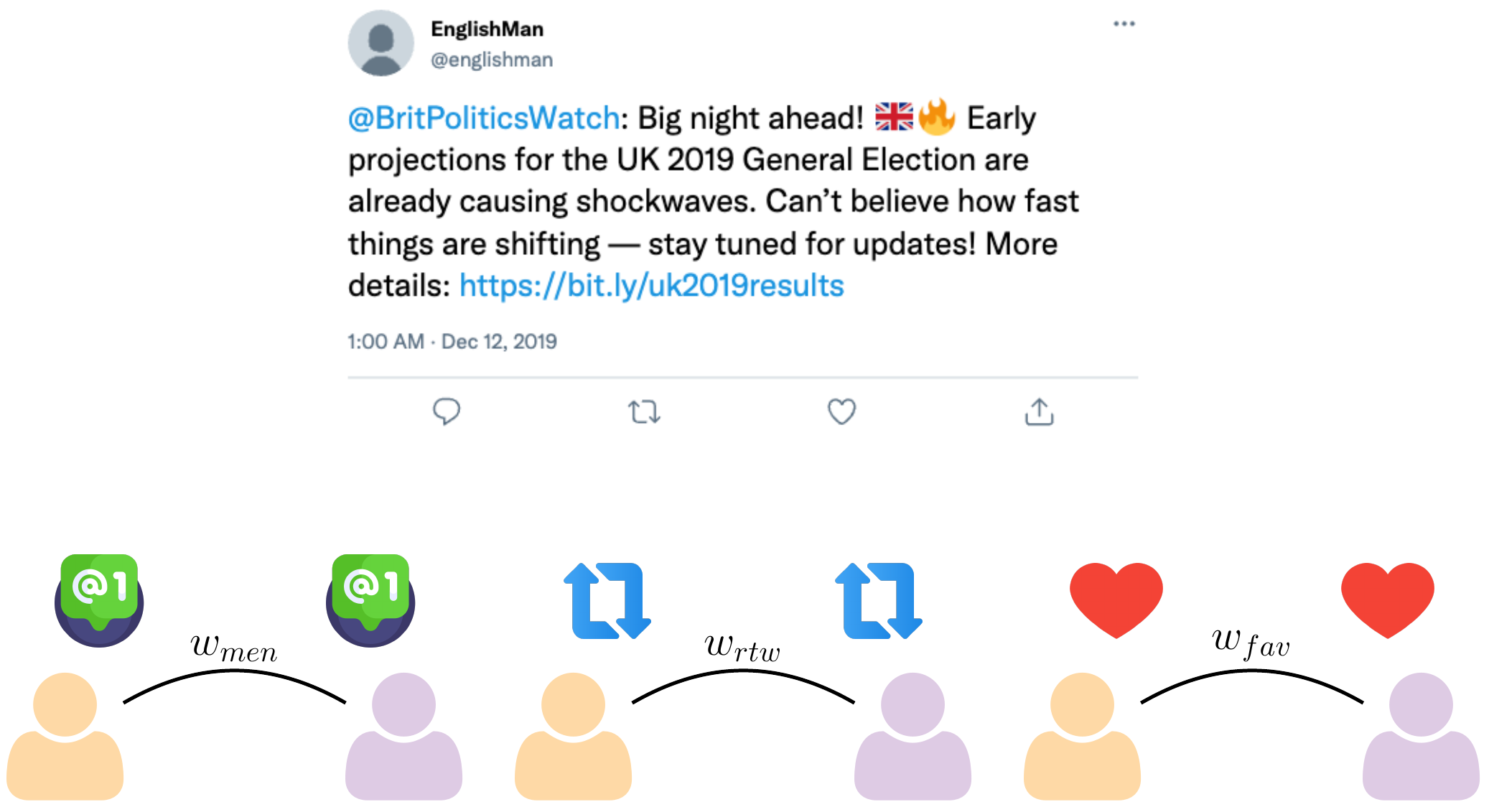}
    \caption{Coordinated online behavior is inherently multimodal, as users can coordinate through different types of actions.}
    \label{fig:multimodal_cb}
\end{figure}

Understanding and detecting coordinated online behavior is critical for maintaining the integrity of digital platforms. Researchers and policymakers alike seek to develop methodologies to identify patterns of coordination, disentangle genuine collective action from inauthentic manipulation, and mitigate the harmful consequences of coordinated inauthentic behavior. \\

\textbf{Detection approaches.} Various computational techniques, including network-based analysis~\cite{cima2024coordinated,weber2021amplifying} and machine learning models~\cite{sharma2021identifying,zhang2021vigdet} have been employed to unveil hidden networks of coordination. 
Users may engage in a wide range of coordinated actions, which complicates their detection. For example, users might coordinate through synchronized retweets to amplify specific content~\cite{loru2023influence}, through likes, or by using identical hashtags to dominate trending topics~\cite{luceri2023unmasking,ng2023coordinated}. More covert forms of coordination can include strategic timing of posts~\cite{pacheco2021uncovering}, the use of similar narratives across multiple accounts, or the deliberate targeting of specific audiences. These varied manifestations highlight the inherent multimodal nature of coordination in social media, where the modality of the actions plays a crucial role in shaping the dynamics and outcomes of coordinated behavior. Figure~\ref{fig:multimodal_cb} shows how coordinated online behavior is inherently multimodal, as users can coordinate through different types of actions. Therefore, it is crucial to design and provide a robust methodology capable of uncovering coordinated behaviors that manifest through diverse user actions. \\

\textbf{Network-based approaches and multimodality.}
While network-based methods have proven effective in detecting such behavior, most approaches either focus on a single type of action or treat different actions separately, resulting in the construction of multiple single-layer networks that miss the opportunity to analyze them in an integrated way. 
Although some studies have begun to explore multimodal interactions, they fail to fully leverage the richness and complexity that such diversity offers~\cite{magelinski2020detecting,magelinski2022synchronized}. Many of these approaches flatten the layers into a single aggregated network, thereby losing the distinct characteristics of each modality and the interplay between them~\cite{dey2024coordinated,luceri2023unmasking}. The current landscape reveals two key limitations: first, existing multimodal analyses do not truly exploit the potential of integrating different types of actions; second, there is a lack of systematic investigation into how and whether multimodality offers advantages over analyzing each type of action independently. As a result, we lack not only a clear understanding of the value of multimodality in this context but also effective frameworks or methods to operationalize it properly. Addressing these gaps is essential for advancing our ability to detect and interpret coordinated behavior in a more comprehensive and meaningful way. \\

\textbf{Our work.}
For this reason, it is important to assess the potential benefits of adopting a comprehensive approach that accounts for the multimodal nature of coordinated behavior. In this work, we propose different ways to operationalize the detection of multimodal coordinated behavior by incorporating various types of actions and, consequently, different forms of coordination. By integrating multiple modalities -- such as retweets, mentions, hashtag usage, and posting patterns -- these operationalizations provide a holistic perspective on how users coordinate their efforts. The objective is not only to capture the diversity of coordinated actions but also to detect subtle and complex behaviors that might be overlooked when focusing on a single modality. By bridging these different dimensions of coordination, we enable an enhanced understanding and identification of coordinated efforts in social media, contributing to more robust and reliable analysis frameworks. \\

\begin{figure}[!htbp]
    \centering
    \includegraphics[width=0.7\textwidth]{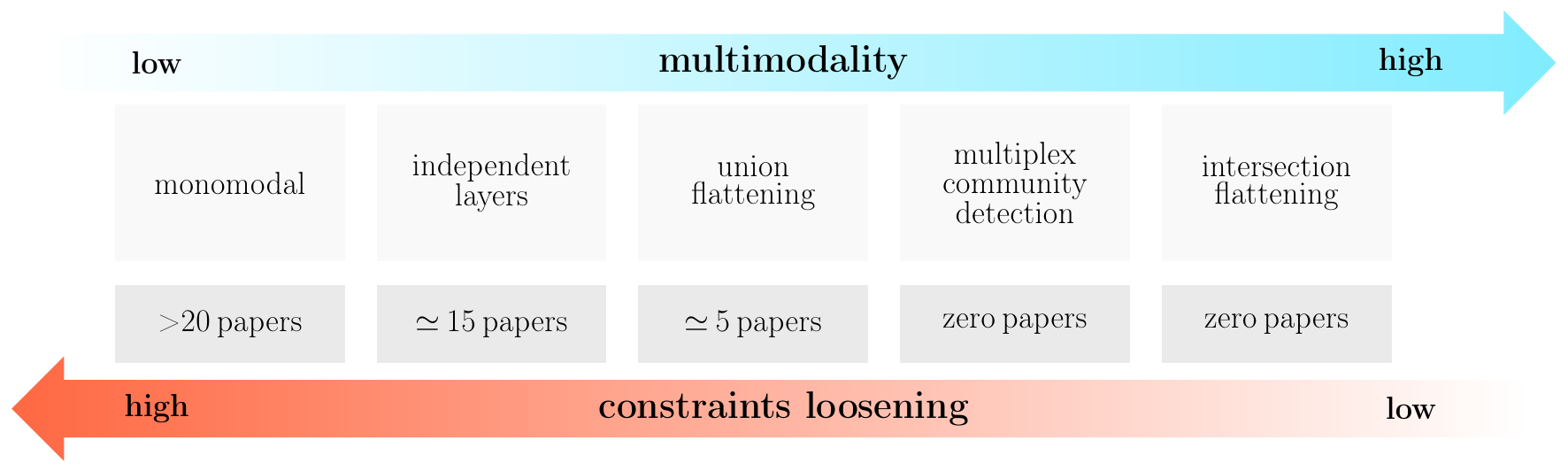}
    \caption{Trade-off between multimodality and constraints loosening in coordinated behavior detection. Increasing the level of multimodality enhances the complexity and richness of the analysis but may also restrict the detection to only the most tightly coordinated actors.}
    \label{fig:multimodality_tradeoff}
\end{figure}

\textbf{Multimodality trade-off.}
Our framework includes several multimodal detection approaches for coordinated behavior, and Figure~\ref{fig:multimodality_tradeoff} presents examples of possible ways to operationalize multimodality, along with the number of studies in the literature that have adopted each approach. For example, the approach frequently used in the literature -- flattening the layers into a single network -- could be one solution, albeit perhaps not the most effective. Another approach, which considers different coordination networks and actions independently, as is often done, represents the weakest form of multimodality operationalization, as it overlooks the interplay between modalities. On the other end of the spectrum, the strongest approach forces the model to recognize users as coordinated only if they demonstrate coordination across all types of actions. This stricter operationalization fully embraces multimodality but at the cost of increased selectivity. As we can anticipate, there is an inherent trade-off in leveraging multimodality. While enriching the model by integrating more modalities allows for the detection of complex and nuanced forms of coordinated behavior, increasing the intersection of multimodal conditions too much may overly constrain the analysis, potentially limiting the detection to only a few instances of coordination. Figure~\ref{fig:multimodality_tradeoff} illustrates this trade-off, highlighting the balance between the level of multimodality and constraints loosening in coordinated behavior detection. \\

\textbf{Main findings.}
Experiments conducted on two Twitter/X datasets demonstrate the potential benefit of using our multimodal framework in capturing coordinated behavior across diverse types of user actions. By contrasting monomodal and multimodal analyses, we show that relying on a single co-action can obscure important patterns, whereas integrating multiple actions reveals a richer and more complete picture of coordination. Identifying in advance which co-action is most informative remains challenging, further motivating the need for multimodality to avoid overlooking meaningful signals. Our results also highlight that flattening the network may lead to a substantial loss of structural information, reducing the ability to detect coordinated communities. In contrast, the multimodal approach preserves nearly all communities identified by monomodal methods while also uncovering additional structures that would otherwise remain hidden.\\

In summary, this study not only expands the toolbox of methods for the detection of multimodal coordinated behavior, but also provides a comprehensive exploration of how to effectively operationalize multimodality, striking a balance between complexity and detectability.

\subsection{Research focus}
Understanding coordinated online behavior requires careful analysis of interactions across different digital modalities. Existing approaches typically fall into two broad categories: monomodal methods, which analyze coordination based on a single type of interaction (e.g., retweets, hashtags, or URLs sharing), and multimodal methods, which integrate multiple modalities to capture a more comprehensive picture of coordination. However, the comparative effectiveness of these approaches remains an open question. This study explores the contributions of monomodal and multimodal methods to the detection and analysis of coordinated online behavior, with a particular focus on whether different modalities offer complementary insights or largely redundant information. The following research questions guide our investigation:
\begin{itemize}
    \item \textbf{RQ1}:
\textit{What is the unique contribution of each individual data modality to the detection and analysis of coordinated online behavior?}
    This RQ aims to determine whether monomodal approaches -- each focused on a single type of interaction, such as synchronized retweets, hashtags, or URLs sharing -- produce similar or differing results between one another. By comparing the outcomes of analyses performed separately on each modality, we assess whether these modalities capture redundant or complementary aspects of coordination. For example, if all modalities provide similar information, their individual analyses would converge on consistent findings. However, if different modalities reveal distinct patterns of coordination, this would indicate that each provides unique and potentially critical information for understanding coordinated behavior.
    \item \textbf{RQ2} \textit{How do multimodal approaches contribute to the analysis of coordinated online behavior compared to monomodal approaches?} Additionally, \textit{how do different implementations of multimodality affect the results?}
    Here, we explore the added value of multimodal approaches in detecting and analyzing coordinated online behavior compared to monomodal analyses. While monomodal approaches focus on a single type of interaction, multimodal approaches leverage multiple modalities simultaneously, potentially offering a more comprehensive view of coordination patterns. 
    However, multimodality is a flexible concept that encompasses various ways of combining modalities. Different strategies for integrating modalities may highlight different aspects of coordinated behavior and yield varying analytical outcomes. This question thus addresses two key points: the overall benefit of multimodal analyses relative to monomodal ones, and the impact that different implementations of multimodality have on the results. 

\end{itemize}

To address the research questions outlined above, we conduct our analysis on two Twitter datasets. One dataset offers a rich and appropriate context for studying coordinated online behavior due to the platform’s public and interaction-driven nature. It has been previously used in the study of coordination dynamics, although not in a multimodal setting~\cite{nizzoli2021coordinated,hristakieva2022spread,loru2023influence}. Conversely, the second dataset includes a ground-truth labeling, enabling a validation of the ability of multimodal approaches not only to identify more structures, but also that these structures capture actual coordination.

While coordinated behavior can be evaluated through various downstream tasks, the focus of this work is methodological: we aim to compare network science–based multimodal formulations and assess how each data modality contributes to detecting patterns associated with coordinated behavior. The coordinated communities identified through our methods could be used in downstream analyses -- such as studying propaganda dissemination~\cite{hristakieva2022spread}, political leaning~\cite{nizzoli2021coordinated}, or bot likelihood~\cite{hristakieva2022spread}. However, these analyses lie beyond the scope of this work. Our aim is to systematically investigate the added value of multimodality within network-based frameworks for coordinated behavior detection.

\subsection{Contributions}
This work advances the analysis of coordinated online behavior by examining the role of different data modalities, introducing a novel method for detecting multimodal coordination, and comparing monomodal and multimodal approaches. Additionally, we explore different levels of multimodal integration, highlighting key trade-offs in capturing coordination patterns. Below, we outline our main contributions:
\begin{itemize}
    \item We perform the first comparative analysis of operationalizations for detecting multimodal coordinated behavior, including an approach that fully leverages the multimodal structure of a multiplex coordination network and captures interdependencies among different modalities through a community detection algorithm specifically designed for multiplex structures.
    
    \item We analyze the degree of overlap and divergence in the information provided by different data modalities when used individually in monomodal approaches, revealing how each modality contributes unique or redundant insights. This lays the groundwork for understanding how combining multiple modalities enhances the analysis of coordinated online behavior.

    \item We compare monomodal and multimodal approaches to evaluate the advantages and limitations of integrating multiple data modalities, providing insights into the overall contribution of multimodal analyses for detecting and understanding coordinated online behavior.
    
    \item We analyze different implementations of multimodality, ranging from weakly integrated approaches with fewer constraints to strongly integrated ones that impose stricter requirements. This exploration sheds light on the trade-off between capturing broader, more inclusive patterns of coordination and identifying only the most tightly coordinated actors. Our findings clarify how varying levels of integration affect the results and offer guidance on balancing these competing priorities in robust multimodal analyses.
\end{itemize} \section{Related work}
\label{sec:mul_related_work}
Coordination detection methods can be classified into two main categories depending on their underlying approach: network science or machine learning. Network science approaches can be categorized into single-layer or multi-layer ones. The following sections discuss the existing solutions in each category.

\subsection{Machine learning-based approaches}

This category of methods applies data mining and machine learning techniques to extract patterns and insights from various data types, including text~\cite{stampe2024benchmarking}, images~\cite{stampe2023towards}, audio~\cite{mariconti2019you}, and interactions~\cite{sharma2021identifying,zhang2021vigdet}. However, only a few of these approaches extract multimodal features from the data or leverage multiple types of user activity. Despite the diversity of machine learning methods -- and the absence of a unifying framework comparable to that of network-based approaches -- they can generally be grouped into two broad categories: unsupervised methods, which cluster users based on behavioral similarities, and supervised methods, which classify users as coordinated or non-coordinated based on labeled examples.

Among unsupervised approaches, some methods model user activity as temporal point processes. For instance,~\citet{sharma2021identifying} introduced \textsc{AMDN-HAGE}, a generative model that combines activity traces (\textsc{AMDN}) with user group estimation (\textsc{HAGE}) to detect synchronized user groups -- assuming all detected coordination to be malicious. Similarly,~\citet{zhang2021vigdet} jointly learn user-group distributions by leveraging embeddings and temporal logic, clustering users through an expectation-maximization algorithm. In a different direction, \citet{erhardt2023detecting,erhardt2024hidden} represent user behaviors as interacting Markov chains within a discrete-time stochastic model to capture coordination dynamics. While~\citet{smith2025unsupervised} use Bayesian inference to identify groups of accounts that share similar
account-level characteristics and target similar narratives.

Supervised methods, on the other hand, rely on labeled data to identify coordinated users.~\citet{zhang2023capturing} address cross-platform coordination by using an auxiliary platform with known coordinated users to train neural time series encoders, followed by a multi-layer perceptron to detect similar behavior on a target platform.~\citet{mariconti2019you} detect coordinated hate attacks on YouTube by extracting features from video metadata, audio, and thumbnails, and feeding them into an ensemble classifier trained on content that had previously been targeted by coordinated raids. \

Ground-truth information is rarely available in coordinated behavior detection tasks, which limits the applicability of supervised learning approaches. Consequently, recent research has increasingly explored unsupervised machine learning methods -- often incorporating multimodal information -- to model user activity and detect coordinated patterns~\cite{zhang2023capturing, zhang2021vigdet, sharma2021identifying}. These methods typically rely on embedding the temporal activity of users into latent spaces that capture behavioral similarities. However, such approaches suffer from a lack of interpretability, making it difficult to understand or explain the mechanisms underlying detected coordination. This is a limitation, as in many cases the goal extends beyond merely identifying coordinated users to include the characterization of the behaviors and interactions that define their coordination~\cite{mannocci2024detection}.

\subsection{Network science-based single-layer approaches}
Most of the detection methods in the literature exploit the so-called coordination network, a graph where nodes represent users and edges represent a co-action, i.e., an action of the same type performed by two users. For instance, two users retweeting the same tweet perform a co-retweet, or posting a post with the same hashtag, a co-hashtag. Generally, the weight of an edge is proportional to the number of co-actions between the two users. Communities of users can be detected in the coordination network to identify groups of users that are more densely connected to each other than to the rest of the network. These communities are interpreted as groups of users that are coordinating their actions~\cite{mannocci2024detection}. The majority of the works in literature are \textit{monomodal} approaches, as they focus on a single type of co-action, such as \textit{co-retweet}~\cite{cima2024coordinated,nizzoli2021coordinated}, \textit{co-hashtag}~\cite{wang2023evidence}, and \textit{co-URL}~\cite{yang2025coordinated}. Many others construct several coordination networks, one for each co-action, to perform independent analysis and compare the results~\cite{pante2025beyond}. For instance, \citet{weber2021amplifying} analyze co-retweet, co-URL, co-hashtag, co-mention, and co-reply, and \citet{ng2023you} construct a co-URL and co-tweet coordination network, i.e., respectively, users posting similar posts and posts with the same URLs.

Although they perform analysis on different types of co-actions, they do not consider the interdependence between them. Indeed, a model based on the construction of a coordination network for each type of co-action does not capture the interplay between different types of co-actions. For instance,~\citet{dash2024decoding} refer to their analysis as a multimodal analysis, even if they detect the coordinated communities in each layer independently. With respect to Figure~\ref{fig:multimodality_tradeoff}, this work falls within the \textit{independent layers} approach.

\subsection{Network science-based multi-layer approaches}
As already discussed, online coordination is a complex phenomenon that can involve different types of actions, possibly simultaneously.
In this context, some works have proposed to model the coordinated behavior as a multiplex coordination network~\cite{magelinski2020detecting,magelinski2022synchronized}.
At a high level, a multiplex network is a type of network that consists of multiple layers, where each layer represents a different type of relationship or interaction between the same set of nodes. In the context of coordinated behavior, each layer represents a different type of co-action, and the multiplex network can capture the interplay between different types of co-actions. However, modeling the coordination as a multiplex network is not sufficient to capture the multimodality of coordinated behavior if the method used to detect coordinated communities does not fully exploit the multiplex nature of the network~\cite{dey2024coordinated,emeric2023interpretable,ng2022combined,ng2023you,graham2024coordination}. 

Indeed, most of the works declaring to exploit a multiplex network, in fact, flatten the multiplex network into a single-layer network before applying the community detection method -- \textit{union flattening} in Figure~\ref{fig:multimodality_tradeoff}. This is done by summing the weights of the edges in the different layers~\cite{ng2022combined}, by considering the union of the edges in the different layers~\cite{ng2023you,dey2024coordinated,luceri2023unmasking}, or by creating a multigraph, where all the layers are merged into a single layer~\cite{graham2024coordination}. This flattening step is motivated by the fact that most of the community detection methods are designed to work on single-layer networks and are not able to handle multiplex networks. However, this flattening step may lead to a loss of information, as it does not take into account the different types of co-actions. 

A couple of works leverage a multi-view clustering algorithm, which can handle multiplex networks without flattening them~\cite{magelinski2020detecting,magelinski2022synchronized}. However, the leveraged algorithm, proposed by~\citet{cruickshank2020multi}, optimizes the modularity in each layer, not considering the inter-layer modularity. This means that the clusters are not allowed to span between views or layers, and the modularity can be reduced to just the sum of the intra-layer modularities. This approach is different from the method proposed in this work, which allows communities to span multiple layers. Finally, a work by~\citet{emeric2023interpretable} proposes a method to detect coordinated behavior in multiplex networks by applying \louv on each layer and an iterative probabilistic voting consensus algorithm to achieve consensus clustering~\cite{emeric2023interpretable}. Also, this work does not allow communities to span multiple layers and first optimizes each layer independently.

Finally, the main limitation of the aforementioned works is not only the lack of a method that can fully exploit the multimodal nature of a multiplex coordination network, but also the lack of a comprehensive definition and evaluation of what multimodal coordinated behavior is and how its detection can be operationalized. 
To address this gap, we present the first comparative analysis of different ways to operationalize the detection of multimodal coordinated behavior, including an approach that fully leverages the multimodal structure of a multiplex coordination network. We also investigate the contribution of the multimodality compared to the monomodal approaches. Indeed, it is still an open question whether different modalities of co-actions provide distinct information for the analysis of coordinated behavior, or whether they yield equivalent results in terms of coordinated community discovered.

\section{Methodology}
\label{sec:mul_method}

\begin{figure}[!htbp]
    \centering
    \includegraphics[width=0.8\textwidth]{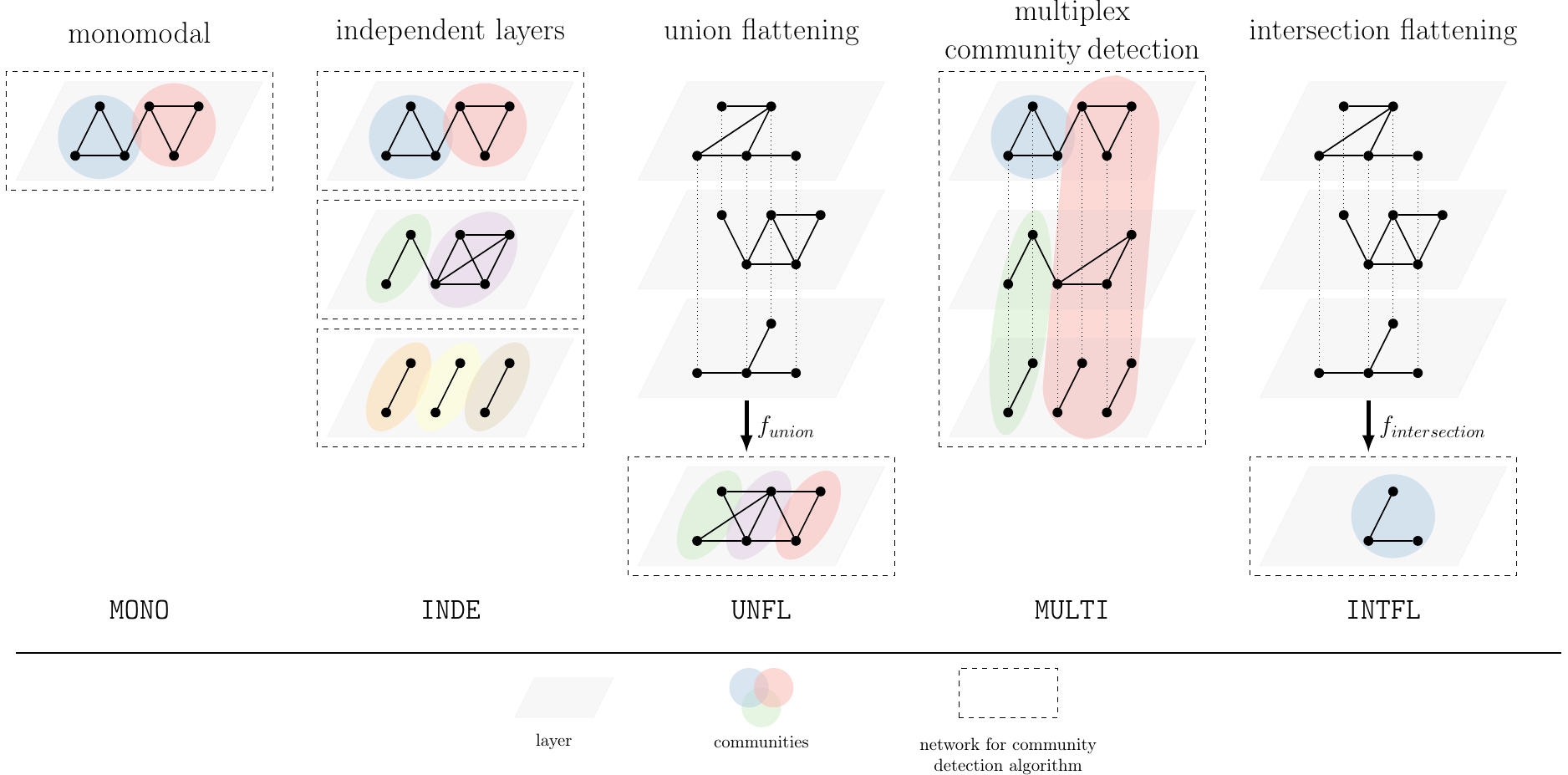}
    \caption{Different operationalizations of multimodality. The colored shapes are the detected communities, and the dashed boxes show on which network the community detection algorithm is performed.}
    \label{fig:multimodality}
\end{figure}

This work aims to investigate the role of multimodality in detecting coordinated behavior by systematically comparing monomodal and multimodal approaches. We first assess the contribution of monomodal methods, analyzing how each type of user action independently reveals coordination patterns. We then examine various ways to operationalize multimodality and assess how each contributes beyond what monomodal approaches can achieve. 
The framework proposed in this work builds on a network science approach, following the standard conceptual model for coordinated behavior detection~\cite{mannocci2024detection}.
This section is structured into three parts: the foundational network science framework behind our methodology, the detection methods for both monomodal and multimodal approaches, and the methodology for conducting the analysis needed to address the research questions.

\subsection{Preliminaries}
The analytical framework for coordinated behavior detection exploiting network science-based methods can be structured in four steps: user selection, coordination network construction, network filtering, and community detection~\cite{mannocci2024detection}.
The \textit{user selection} step identifies a subset $U' \subseteq U$ of users based on criteria aligned with the analysis goals, often focusing on the most active accounts, since a small fraction of users generates the majority of activity on social platforms~\cite{pei2014searching}. Common strategies include selecting super-producers, who post frequently~\cite{magelinski2020detecting,pacheco2021uncovering}, or super-spreaders, whose content is widely reshared~\cite{nizzoli2021coordinated,dimarco2024posthoc}.
Then, the second step involves the construction of the \textit{coordination network} that in the most general case is a multiplex network $G(V, E, W, L)$~\cite{magnani2021community}. To build $G$, one must define the types of co-actions to consider (e.g., re-shares, mentions, follows, etc.). When multiple types of co-actions are used, $G$ is a multiplex network with $|L|$ layers, where $L$ is the set of layers, each corresponding to a selected co-action. However, the majority of existing works leverage a single co-action, i.e., the number of layers is $|L|=1$ and $G$ is a single-layer network. Each layer $l \in L$ is a monomodal network in $G$ and is an undirected weighted graph $G^l(V^l, E^l, W^l )$, where $V^l \subseteq U'$, $E^l$ and $W^l$ respectively denote the set of nodes, edges, and weights of layer $l$. Given a layer $l$, an edge $e^l_{jk}$ is created if there exists a co-action of type $l$ between two users $(u_j, u_k)$. The edge weight $w^l_{jk}$ is obtained via a similarity function that computes pairwise user similarities. 
\textit{Network filtering} is an optional but important step used to retain only meaningful coordination signals while reducing network size for more efficient analysis. A common strategy removes edges with weights below a threshold and the resulting isolated nodes~\cite{luceri2023unmasking,emeric2023interpretable}. Another approach filters based on the timing of co-actions, keeping only those occurring within specific time windows -- either adjacent~\cite{vargas2020detection,weber2021amplifying} or overlapping~\cite{magelinski2022synchronized,pacheco2021uncovering}.
The final step detects coordinated communities through a \textit{community detection} algorithm on the coordination network. Multiplex networks are either flattened prior to detection~\cite{interdonato2020multilayer,weber2021amplifying} or analyzed using multiplex-specific algorithms~\cite{magnani2021community}.

\subsection{Operationalizations of multimodality}
\label{sec:op_mul}
Multimodality can be operationalized in various ways. In the following, we formalize the monomodal and several multimodal approaches that will be compared in the experimental section. Figure~\ref{fig:multimodality} illustrates these operationalizations, emphasizing the type of network used for community detection, as this step primarily distinguishes the different methods. \\

\textbf{Monomodal.} This implementation does not operationalize the multimodality, since coordinated behavior is detected by performing the community detection on a single-layer network, representing a single type of co-action. However, it plays a crucial role as a \textit{baseline} for evaluating and comparing multimodal approaches. The greatest disadvantage of this approach is that it analyzes only one type of co-action, leading to a loss of information. Formally, given a monomodal approach $l \in L$, a community detection algorithm produces a set of communities $C = \{C^l_1, C^l_2, \dots, C^l_{k}\}$, where each $C^l_i \subseteq G^l$. In the rest of the work, we refer to this approach as \mono. \\ 

\textbf{Independent layers.} This implementation detects coordinated behavior by identifying communities in each layer independently and then performing comparisons  between the detected communities. So it leverages multiple monomodal approaches, focusing on each type of co-action independently. This is the most relaxed form of multimodality that completely loses the interplay between the different types of co-actions. For each modality $l \in L$, a community detection algorithm identifies a set of communities, similarly to the monomodal approach. In the rest of the work, we refer to this approach as \ind. \\

\textbf{Union flattening.} Multimodality is operationalized by flattening the multiplex network into a single-layer network, by considering the union of the edges and nodes of the different layers. The aggregation of the weights can be done in different ways, for instance, the sum of the weights of the edges or the maximum weight of the edges. The choice of the aggregation method depends on the characteristics of the network and the purpose of the analysis. 
This operationalization implements a logical OR, requiring that edges exist in at least one layer of the network. For this reason, we also refer to it as \textit{union flattening}, which corresponds to considering users as coordinated if they engage in \textit{at least one} type of co-action. Formally, the multiplex network is collapsed into a single-layer network $G^{\cup} = (V, E, W^{\cup})$, where $E$ is the union of all intra-layer edges and $W^{\cup}$ is the set of aggregated weights across the layers. A community detection algorithm is then applied on $G^{\cup}$. In the rest of the work, we refer to this approach as \unfl. \\

\textbf{Multiplex community detection.}
Multimodality is operationalized by performing a community detection algorithm designed for multiplex networks. In the literature, several algorithms have been proposed for this purpose~\cite{magnani2021community,de2015identifying}, with the advantage of leveraging inter-layer connections, which are implicit in multiplex networks. In such networks, two nodes representing the same user across different layers are implicitly linked, allowing the algorithm to capture multimodal coordination. Formally, a community detection algorithm is applied to the full multiplex network, yielding a set of communities that might span multiple layers. In the rest of the work, we refer to this approach as \mul. \\

\textbf{Intersection flattening.}
Multimodality is operationalized by flattening the multiplex network into a single-layer network, by considering the intersection of the edges and nodes of the different layers. As well as the union flattening, the aggregation of the weights can be done in different ways. This operationalization represents the strictest form of multimodality and implements a \textit{logical AND}, requiring that edges appear in all layers of the network. Accordingly, we refer to it as \textit{intersection flattening}, as it considers users coordinated only if they engage in \textit{all} types of co-action.
Formally, the multiplex network is collapsed into a single-layer network $G^{\cap} = (V, E^{\cap})$, where $E^{\cap}$ is the intersection of all intra-layer edges. A community detection algorithm is then applied on $G^{\cap}$. In the rest of the work, we refer to this approach as \infl.

\subsection{Communities comparison}
In this section, we present the methodology used to address both research questions. This general approach applies to the analysis of the \textit{independent layers} operationalization -- by assessing the contribution of each monomodal approach -- and to the comparison between multimodal and monomodal approaches. To compare two operationalizations, our method measures the overlap between the communities they produce. Based on the degree of overlap, we identify whether communities and nodes are detected by only one of the two approaches or by both. This allows us to label communities and nodes as \textit{gained}, \textit{lost}, or \textit{common}.

\subsubsection{Overlapping communities}
\label{sec:mul_overlapping}

To quantify the degree of overlapping between the communities identified by two different approaches $A$ and $B$, we define the \textit{overlap matrix} $O$. 
Given the sets of communities detected by the two approaches  $C^A=\{C_1^A, \ldots, C^A_{k}\}$ and $C^B=\{C_1^B, \ldots, C^B_{k^\prime}\}$, we define the \textit{overlap matrix} $O = [o_{ij}] \in \mathbb{R}^{k^\prime \times k}$ where $o_{ij}$ denotes the \textit{harmonic mean overlap} at row $i$ and column $j$. We can observe, that the size of the two sets $C^A$ and $C^B$ might not be the same, i.e, $k^\prime \leq k$. Hence, $\forall i=1, \ldots, k^\prime, j=1, \ldots, k$ we define $o_{ij}$ as:
\begin{equation}
    o_{ij}^{AB} = \frac{2 \cdot  r_{ij}^{AB} \cdot  r_{ji}^{AB}}{ r_{ij}^{AB} +  r_{ji}^{AB}} ~~~~~~,
\end{equation}
where $r$ is defined as the \textit{relative overlapping}  of a community $i$ with respect to a community $j$. We remark that $r_{ij}^{AB}$ is different from $r_{ji}^{AB}$ and is defined as:
\begin{equation}
    r_{ij}^{AB} = \frac{|int_{ij}^{AB}|}{|C_i^A|}~~~~~~,
\end{equation}
where $int_{ij}^{AB} = C_i^A \cap C_j^B$ is the \textit{intersection} between two communities.

In case we need to compare \mono against \mul, we account for the fact that multiplex communities may span multiple layers. To this end, for each multiplex community $C$ we select only the subgraph  $C|_l$ involving the subset of nodes belonging to the specific layer $l$, which corresponds to the \mono approach, i.e., we select $C|_l = C_i \cap V^l$.

\subsubsection{Lost, common, and gained communities}
\label{sec:mul_com}
In this section, we analyze the overlap matrices computed in the previous step to classify communities as lost, common, or gained. To address the first research question, we examine whether the results of the \ind approaches differ or largely overlap, assessing the uniqueness of each modality. Similarly, we compare \unfl and \mul against \ind. The classification of communities as lost, gained, or common is inherently perspective-dependent, as it relies on which approach is chosen as the reference. When comparing two community detection methods -- such as a \mono and a multimodal one -- the same set of unmatched communities may be considered either lost or gained depending on the direction of the comparison. In our analysis, we adopt the perspective of the multimodal approach as the reference, since our goal is to evaluate its contribution relative to the monomodal baselines. By doing so, we can properly interpret which communities are newly identified by the multimodal view, which ones are retained from the monomodal ones, and which are no longer detected. This choice of perspective ensures that the labeling captures the added value of multimodal integration. Hence, in the following analysis, we designate the multimodal approaches as the reference method, denoted by approach $B$. Instead, when comparing the \ind approaches, the choice of the reference method is less critical, although a \mono approach must still be selected as the baseline for consistency. For instance, in this work, we adopt the co-retweet modality as the reference.

Formally, the first step is the execution of the Hungarian algorithm~\cite{kuhn1955hungarian}, which provides the subset of the couple of communities maximizing the sum of the harmonic mean overlap. Formally, let $\mathcal{P}=\{1, \dots, k\} \times \{1, \dots, k^\prime\}$ be the Cartesian product of the communities of the approach $A$ and $B$, the Hunagrian algorithm finds an assignment $\pi^*$ such that the total overlap is maximized, i.e., $\pi^* = \arg\max_{\pi \in \mathcal{P}} \sum_{(i, j) \in \pi} o_{ij}$, where $\pi^* \subseteq \mathcal{P}$ is a one-to-one matching. We must note that $\pi^*$ is a subset, since if $k > k^\prime$, not all communities of $C^B$ are matched. Each pair $(i, j) \in \pi^*$ represents a match between community $A_i$ and community $B_j$. Hence, we define the non-matching communities of approach 
$A$ as $A_{\text{unmatched}} = \left\{ i \in \{1, \dots, k\} \;\middle|\; \nexists j \in \{1, \dots, k^\prime\} \text{ such that } (i, j) \in \pi^* \right\}$ and similarly, for approach $B$, $B_{\text{unmatched}} = \left\{ j \in \{1, \dots, k^\prime\} \;\middle|\; \nexists i \in \{1, \dots, k\} \text{ such that } (i, j) \in \pi^* \right\}$.

Given a threshold $\theta \in [0, 1]$, we define two functions $\lambda^{lost}$ and $\lambda^{gain}$ to label respectively the sets of communities of $C^A$ and $C^B$ as follows:

\begin{equation}
    \lambda^{lost}(C_i^A, O) =
    \begin{cases}
        \text{common} & \text{if } \exists j \text{ s.t. } (i, j) \in \pi^* \text{ and } o_{ij}^{AB} \geq \theta \\
        \text{lost} & \text{if } \exists j \text{ s.t. } (i, j) \in \pi^* \text{ and } o_{ij}^{AB} < \theta \\
        \text{lost} & \text{if } i \in A_{\text{unmatched}}
    \end{cases}~~~~~~,
\end{equation}

\begin{equation}
    \lambda^{gain}(C_j^B, O) =
        \begin{cases}
        \text{common} & \text{if } \exists i \text{ s.t. } (i, j) \in \pi^* \text{ and } o_{ij}^{AB} \geq \theta \\
        \text{gained} & \text{if } \exists i \text{ s.t. } (i, j) \in \pi^* \text{ and } o_{ij}^{AB} < \theta \\
        \text{gained} & \text{if } j \in B_{\text{unmatched}}
    \end{cases}~~~~~~.
\end{equation}

\subsubsection{Lost, common, and gained nodes}
\label{sec:mul_nodes}
The community labeling methodology described in Section~\ref{sec:mul_com} effectively summarizes the overlap information, highlights differences between approaches, and offers a more intuitive interpretation of the results. However, it relies on an arbitrary threshold $\theta$ to distinguish between lost, common, and gained communities, which may influence the outcomes. To address this limitation, we also propose a node-level labeling method that classifies nodes as lost, common, or gained without requiring any threshold. While this approach avoids the arbitrariness of $\theta$, it only provides insights at the node level and does not capture information about entire communities.

Let $\pi^*$ be the optimal assignment obtained via the Hungarian algorithm, $A_{\text{unmatched}}$ the non-matching communities of approach $A$, and $B_{\text{unmatched}}$ for approach $B$. We define a function $\lambda^{node}$ to label the set of nodes $V$ as follows:
\begin{equation}
    \lambda^{node}(V, C_i^A, C_j^B, O) =
        \begin{cases}
            \text{common} & \text{if } \exists (i, j) \in \pi^* \text{ such that } v \in C_i^A \cap C_j^B \\
            \text{lost} & \text{if } \exists (i, j) \in \pi^* \text{ such that } v \in C_i^A \setminus C_j^B \\
            \text{gained} & \text{if } \exists (i, j) \in \pi^* \text{ such that } v \in C_j^B \setminus C_i^A \\
            \text{lost} & \text{if } \exists i \in A_{\text{unmatched}} \text{ such that } v \in C_i^A \\
            \text{gained} & \text{if } \exists j \in B_{\text{unmatched}} \text{ such that } v \in C_j^B
        \end{cases}~~~~~~.
\end{equation}

This classification applies to the same three comparison scenarios as defined for communities.
 
\section{Experimental setting}
\label{sec:mul_experiments}
In this section, we present the dataset utilized for the experiments, detail the preprocessing steps applied, and describe the implementation choices adopted for the method.\footnote{\url{https://zenodo.org/records/17709815}}

\subsection{Data}
The proposed method can be applied to any dataset that contains multiple types of social media actions, such as sharing, replying, or publishing posts with hashtags. 
In this study, we use two datasets: \uk from~\citet{nizzoli2021coordinated} and \russia from~\citet{seckin2025labeled}. \\

\textbf{\uk dataset.}
\uk dataset is collected via Twitter Streaming API during the 2019 UK General Election. The collection spans from 12 November to 12 December 2019, covering the month leading up to the election. It includes all tweets mentioning at least one of the most popular election-related hashtags, as well as all tweets from the official accounts of the two main parties and their leaders, along with the interactions (retweets and replies) those tweets received. The final dataset includes  a  total of  \num{11264820} tweets published  by  \num{1179659} distinct users.

This dataset is publicly available for research purposes\footnote{\url{https://doi.org/10.5281/zenodo.4647893}} and has been widely adopted in the study of coordinated online behavior~\cite{hristakieva2022spread,nizzoli2021coordinated,loru2023influence}. It has been shown to include several coordinated communities, making it an appropriate benchmark for our analysis. The data contains user interactions, retweet relationships, and temporal information. Previous research has primarily analyzed this dataset through the co-retweet network. In contrast, we extend the analysis to a multimodal representation that integrates multiple types of user interactions, enabling a deeper detection of coordination patterns beyond retweet activity alone. \\

\textbf{\russia dataset.}
The \russia dataset is part of a collection of 26 Twitter campaigns involving information operations (IOs), each containing posts verified by a social media platform as part of coordinated activity, along with accounts that discussed similar topics within the same time frames (control data). In this work, we focus on the largest campaign in terms of users, namely ``Russia\_1”. Posts and user information belonging to the IO campaign were collected from a public archive of state-sponsored information operations, made available through the platform’s transparency website. The authors then retrieved all hashtags used by IO accounts within each campaign to identify control accounts (\textit{ctrl}) discussing similar topics. These hashtags were used as search queries via the platform’s API to collect accounts that posted on the same dates and used the same hashtags as the IO accounts. Finally, the daily timelines of the control accounts were reconstructed by extracting up to 100 messages posted on the same dates as the IO accounts. The \russia dataset spans the period from 2009 to 2018 and includes, respectively, \num{3293} IO accounts and \num{31303} control accounts, with \num{1826344} and \num{1823859} posts. This dataset is publicly available for research purposes.\footnote{\url{https://zenodo.org/records/14189193}}

The \russia dataset presents several features that make it particularly interesting for our analysis. First, it provides two labeled groups that can be leveraged as ground-truth for coordinated behavior detection: IO accounts are considered coordinated, while control accounts are labeled as non-coordinated. Although this labeling is an approximation -- since IO campaigns typically rely on a coordinated core of malicious accounts to be effective~\cite{starbird2019disinformation} -- it is valuable in a domain where reliable ground-truth is scarce. This labeling strategy has already been adopted in previous studies~\cite{vargas2020detection, cima2024coordinated}.
Secondly, compared to the \uk dataset, which covers approximately one month, the \russia dataset spans nine years. Conducting experiments on datasets with such different temporal scales allows for an analysis of coordination patterns over both short and long time frames. While most studies in the literature focus on short- or medium-term coordination, the examination of long-term coordination is equally meaningful, as it captures the dynamic and evolving nature of coordinated behavior over time.

\subsection{Preprocessing}
Some data preprocessing steps are necessary for preparing the datasets for the proposed methods. In \uk dataset, we filter out the hashtags and mentions adopted for the data collection, as they would alter the results of the analysis, since we are searching for common actions. So the following hashtags have been filtered out: GE2019, GeneralElection19, GeneralElection19, GeneralElection2019, VoteLabour, VoteLabour2019, ForTheMany, ForTheManyNotTheFew, ChangeIsComing, RealChange, VoteConservative, VoteConservative2019, BackBoris, GetBrexitDone.
Likewise, the following mentions have been filtered out: jeremycorbyn, UKLabour, BorisJohnson, Conservatives. Instead, for the \russia dataset, information about the specific hashtags used by IO accounts to identify and collect control accounts is not publicly available.

In the literature, the domain is often considered instead of the full URL, as it provides a more generalized and meaningful representation of the source~\cite{giglietto2020takes}. Following this approach, for \uk dataset we extracted the domain from the URLs to focus on this higher-level identifier. 
Subsequently, the most frequently shared URLs are manually reviewed to filter out\footnote{The complete list of domain URLs removed is:
    \begin{inparaenum}[\itshape a\upshape)]
            \item \url{twitter.com},
            \item \url{cards.twitter.com},
            \item \url{youtube.com},
            \item \url{instagram.com}, 
            \item \url{open.spotify.com},
            \item \url{google.com},
            \item \url{reddit.com},
            \item \url{play.google.com},
            \item \url{bing.com},
            \item \url{google.co.uk}.
      \end{inparaenum}} those that, based on qualitative assessment within the specific context of the dataset, are not considered plausible targets of coordination, aligning with established methodologies and enhancing the interpretability of the analysis~\cite{luceri2023unmasking}.
In the \russia dataset, URLs have been hashed for privacy reasons, preventing the extraction of their domains. Therefore, in this case study, we rely on the complete hashed URL strings instead.

\subsection{Implementation choices}
In the following, we show the implementation choices regarding the framework presented in Section~\ref{sec:mul_method}. \\

\textbf{Network-science method.}
In the \textit{user selection} step, we aim to maximize the number of users while ensuring sufficient overlap across different actions, without incurring high computational costs. To this end, for \uk dataset, for each action type $l$, we select the top 5\% most active users, denoted as $U^\prime_l$, and define the final user set $U^\prime$ as the union of these subsets. While this does not guarantee that each user is active in all modalities, it enables us to retain more users compared to prior studies~\cite{nizzoli2021coordinated,dimarco2024posthoc}, while ensuring substantial inter-layer overlap and a shared user base for multimodal comparisons. Conversely, the \russia dataset is smaller in scale and computationally feasible to process. Thus, we retained all users for the analysis of such dataset.

The second step is the \textit{coordination network construction}. We focus on the most common actions (co-actions) on Twitter/X -- such as, co-retweet (\rtw), co-reply (\rpl), co-mention (\men), co-hashtag (\hash), co-URL (\urld). 
For both datasets, we build a multiplex network with five layers, each corresponding to a different co-action. Every user is represented by a vector capturing their activity: retweeted post IDs, replied-to tweet IDs, mentioned user IDs, hashtags, and shared URLs. In each layer, we calculate the edge weight between pairs of users using the cosine similarity of their TF-IDF-weighted vectors~\cite{mannocci2024detection}. This approach reduces the influence of widely shared or viral content and emphasizes the significance of less common items~\cite{nizzoli2021coordinated,dimarco2024posthoc}. We also compute the absolute overlapping between the vectors representing the users to have a measure of the number of common actions generating the edge. This measure is useful to understand the significance of the edge.

Moreover, in this work, for both datasets, we consider an \textit{overlapping} time window, evenly distributed in time. Specifically, for \uk of 6 hours with a shift of 5 hours and consequently a 1-hour overlap. This choice is motivated by the fact that we want to capture both spontaneous and malicious coordination, without focusing on a specific type of behavior. In this way, according to the period of reference (one month), we obtain around 147 time windows (the precise number can vary according to the action). 
In the case of the \russia dataset, we applied a 7-day time window with a 6-day shift, leading to a 1-day overlap and approximately 470 time windows. Given that this dataset covers 9 years, we used broader temporal windows to capture longer-term coordination dynamics.
In this work, we build a coordination network for each time window and then we merge them into a single network by averaging the weights of edges across time windows and summing those that appeared in multiple windows. The final coordination network is a multiplex network with 5 layers, one for each type of action.

\begin{table}[!t]
    \caption{Number of nodes and edges in the \uk coordination network before and after the filtering steps. The thresholds are set on the number of common actions generating the edge ($th_a$) and on the edge weights ($th_w$).}
    \label{tab:filtering_action_uk}
	\centering
    \setlength{\tabcolsep}{6pt}
    \scalebox{0.7}{
	\begin{tabular}{lllllllll}
	    \toprule
        & \multicolumn{2}{c}{\textit{no filter}} &  & \multicolumn{2}{c}{\textit{filtered on actions}} &  & \multicolumn{2}{c}{\textit{filtered on weight}} \\
        \cmidrule{2-3} \cmidrule{5-6} \cmidrule{8-9}
		\textbf{layer} & \textbf{nNodes} & \textbf{nEdges} & $\boldsymbol{th_a}$  & \textbf{nNodes} & \textbf{nEdges}& $\boldsymbol{th_w}$  & \textbf{nNodes} & \textbf{nEdges} \\
		\midrule
        \rtw & \num{58473} & \num{379668222} & 26  & \num{19311} & \num{3509326}  & 0.14  & \num{17200} & \num{1426822} \\ 
        \rpl  & \num{33533} & \num{38174756} & 4  & \num{17061} & \num{2663906} & 0.43 & \num{16186} & \num{1453235} \\
	    \urld & \num{43300} & \num{117456182} & 3  & \num{19695} & \num{4601506} & 0.88 & \num{18159} & \num{1494889} \\
        \men & \num{61177} & \num{1146693369} & 48 & \num{19929} & \num{6120082} & 0.16  & \num{15904} & \num{1298155} \\
        \hash & \num{58666} & \num{382965201} & 14	& \num{19973} & \num{3042391} & 0.25 & \num{18225} & \num{1478510} \\

		\bottomrule
	\end{tabular}
    }
\end{table}

\begin{table}[!t]
    \caption{Number of nodes and edges in the \russia coordination network before and after the filtering steps. The thresholds are set on the number of common actions generating the edge ($th_a$) and on the edge weights ($th_w$). The last three columns report the percentages of nodes filtered for the overall dataset and separately for the control (\textit{ctrl}) and coordinated (\textit{coord}) groups.}
    \label{tab:filtering_action_iorussia}
	\centering
    \setlength{\tabcolsep}{6pt}
    \scalebox{0.7}{
	\begin{tabular}{llllllllllll}
	    \toprule
        & \multicolumn{2}{c}{\textit{no filter}} &  & \multicolumn{2}{c}{\textit{filtered on actions}} &  & \multicolumn{2}{c}{\textit{filtered on weight}} & \multicolumn{3}{c}{\textit{filteringRates (\%)}} \\
        \cmidrule{2-3} \cmidrule{5-6} \cmidrule(r{2pt}){8-9} \cmidrule(l{2pt}){10-12}
		\textbf{layer} & \textbf{nNodes} & \textbf{nEdges} & $\boldsymbol{th_a}$  & \textbf{nNodes} & \textbf{nEdges} & $\boldsymbol{th_w}$  & \textbf{nNodes} & \textbf{nEdges} &  \textbf{\%overall} & \textbf{\%ctrl} & \textbf{\%coord} \\
		\midrule

        \rtw & \num{8090} & \num{104478} & 2  & \num{4604} & \num{55288}  & 0.14  & \num{3397} & \num{27566} &  76.85 & 81.63 & 41.21 \\

        \rpl  & \num{1409} & \num{5520} & 2  & \num{436} & \num{1652} & 0.58 & \num{203} & \num{826} &  97.66 & 99.57 & 66.40 \\
        
        \urld & \num{19603} & \num{3515318} & 2  & \num{2533} & \num{11393} & 0.27 & \num{2327} & \num{5698} &  88.14 & 89.51 & 77.35 \\
        
        \men & \num{11263} & \num{179360} & 2 & \num{6920} & \num{98862} & 0.12  & \num{4874} & \num{49478} &  74.02 & 78.68 & 34.71 \\
        
        \hash & \num{21038} & \num{1358160} & 7	& \num{4367} & \num{67102} & 0.25 & \num{3169} & \num{33512} &  85.77 & 90.32 & 43.14 \\

	\bottomrule
	\end{tabular}
    }
\end{table} 
\textit{Network filtering} step aims to filter out insignificant edges to highlight meaningful coordination, while preserving as much user overlap across layers as possible. For \uk dataset, first, edges generated by few common actions are removed, using action-specific thresholds chosen to retain up to \num{20000} nodes per layer, balancing coverage and computational cost. Then, a second filter is applied to edge weights, using the median weight in each layer as a threshold to retain a representative set of edges. This two-step process ensures both relevance and scalability, as detailed in Table~\ref{tab:filtering_action_uk}.

Similarly, for the \russia dataset, we applied the same two filtering procedures. In the first step, which relies on common actions, a lower threshold was adopted to account for the smaller network scale and to preserve an adequate number of nodes. For the second step, based on edge weights, we used the same criterion of \uk, applying the median weight within each layer as the filtering threshold. Table~\ref{tab:filtering_action_iorussia} summarizes the details of the \russia network filtering. Compared to the \uk dataset, it additionally reports the percentages of nodes filtered for the overall dataset and separately for the control (\textit{ctrl}) and coordinated (\textit{coord}) groups. We observe that the filtering process removes a larger proportion of control users than coordinated users.
\\

\textbf{Multimodal operationalizations.}
For all the multimodal operationalizations, we adopted two base methods, namely \louv~\cite{blondel2008fast} and \info~\cite{rosvall2008maps}. The implementation choices for the multimodal operationalizations described in Section~\ref{sec:op_mul} are the following:
\begin{itemize}
    \item \ind is implemented by performing \louv~\cite{blondel2008fast} and \info~\cite{rosvall2008maps} algorithms on each layer independently. For \louv, the resolution parameter is set to 1.
    \item \unfl is implemented by flattening the multiplex network into a single-layer network, with \textit{3 different strategies} of weight aggregation: 
    \begin{itemize}
        \item \unfl (nw): unweighted edges~\cite{dey2024coordinated,luceri2023unmasking,ng2023you},
        \item \unfl (ec): edge weight equal to the number of layers in which the edge is present,
        \item \unfl (sum): edge weight equal to the sum of the weights of the edges in the different layers~\cite{ng2022combined}.
    \end{itemize}
    We then apply the \louv (setting the resolution parameter to 1) and \info algorithms on the flattened network.
    
    \item \mul is implemented by applying the \glouv~\cite{mucha2010community} and \ginfo~\cite{de2015identifying} algorithms on the multiplex network. The first one adapts \louv for multiplex networks and maximizes modularity across all layers, accounting for both intra-layer and inter-layer structures. A key feature is the \textit{coupling parameter} $\omega$, which regulates the weight of inter-layer connections. This approach enables communities to span layers without requiring presence in all of them, offering a balanced form of multimodality. In this study, $\omega$ is fixed to enable fair comparison across different multimodal operationalizations. We set the resolution parameter $\gamma=1$ and the coupling parameter $\omega=0.1$. 
    Similarly, \ginfo adapts \info for multiplex networks. In this case, the weight of inter-layer connections is set to the default value $w_{inter}=0.15$.
    \item \infl is implemented by flattening the multiplex network into a single-layer network, considering the intersection of the edges and nodes of the different layers and the weight equal to the sum of the weights of the edges in the different layers. We apply \louv (setting the resolution parameter to 1) and \info on the flattened network.
\end{itemize}

\section{Results}
\label{sec:mul_results}
In this section, we present the results of the experiments conducted to evaluate the proposed approach. In the first section, we compare the layers of the coordination network, while in the second part, we analyze the different ways of operationalizing the multimodal coordinated behavior.  

\subsection{RQ1: Contribution of modalities}
\label{sec:co_action_comparison}
The first research question seeks to determine whether different modalities contribute equally to the understanding of coordinated behavior. If distinct modalities highlight different users or communities, it is evident that they convey unique information. However, even when the same users or communities emerge across modalities, their properties and characteristics may vary significantly. This variation could reveal unique patterns, roles, or dynamics within specific layers, underscoring the importance of analyzing the interplay between modalities to fully capture the multifaceted nature of coordination.

\subsubsection{Layers comparison}

\begin{table}[!t]
    \caption{Statistics of the \uk coordination network after applying the filtering steps.}
    \label{tab:info_network_uk}
	\centering
    \setlength{\tabcolsep}{6pt}
    \scalebox{0.8}{
	\begin{tabular}{clrrrrrrrrrr}
        \toprule
        & & \multicolumn{5}{c}{\textit{network's weight statistics}} & & & &\\
       \cmidrule{3-7}
        & \textbf{approach} & \textbf{mean} & \textbf{median} & \textbf{stdDev} & \textbf{max} & \textbf{min} & \textbf{nNodes} & \textbf{nEdges} & \textbf{nCC$^\dagger$} & \textbf{nCLV$^\ddagger$} & \textbf{nCIM$^{\ddagger\ddagger}$}\\
        \midrule
        \multirow{5}{*}{\rotatebox[origin=c]{90}{\textit{monomodal}}} & \rtw & 0.189 & 0.171 & 0.060 & 1.000 & 0.126 & \num{17891} & \num{1760094} & 64 & 100 & 65\\
        & \rpl & 0.610 & 0.584 & 0.121 & 1.000 & 0.451 & \num{16071} & \num{1334039} & 6 & 23 & 211 \\
        & \urld & 0.923 & 0.936 & 0.074 & 1.000 & 0.791 & \num{18741} & \num{2302080} & 117 & 134 & 351\\
        & \men & 0.173 & 0.151 & 0.072 & 1.000 & 0.111 & \num{18444} & \num{3047793} & 61 & 95 & 145\\
        & \hash & 0.362 & 0.335 & 0.101 & 1.000 & 0.246 & \num{18300} & \num{1520067} & 126 & 186 & 127\\
        
        \midrule
        \multirow{5}{*}{\rotatebox[origin=c]{90}{\textit{multimodal}}} & \unfl (nw) &  \multicolumn{1}{c}{-} &  \multicolumn{1}{c}{-} &  \multicolumn{1}{c}{-} &  \multicolumn{1}{c}{-} &  \multicolumn{1}{c}{-} & \num{39646} & \num{7303637} & 106 & 175 & 366 \\
         & \unfl (ec) & 1.364 & 1.000 & 0.688 & 5.000 & 1.000 & \num{39646} & \num{7303637} & 106 & 154 & 382 \\
         & \unfl (sum) & 0.596 & 0.537 & 0.371 & 3.971 & 0.111 & \num{39646} & \num{7303637} & 106 & 169 & 425 \\
         & \mul & \multicolumn{1}{c}{-} &  \multicolumn{1}{c}{-} &  \multicolumn{1}{c}{-} &  \multicolumn{1}{c}{-} &  \multicolumn{1}{c}{-} &  \multicolumn{1}{c}{-} &  \multicolumn{1}{c}{-} &  \multicolumn{1}{c}{-} & 180 & 124\\
         & \infl & 2.245 & 2.219 & 0.235 & 3.849 & 1.826 & 257 & 375 & 19 & 27 & 43\\
		\bottomrule
        \multicolumn{10}{l}{{\small $\dagger$ nCC: number of connected components}, {\small $\ddagger$ nComLV: number of communities with \louv}} \\
        
        \multicolumn{10}{l}{{\small $\ddagger\ddagger$ nComIM: number of communities with \info}}
	\end{tabular}
    }
\end{table}
 
\begin{table}[!t]
    \caption{Statistics of the \russia coordination network after applying the filtering steps.}
    \label{tab:info_network_iorussia}
	\centering
    \setlength{\tabcolsep}{6pt}
    \scalebox{0.8}{
	\begin{tabular}{clrrrrrrrrrr}
        \toprule
        & & \multicolumn{5}{c}{\textit{network's weight statistics}} & & & &\\
       \cmidrule{3-7}
        & \textbf{approach} & \textbf{mean} & \textbf{median} & \textbf{stdDev} & \textbf{max} & \textbf{min} & \textbf{nNodes} & \textbf{nEdges} & \textbf{nCC$^\dagger$} & \textbf{nCLV$^\ddagger$} & \textbf{nCIM$^{\ddagger\ddagger}$}\\
        \midrule
        \multirow{5}{*}{\rotatebox[origin=c]{90}{\textit{monomodal}}} & \rtw & 0.326 & 0.236 & 0.233 & 1.000 & 0.137 & \num{3397} & \num{27566} & 441 & 453 & 441\\
        
        & \rpl & 0.774 & 0.750 & 0.132 & 1.000 & 0.584 & \num{203} & \num{826} & 18 & 21 & 26 \\
        
        & \urld & 0.534 & 0.495 & 0.208 & 1.000 & 0.268 & \num{2327} & \num{5698} & 450 & 455 & 449\\
        
        & \men & 0.251 & 0.191 & 0.180 & 1.000 & 0.117 & \num{4874} & \num{49478} & 663 & 679 & 663\\
        
        & \hash & 0.431 & 0.362 & 0.187 & 1.000 & 0.246 & \num{3169} & \num{33512} & 348 & 361 & 348\\
        
        \midrule
        \multirow{5}{*}{\rotatebox[origin=c]{90}{\textit{multimodal}}} & \unfl (nw) &  \multicolumn{1}{c}{-} &  \multicolumn{1}{c}{-} &  \multicolumn{1}{c}{-} &  \multicolumn{1}{c}{-} &  \multicolumn{1}{c}{-} & \num{6921} & \num{77972} & 762 & 785 & 761 \\
        
         & \unfl (ec) & 1.502 & 1.000 & 0.721 & 5.000 & 1.000 & \num{6921} & \num{77972} & 762 & 785 & 762 \\
         
         & \unfl (sum) & 0.507 & 0.408 & 0.410 & 4.000 & 0.117 & \num{6921} & \num{77972} & 762 & 790 & 763 \\
         
         & \mul & \multicolumn{1}{c}{-} &  \multicolumn{1}{c}{-} &  \multicolumn{1}{c}{-} &  \multicolumn{1}{c}{-} &  \multicolumn{1}{c}{-} &  \multicolumn{1}{c}{-} &  \multicolumn{1}{c}{-} &  \multicolumn{1}{c}{-} & 798 & 762\\
         
         & \infl & 1.621 & 1.617 & 0.058 & 1.680 & 1.564 & 6 & 3 & 19 & 3 & 3\\
	\bottomrule
        
         \multicolumn{10}{l}{{\small $\dagger$ nCC: number of connected components}, {\small $\ddagger$ nComLV: number of communities with \louv}} \\

          \multicolumn{10}{l}{{\small $\ddagger\ddagger$ nComIM: number of communities with \info}}
	\end{tabular}
    }
\end{table}

\textbf{Statistics comparison.}
Table~\ref{tab:info_network_uk} and Table~\ref{tab:info_network_iorussia} respectively for \uk and \russia report key statistics for each implemented method, including edge weight statistics, the number of nodes, edges, connected components, and communities. \unfl (nw) contains empty values for the weight statistics because the flattened network is unweighted. Similarly, the \mul approach also presents mostly empty values across the statistics, as it operates on a multiplex network composed of five separate layers -- meaning that the relevant statistics are computed individually for each layer. In both datasets, among the layers, \urld and \rpl exhibit the highest mean edge weights, while \men has the lowest. \men has the highest number of edges, whereas \rpl has the fewest. Interestingly, \rpl has only 6 (\uk) and 18 (\russia) connected components, with one large connected component and a few small isolated ones. However, for all layers, a large connected component exists. 

The similar number of communities detected by \louv and \info on \russia dataset is likely due to its more strongly coordinated structure: state-linked IO campaigns tend to form dense, well-separated clusters, which may lead different detection strategies to converge on comparable partitions. In contrast, the \uk dataset is more heterogeneous, with diverse user types and short-lived interaction bursts during the election period. This weaker structural cohesion likely causes \louv to merge loosely connected regions, while \info captures finer-grained distinctions and consequently identifies many more communities, though we cannot be certain that this is the only factor driving the difference.

Therefore, in both datasets, the upper part of the tables compares different single-layer approaches, while the lower part presents the results of the multimodal approaches. This analysis offers an initial comparison of the different operationalizations of multimodality. Among them, \infl results in the lowest number of nodes and edges, whereas \unfl captures the highest. \infl significantly reduces the network size since it retains only edges present in all layers, identifying the fewest communities. In addition, the mean edge weight is significantly higher with respect to the other flattening approaches. On one hand, this suggests that \infl may capture extremely clear coordinated behaviors, exploiting all modalities at the same time. On the other hand, it is very selective, and it is the strictest operationalization of multimodality, limiting the amount of captured information compared to other methods. For this reason, our focus will shift to comparing \ind, \unfl, and \mul approaches. \\

\begin{figure}[!htbp]
    \centering
    \includegraphics[width=\textwidth]{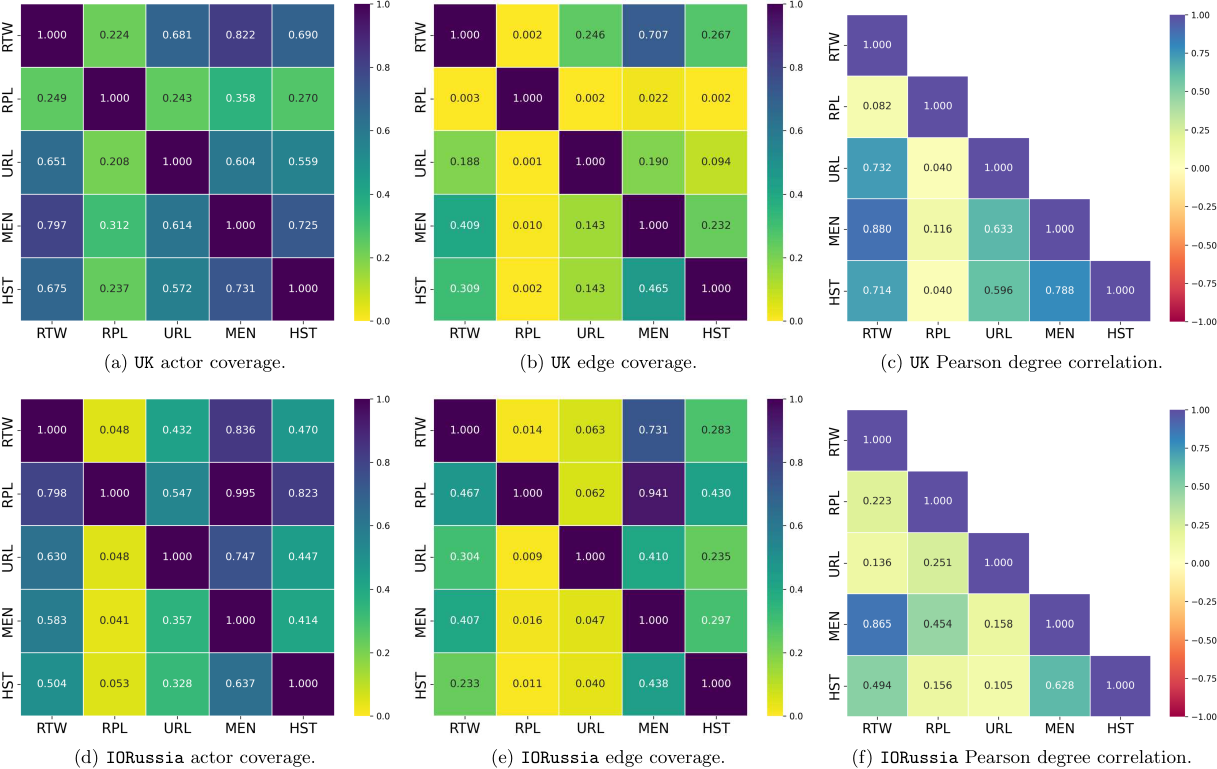}
    \caption{Metrics comparison between the layers of the UK and Russia coordination network. The computed metrics are on the actors and edges coverage, and Pearson correlation.}
    \label{fig:layers_comparison_uk}
\end{figure}

\textbf{Metrics comparison.}
In this analysis, we leverage metrics recommended by~\citet{brodka2018quantifying} for comparing the layers of a multiplex network: actor coverage, edge coverage, and Pearson correlation. Figure~\ref{fig:layers_comparison_uk} presents heatmaps comparing the layers, or modalities, of the coordination network. Each heatmap displays a pairwise comparison between two layers, with rows and columns representing the layers and cells containing the computed metrics.

    \textit{Actor coverage.} The coverage metrics assess the percentage of actors and edges shared between two layers, offering insights into the degree of overlap. These metrics are directional, meaning that the coverage $cov(i, j) \ne cov(j, i)$. Specifically, the value in cell $(i, j)$ represents the percentage of actors or edges in layer $i$ that are also present in layer $j$. The coverage values range from $0$ to $1$, where $1$ indicates complete overlap between the actors or edges in layer $i$ and those in layer $j$.  

    Rnoesults for \uk, shown in Figure~\ref{fig:layers_comparison_uk}a, indicate that \men has the largest values with both \rtw and \hash in both directions. This implies that the actors involved in \men are largely present in \rtw and \hash, and vice versa. Conversely, \rpl exhibits the smallest actor coverage with all other layers, signifying that actors in \rpl are less involved in the other layers. \urld, excluding \rpl, shows a moderate level of actor coverage with all other layers. These findings suggest that \rpl is the most distinct layer in terms of actor involvement, while \rtw and \men appear to be quite similar. In particular, the actor coverage values are comparable in both directions, reflecting the similar number of actors present in these two layers.
    
   Instead, results for \russia, depicted in Figure~\ref{fig:layers_comparison_uk}d, show that \rpl has high values of coverage with all the layers, except \urld, meaning that most of the actors involved in the other layers are also present in \rpl. But this was expected, since \rpl is very small in terms of the number of actors (see Table~\ref{tab:info_network_iorussia}). Indeed, in the other direction, \rpl has very small values of coverage with all the layers.  Also \men have high values of coverage in one direction with all the other layers. As for \uk, \men and \rtw still exhibit high actor coverage in both directions, indicating a strong overlap of actors between these two layers.

    \textit{Edge coverage.} Results for \uk, shown in Figure~\ref{fig:layers_comparison_uk}b, mirror those of actor coverage but exhibit some directional variations. For example, \rtw has an edge coverage of $0.707$ with \men, while \men has only $0.409$ with \rtw. This disparity arises because co-mention contains twice as many edges as \rtw, leading to greater representation of \rtw edges within \men. \hash also shares a significant proportion of edges with \men. In contrast, \rpl and \urld show the smallest edge coverage with all other layers. For \rpl, this is expected due to its low actor coverage. However, the small edge coverage of \urld is surprising, given its moderate actor coverage. This discrepancy might result from the unique nature of URL-sharing actions compared to other actions. Results for \russia, shown in Figure~\ref{fig:layers_comparison_uk}e, are quite similar to those of \uk. Indeed, \rtw and \rpl have high values of edge coverage with \men, but not vice versa,  due to the different number of edges in the layers.

    \textit{Pearson degree correlation.} The Pearson correlation evaluates the linear relationship between the degree of actors in two layers, with values ranging from $-1$ to $1$. A correlation of $-1$ indicates that high-degree actors in one layer are low-degree in the other and vice versa, while a correlation of $1$ indicates that high-degree (or low-degree) actors in one layer are also high-degree (or low-degree) actors in the other. Notably, this metric focuses on the number of incident edges for each actor-layer pair, without considering the specific adjacent actors.
    
    \uk results, shown in Figure~\ref{fig:layers_comparison_uk}c, reveal a strong similarity in actor degree distribution among \rtw, \men, and \hash, with the strongest correlation observed between \rtw and \men. In contrast, the weakest correlation is between \rpl and \urld, further highlighting their distinctiveness. 
    Similarly, \russia results, shown in Figure~\ref{fig:layers_comparison_uk}f, indicate that \rtw, \men, and \hash maintain a strong correlation in actor degree distribution, with \rtw and \men again exhibiting the highest correlation.
    \\
    
    These preliminary findings highlight on both datasets that \rpl, and to a lesser extent \urld, stand out as distinct layers, while \rtw, \men, and \hash show considerable similarity. This result underscores the complementary nature of the different layers: although they provide diverse information, actors tend to maintain similar roles across layers.

\subsubsection{Communities comparison} 
\label{sec:communities_comparison}

% \begin{figure}[!t]
%     \centering

%     \begin{subfigure}[b]{0.35\textwidth}
%         \centering
%         \includegraphics[width=1\textwidth]{./figure_5a.png}
%         \caption{\textit{NMI} \uk dataset.}
%         \label{fig:nmi_louvain_uk}
%     \end{subfigure}
% \hspace{0.5cm}
%     \begin{subfigure}[b]{0.35\textwidth}
%         \centering
%         \includegraphics[width=1\textwidth]{./figure_5b.png}
%         \caption{\textit{NMI} \russia dataset.}
%         \label{fig:nmi_louvain_iorussia}
%     \end{subfigure}

%     \caption{Normalized Mutual Information (NMI) heatmap comparing the communities detected with the \louv algorithm, considering only those with size greater than 200 for \uk, and 70 for \russia, across all pairs of co-actions.}
%     \label{fig:nmi_louvain}
% \end{figure}

\begin{figure}[!htbp]
    \centering
    \includegraphics[width=0.7\textwidth]{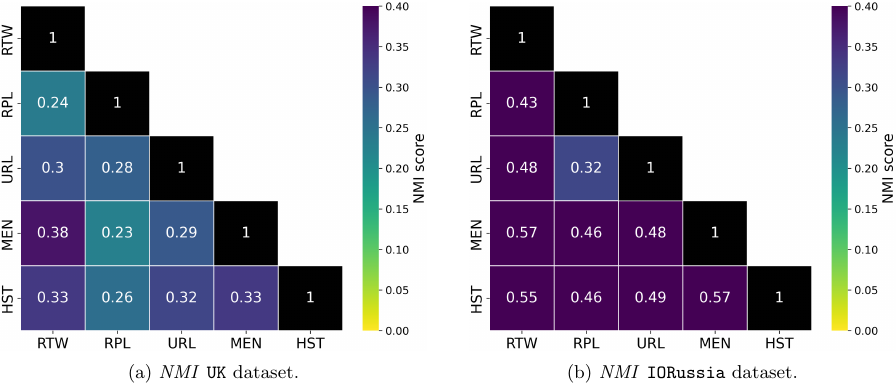}
    \caption{Normalized Mutual Information (NMI) heatmap comparing the communities detected with the \louv algorithm, considering only those with size greater than 200 for \uk, and 70 for \russia, across all pairs of co-actions.}
    \label{fig:nmi_louvain}
\end{figure}

 \textbf{Normalized Mutual Information.} 
We begin with a quantitative analysis to evaluate the similarity between the communities identified by different monomodal approaches. To achieve this, we use Normalized Mutual Information ($NMI$), a metric designed to compare two partitions of a set of nodes~\cite{liu2019evaluation}. $NMI$ quantifies the extent to which the community structures from different modalities align, providing a standardized measure of agreement between the partitions. This analysis allows us to assess how closely the community detection results from different monomodal approaches correspond to each other.
From now on, we consider the communities with more than 200 nodes for \uk and 70 nodes for \russia, for all couples of co-actions, and results obtained with \info are reported in Appendix~\ref{sec:app_infomap} for completeness.

Results for both datasets are reported in Figure~\ref{fig:nmi_louvain}. Specifically, Figure~\ref{fig:nmi_louvain}a shows results for \uk, which confirm that \rtw and \men are quite similar in terms of communities detected, reaching the maximum value equal to $NMI = 0.38$. The smallest values are reached by all co-actions with \rpl, which is in line with the previous results, showing that \rpl is the most different co-action in terms of detected communities. \russia results, shown in Figure~\ref{fig:nmi_louvain}b, present high values of NMI between all combinations of co-actions, except those between \rpl and \urld, which are lower. In general, the NMI values for \russia are higher than those for \uk, indicating a greater similarity in community structures across modalities in \russia. This suggests that coordinated behavior in \russia is more consistently captured across different modalities compared to the \uk.
\\

% \begin{figure}[!t]
%     \centering
%     \begin{minipage}{0.8\textwidth}
%         \centering
%         \includegraphics[width=0.5\linewidth]{./figure_6_legend_horizontal.png}
%         \label{fig:legend_starplot1b}
%     \end{minipage}

%     \begin{subfigure}[b]{0.35\textwidth}
%         \centering
%         \includegraphics[width=1\textwidth]{./figure_6a.png}
%         \caption{\rtw vs. \ind \textit{communities} (\uk).}
%         \label{fig:communities_flux_co_retweet_louvain_uk}
%     \end{subfigure}
% \hspace{0.5cm}
%     \begin{subfigure}[b]{0.35\textwidth}
%         \centering
%         \includegraphics[width=1\textwidth]{./figure_6b.png}
%         \caption{\rtw vs. \ind \textit{nodes} (\uk).}
%         \label{fig:users_flux_co_retweet_louvain_uk}
%     \end{subfigure}

%     \begin{subfigure}[b]{0.35\textwidth}
%         \centering
%         \includegraphics[width=1\textwidth]{./figure_6c.png}
%         \caption{\rtw vs. \ind \textit{communities} (\russia).}
%         \label{fig:communities_flux_co_retweet_louvain_iorussia}
%     \end{subfigure}
% \hspace{0.5cm}
%     \begin{subfigure}[b]{0.35\textwidth}
%         \centering
%         \includegraphics[width=1\textwidth]{./figure_6d.png}
%         \caption{\rtw vs. \ind \textit{nodes} (\russia).}
%         \label{fig:users_flux_co_retweet_louvain_iorussia}
%     \end{subfigure}

%     \caption{Bar charts for \louv approach showing the number of lost, common, and gained communities and nodes switching from \ind to \rtw.}
%     \label{fig:flux_co_retweet_louvain}
% \end{figure}

\begin{figure}[!htbp]
    \centering
    \includegraphics[width=0.7\textwidth]{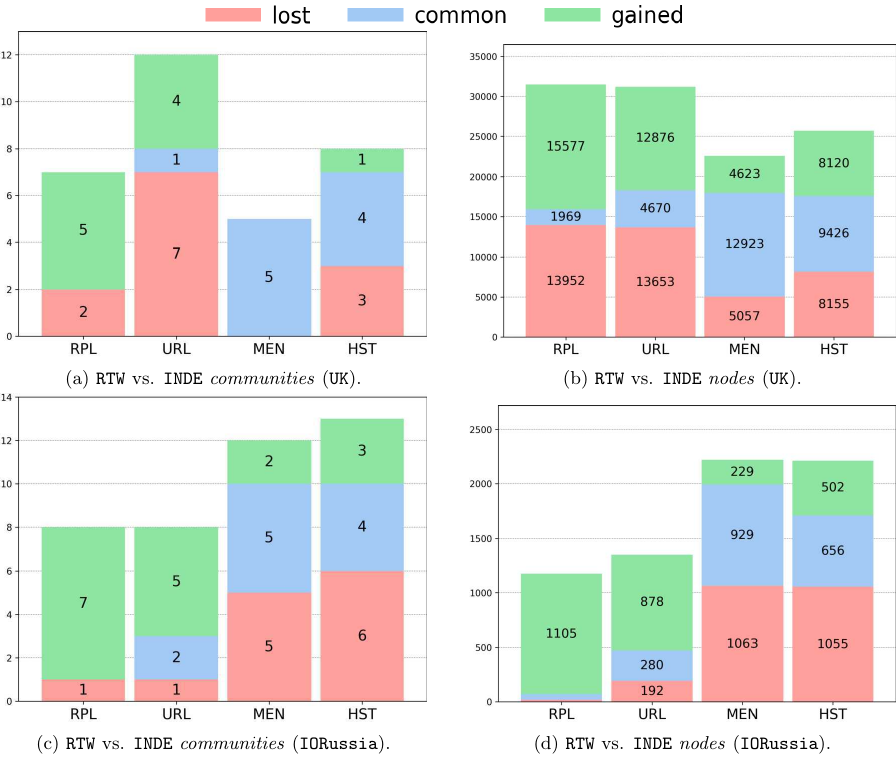}
    \caption{Bar charts for \louv approach showing the number of lost, common, and gained communities and nodes switching from \ind to \rtw.}
    \label{fig:flux_co_retweet_louvain}
\end{figure}

\textbf{Lost, common, and gained communities and nodes.} 
Based on the methodology described in Section~\ref{sec:mul_overlapping}, we compute the overlapping of all possible combinations of communities, with size greater than 200 nodes for \uk and 70 nodes for \russia. 
The matrices obtained from the analysis can be effectively visualized as heatmaps, providing an intuitive representation of the underlying relationships. Appendix~\ref{sec:app_heatmap_uk_louvain} and Appendix~\ref{sec:app_heatmap_uk_infomap} present, for \uk, the heatmaps corresponding to the comparisons performed on the communities detected using the \louv and \info approaches, respectively. Similarly, Appendix~\ref{sec:app_heatmap_iorussia_louvain} and Appendix~\ref{sec:app_heatmap_iorussia_infomap} report the corresponding results for \russia. Specifically, to answer the first research question \rtw vs. \ind, and the second research question \unfl vs. \ind and \mul vs. \ind. A comparison of the overlapping matrices of the three \unfl strategies reveals minimal differences among them. Therefore, we will focus on the \unfl (sum) strategy in subsequent analyses, as it retains the most information. Given the large number of results to present, we report only the aggregated outcomes for brevity.

Hence, following the methodology defined in Section~\ref{sec:mul_com}, we analyze the overlapping matrices of the communities identified by the different approaches, by labeling the communities as lost, common, and gained switching from one approach to another one. We set the threshold $\theta=0.5$, but we obtain similar results for  $\theta=0.7$ and  $\theta=0.9$. A higher threshold reduces the number of common communities, but the patterns observed remain still valid. We focus only on \rtw vs. \ind, as labeling communities with respect to different co-actions is simply a change in perspective, and \rtw is the most common and widely studied action in the literature~\cite{luceri2023unmasking}. 
Similarly, we analyze the lost, common, and gained nodes, according to the methodology defined in Section~\ref{sec:mul_nodes}. Figure~\ref{fig:flux_co_retweet_louvain} presents bar charts summarizing how the communities and nodes identified in the \ind approaches correspond to those found in the \rtw network. Specifically, communities and nodes are classified as lost, common, or gained, based on whether they are no longer detected, consistently identified, or newly discovered in \rtw compared to the \ind approaches. These results compare the different \mono approaches, allowing to understand if different co-actions provide different information. 

Results of \uk with \louv approach are shown in Figure~\ref{fig:flux_co_retweet_louvain}a, which highlights that \rpl and \rtw capture completely different structures: they have no communities in common, with \rtw identifying five new communities and losing the two detected by \rpl. Similarly, \urld and \rtw exhibit almost entirely distinct results, as \rtw loses seven communities compared to \urld and identifies four new ones. In contrast, \rtw and \men fully overlap, detecting the same set of coordinated communities and thus conveying identical information about user coordination. Finally, \rtw and \hash share four communities, while three communities are detected exclusively by \hash. This indicates that \rtw and \hash partially capture complementary aspects of the coordinated behavior, while still identifying some common structures.

Similarly, Figure~\ref{fig:flux_co_retweet_louvain}b shows that \rtw differs substantially from both \rpl and \urld in terms of the coordinated nodes it detects. However, when shifting the focus from communities to individual nodes, no \mono approach clearly emerges as more informative. Rather, the information gained by \rtw is approximately balanced by what it loses, indicating that each modality captures distinct yet equally important facets of coordination. Consistent with previous findings, \rtw remains strongly aligned with \men, while a similar but weaker alignment is observed with \hash.

\russia results for communities and nodes are shown in Figure~\ref{fig:flux_co_retweet_louvain}c and~\ref{fig:flux_co_retweet_louvain}d, respectively. Both with \louv and \info approach, \rtw shares some communities in common with \men and \hash, while it does not share any community with \rpl. Finally, Appendix~\ref{sec:app_infomap} reports results for \info approach. Figure~\ref{fig:flux_co_retweet_infomap} shows similar results for \info approach on \uk, with the only difference that \rtw and \men do not share all the communities in common, but \rtw loses three communities with respect to \men.

Both datasets and both community detection approaches consistently highlight that \rpl is a co-action that differs substantially from \rtw, whereas \men and \hash tend to share more information. However, no strong or systematic overlap emerges across the different co-actions. This implies that none of them can be excluded \textit{a priori} from a multimodal analysis: each co-action may contribute complementary information about coordinated behavior.

\subsubsection{Common communities characterization}
\label{sec:common_com_char}
While we have compared the community results across modalities in terms of node overlap, this does not ensure that the position or role of these communities within the network remains consistent across layers. Even if the same users are identified as part of the communities, their structural properties and interactions may vary depending on the modality. To address this, we compute a set of metrics to characterize the communities previously labeled as common across modalities. This analysis allows us to assess whether these communities exhibit similar structural and functional characteristics across different layers, providing deeper insights into the interplay between modalities. To conduct this analysis, we compute a set of metrics for each community identified as common across modalities (i.e., detected by both \rtw and another \mono approach). The metrics include size, density, average degree, average weight, average clustering coefficient, conductance, and assortativity. These metrics provide a comprehensive characterization of the communities, capturing their structural and functional properties within the network. This approach allows us to evaluate and compare the communities' characteristics across different modalities systematically. \\

% \begin{figure}[!t]
%     \centering
%     \begin{minipage}{0.8\textwidth}
%         \centering
%         \includegraphics[width=0.4\linewidth]{./figure_7_cosine_similarity_legend.png}
%         \label{fig:legend_cosine_similarity}
%     \end{minipage}

%     \begin{subfigure}[b]{0.35\textwidth}
%         \centering
%         \includegraphics[width=1\textwidth]{./figure_7a.png}
%         \caption{\uk dataset -- \louv.}
%         \label{fig:cos_sim_uk_louvain}
%     \end{subfigure}
% \begin{subfigure}[b]{0.35\textwidth}
%         \centering
%         \includegraphics[width=1\textwidth]{./figure_7b.png}
%         \caption{\russia dataset -- \louv.}
%         \label{fig:cos_sim_iorussia_louvain}
%     \end{subfigure}

%     \caption{Cosine similarity between the vectors of the communities detected with \louv in common with \rtw communities. The horizontal line reports the average cosine similarity for each layer's communities.}
%     \label{fig:cosine_similarity}
% \end{figure} 

\begin{figure}[!htbp]
    \centering
    \includegraphics[width=0.7\textwidth]{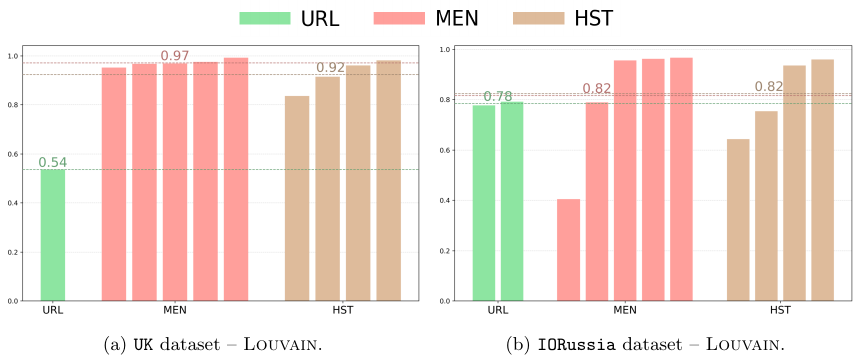}
    \caption{Cosine similarity between the vectors of the communities detected with \louv in common with \rtw communities. The horizontal line reports the average cosine similarity for each layer's communities.}
    \label{fig:cosine_similarity}
\end{figure}

For each pair of common labeled communities (as identified in Figure~\ref{fig:flux_co_retweet_louvain}a), we calculate the cosine similarity between their metrics. Higher cosine similarity values indicate greater similarity between the communities in terms of their structural and functional properties. Additionally, we compute the mean cosine similarity for the common communities within each layer. The results, presented in Figure~\ref{fig:cosine_similarity}, reveal that \men and \hash layers exhibit very high levels of cosine similarity, especially in \uk, suggesting strong alignment in the structural and functional characteristics of the detected communities. This aligns with the finding that \rtw shares a substantial number of communities with both \men and \hash, reinforcing the idea that these modalities capture closely related coordination patterns. In contrast, the \urld layer not only has very few communities in common with \rtw, but the only shared community also exhibits significant structural differences. This suggests that \urld captures a distinct form of coordination, diverging more from retweet-based interactions compared to mention and hashtag. Similar results are obtained also with \info, but \urld layer exhibits higher values.

\begin{figure}[!htbp]
    \centering
    \includegraphics[width=0.7\textwidth]{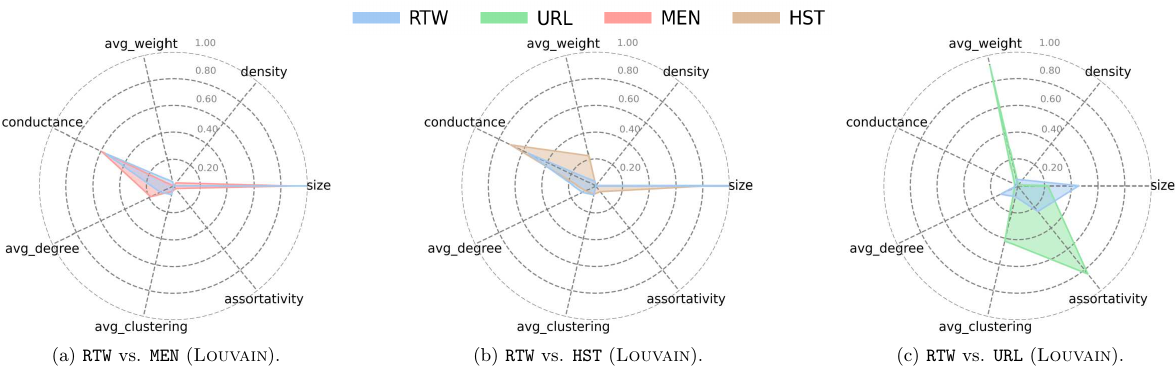}
    \caption{Example of metrics comparison of three couple of common communities between \rtw and \men, \hash, and \urld in \uk dataset.}
    \label{fig:metrics_starplot_uk_louvain}
\end{figure}

A more detailed representation of the common communities' similarity is presented by Figure~\ref{fig:metrics_starplot_uk_louvain}, which visualizes for \uk dataset with \louv approach, as a starplot, the normalized scores between 0 and 1 of the aforementioned metrics for three examples of couples of common communities.  The starplots are color-coded according to the modality to which each community belongs, with a distinct color representing each \mono approach. This visualization allows us to compare the metric scores of matched communities and assess the degree of similarity between them. If two modalities convey similar information, the matched communities will exhibit comparable metric scores, resulting in similar starplot shapes. Conversely, differences in the shapes of the radar plots indicate variations in the information captured by the modalities. This representation effectively highlights the similarities and discrepancies across modes in a clear and intuitive manner. As Figure~\ref{fig:flux_co_retweet_louvain}a shows, \rpl and \rtw do not have communities in common, and for this reason, there is no starplot for comparison. \men common communities exhibit similar values across all metrics, whereas \hash communities are characterized by a high average weight, and their starplot shape slightly differs from that of \rtw. Lastly, \urld, which shares only one common community with \rtw, presents a completely distinct starplot shape, with notably high values for both average weight and assortativity. Similar visualizations are obtained for \uk with \info approach, and with \louv and \info for \russia. They are all shown in Appendix~\ref{sec:app_infomap}.\\

% \begin{figure}[!t]
%     \centering
% \begin{minipage}{0.8\textwidth}
%         \centering
%         \includegraphics[width=0.5\linewidth]{./figure_8_RTW_starplot_legend.png}
%         \label{fig:legend_starplot_uk_louvain}
%     \end{minipage}
%     \vspace{0.3cm}
% \begin{subfigure}[b]{0.3\textwidth}
%         \centering
%         \includegraphics[width=0.9\linewidth]{./figure_8a.png}
%         \subcaption{\rtw vs. \men (\louv).}
%     \end{subfigure}\hfill
%     \begin{subfigure}[b]{0.3\textwidth}
%         \centering
%         \includegraphics[width=0.9\linewidth]{./figure_8b.png}
%         \subcaption{\rtw vs. \hash (\louv).}
%     \end{subfigure}\hfill
%     \begin{subfigure}[b]{0.3\textwidth}
%         \centering
%          \includegraphics[width=0.9\linewidth]{./figure_8c.png}
%          \subcaption{\rtw vs. \urld (\louv).}
%     \end{subfigure}
%     \caption{Example of metrics comparison of three couple of common communities between \rtw and \men, \hash, and \urld in \uk dataset.}
%     \label{fig:metrics_starplot_uk_louvain}
% \end{figure} 

\begin{figure}[!htbp]
    \centering
    \includegraphics[width=0.4\textwidth]{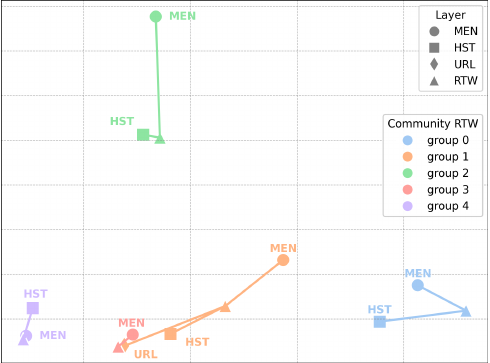}
    \caption{Common communities between \rtw and other co-actions in a two-dimensional space using PCA, for \uk with \louv approach. Each community is represented by a vector of metrics, i.e., size, density, average degree, average weight, average clustering coefficient, conductance, and assortativity. These high dimensional vectors are projected into a two-dimensional space using PCA. Communities are represented with distinct point shapes based on the co-action and are linked to their corresponding \rtw community. Each \rtw community may be associated with multiple communities from different co-actions, with all connected communities sharing the same color for clarity.}
    \label{fig:pca_monomodal_comparison}
\end{figure} 

The starplot visualization allows to have a quick understanding of the differences between the communities, but it is very detailed and does not provide a global view of the differences. Hence, we represent each community as a feature vector of the aforementioned metrics. Then the feature vectors are projected into a two-dimensional space using Principal Component Analysis (PCA). Figure~\ref{fig:pca_monomodal_comparison} shows, for \uk with \louv approach, the scatterplot in two PCA dimensions, where each community is linked with the one in common with \rtw and identified with a different point shape according to the co-action. Each \rtw community may be associated with multiple communities from different co-actions, and the cluster of connected communities are depicted with the same color. The PCA visualization highlights how each cluster of common communities -- same color -- occupies a distinct region of the plot. This spatial separation suggests that common communities of different \ind are structurally similar to each other while remaining distinct from non-common communities. This reinforces the idea that coordination patterns within common communities are preserved across different modalities. However, an exception emerges with the \urld layer, which appears more dispersed and does not conform to the same clustering pattern. This aligns with previous findings that \urld is the most distinct and unconventional layer, capturing a different form of coordination compared to the others. Appendix~\ref{sec:app_infomap} shows the PCA plots for \uk with \info approach, and for \russia with \louv and \info. In these cases, there are some groups of common communities that do not cluster together, indicating that the structural properties of these communities differ more significantly across modalities. This further emphasizes the variability in how different modalities capture coordinated behavior.

\subsection{RQ2: Multimodal approaches comparison}
The second research question seeks to examine how multimodal approaches contribute to the analysis of coordinated online behavior compared to monomodal approaches and how different implementations of multimodality affect the results. Similarly to the first research question, we exploit the definition of lost, common, and gained communities and nodes. 

\subsubsection{Communities comparison}
\label{sec:mul_communities_comparison}

\begin{figure}[!htbp]
    \centering
    \includegraphics[width=0.7\textwidth]{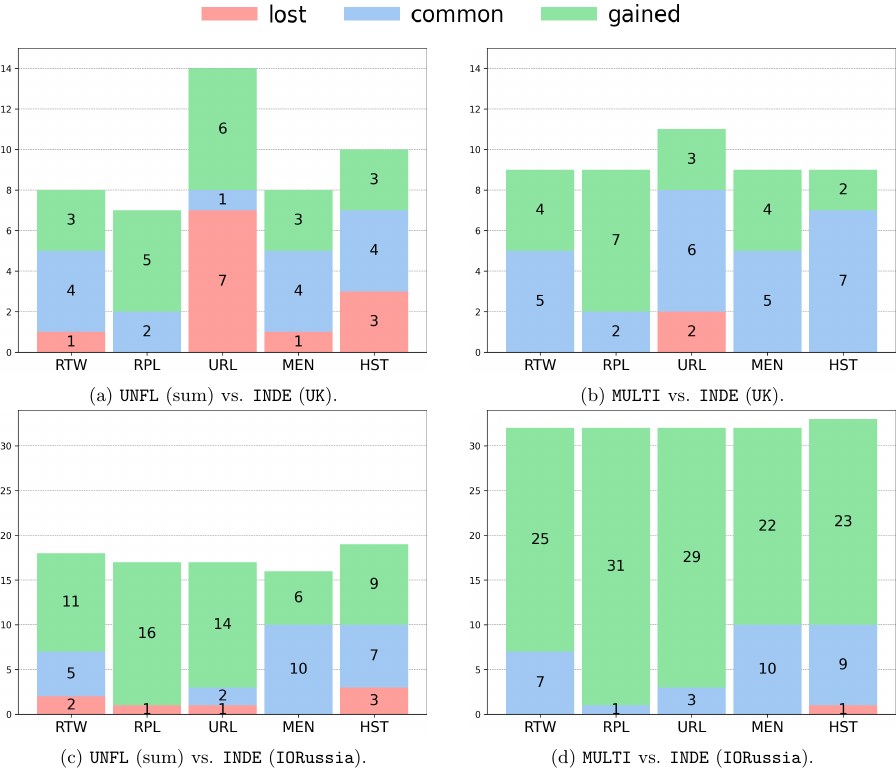}
    \caption{Bar charts showing in the number of lost, common, and gained \textit{communities} switching from \ind to \unfl (sum) and \mul (in \russia and \uk dataset).}
    \label{fig:communities_flux_multimodal_louvain}
\end{figure}
\begin{figure}[!htbp]
    \centering
    \includegraphics[width=0.7\textwidth]{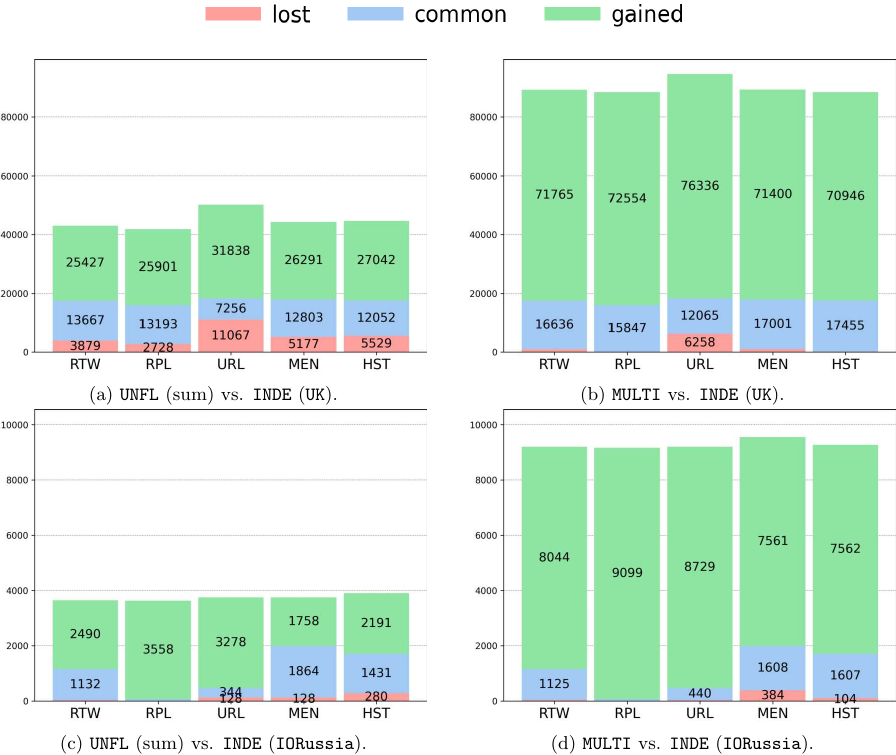}
    \caption{Bar charts showing the number of lost, common, and gained \textit{nodes} switching from \ind to \unfl (sum) and \mul (in \russia and \uk dataset).}
    \label{fig:users_flux_multimodal_louvain}
\end{figure}

We categorize communities as lost, common, or gained when exploiting \unfl and \mul with respect to \ind. We report the results for \uk and \russia with \louv approach in Figure~\ref{fig:communities_flux_multimodal_louvain}, which presents bar charts illustrating the number of communities across these categories, and in Figure~\ref{fig:users_flux_multimodal_louvain} that shows the labelling for the nodes. \\

\textbf{Multimodal approaches comparison.}
Both for \uk and \russia, Figure~\ref{fig:communities_flux_multimodal_louvain} shows that more communities are lost when switching to \unfl than to \mul. While this may seem counterintuitive given that \mul is more restrictive, this result underscores an important limitation of network flattening. Indeed, the process of flattening can lead to the loss of crucial network information and the introduction of noise in the network topology. This highlights the value of the \mul approach, which preserves the structure and information inherent in individual layers, unlike \unfl. These findings also call into question the efficacy of the commonly used \unfl strategy, which is prevalent in the literature~\cite{dey2024coordinated,luceri2023unmasking,graham2024coordination,ng2022combined,ng2023you}. Results obtained with \info (reported in Appendix~\ref{sec:app_infomap}) yield similar outcomes for \russia. Conversely, for \uk, \unfl shows only one lost community, whereas \mul identifies several. This demonstrates that the choice of the community detection algorithm can significantly influence the results. In this case, the \mul approach based on \ginfo does not perform well, as it detects very large communities, collapsing nearly all nodes into a few oversized clusters.\\

\textbf{Union flattening.}
Figure~\ref{fig:communities_flux_multimodal_louvain}a and \ref{fig:communities_flux_multimodal_louvain}c provide insights on the results of \unfl with respect to \ind. Hence, in \uk dataset, \unfl exhibits similar patterns with blocks of common and gained communities and relatively few lost ones with respect to \hash, \men, and \rtw. However, it shows a higher number of lost and gained communities compared with \urld, suggesting that its information is poorly captured in the flattened network. Finally, \unfl has two communities in common with \rpl and no lost communities but a significant number of gained ones, indicating that the detected communities differ substantially from those in the monomodal approach. Similar results are observed in \russia: the \louv approach shows a more stable situation, with only a few lost communities and many gained ones, whereas the \info-based approach loses a substantial number of communities compared to \ind. An exception emerges in \uk, where \unfl loses only one community (with respect to the \men layer) while detecting many common and gained communities relative to \ind. \\

\textbf{Multiplex community detection.}
Figure~\ref{fig:communities_flux_multimodal_louvain}b and~\ref{fig:communities_flux_multimodal_louvain}d demonstrate that \mul has only two and one lost communities with respect to all monomodal approaches. For \uk, the transition from the \rpl layer to the multimodal approach reveals a higher number of gained communities, while other co-actions primarily show common communities and fewer gained ones. This suggests that communities identified using the multimodal \mul approach retain much of the information from the monomodal approach while incorporating additional insights from the interplay between layers. Finally, in \russia, the number of gained communties by \mul approach is very high for all co-actions.\\

\textbf{Discussion.}
This analysis highlights the effectiveness of \mul in preserving existing information while uncovering new communities. Meanwhile, the \unfl is more prone to losing communities, further underscoring its limitations. The results suggest that \mul is more effective in capturing the underlying structure and information of the network, providing a more comprehensive view of coordinated behavior. The node-level analysis reveals similar trends, with \mul outperforming the \unfl in retaining and identifying nodes. An exception arises in the \uk dataset, where \unfl loses only one community and retrieves many common and gained ones, performing better than \mul in this specific case. However, this remains an isolated outcome. Overall, the results consistently show that \mul provides a more reliable and comprehensive view of coordinated behavior, while \unfl generally fails to capture important structural information. This comparison reinforces the importance of multimodal analysis for capturing the complexity of multiplex interactions, and underscores that the choice of community detection algorithm is equally crucial. \\

\begin{figure}[!htbp]
    \centering
    \includegraphics[width=\textwidth]{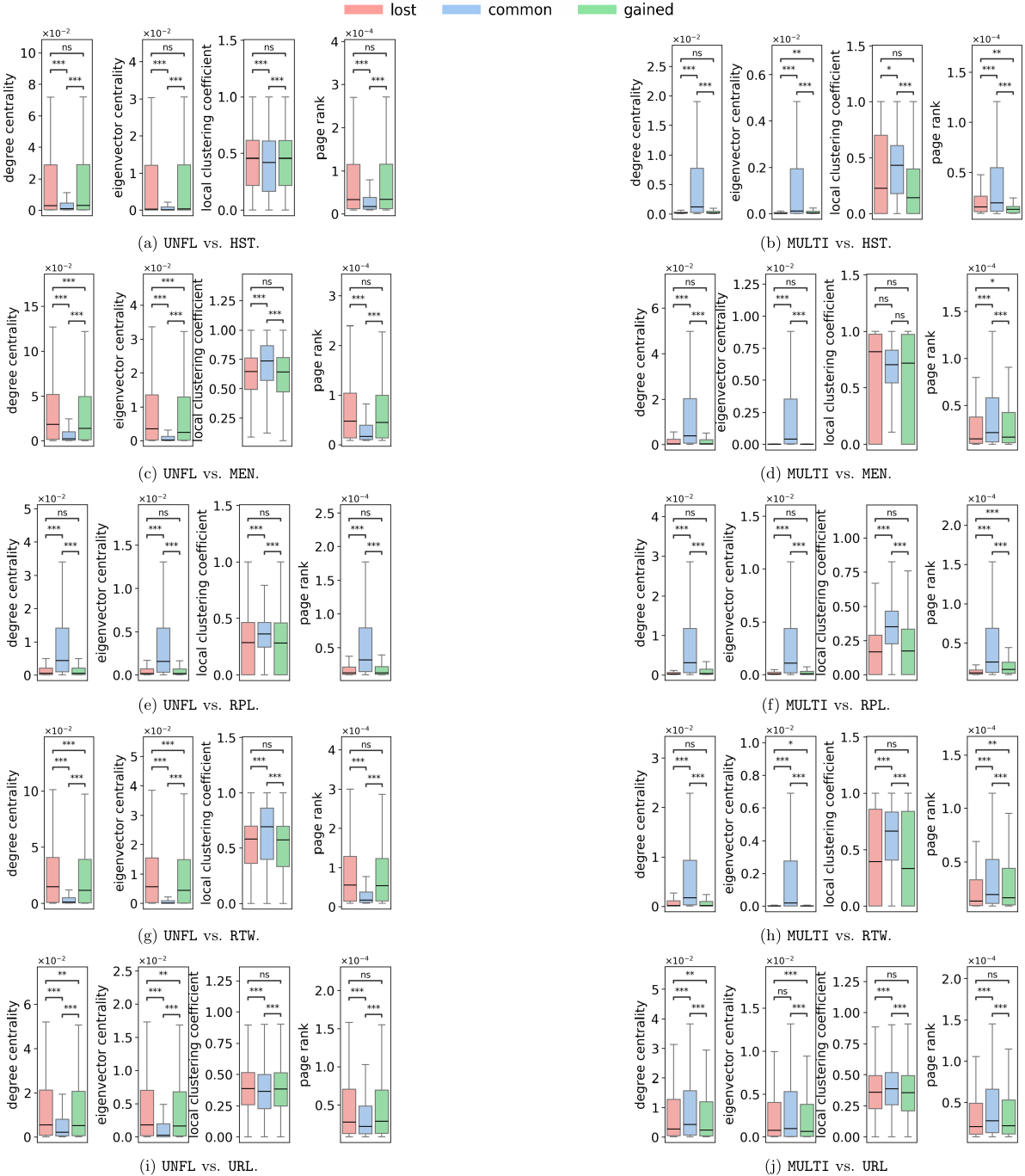}
    \caption{Lost, common, gained nodes characterization for \unfl and \mul (\louv) in \uk dataset. We use the notation \textit{ns} and * to indicate statistical significance levels. Specifically, \textit{ns} denotes that there is no significant difference between the groups, with a p-value in the range $p \geq 0.05$. For statistically significant results, we used the following annotations based on p-value ranges: *: $p \leq 0.05$; **: $p \leq 0.01$; ***: $p \leq 0.001$.}
    \label{fig:node_metrics_boxplot_uk_louvain}
\end{figure}
\subsubsection{Multimodal communities characterization}
\label{sec:mul_com_char}

To assess the effectiveness of multimodal approaches in identifying central nodes, we analyze the lost, common, and gained nodes defined in Section~\ref{sec:mul_nodes} and shown in Figure~\ref{fig:users_flux_multimodal_louvain}. This evaluation aims to determine whether multimodal methods retain or lose key influential nodes. Losing peripheral nodes is not critical, but missing highly connected or bridging nodes would indicate a limitation. Conversely, successfully preserving or discovering new influential nodes would highlight the advantage of multimodality in capturing a more complete network structure.
To conduct this analysis, we compute key centrality and structural metrics for each node including degree centrality, eigenvector centrality, local clustering coefficient, and PageRank. The comparison is performed for both \unfl and \mul. Figure~\ref{fig:node_metrics_boxplot_uk_louvain} illustrates, for \uk with \louv, the distribution of these metrics across lost, common, and gained nodes for each co-action. The results for \russia and for \info approaches are reported in Appendix~\ref{sec:app_infomap} for completeness. \\
We applied the Brunner-Munzel test to compare distributions across the lost, common, and gained node subsets for each metric. Unlike the Mann-Whitney U test, the Brunner-Munzel test does not assume equal variances or distribution shapes, making it more robust to ties, outliers, and variance differences, and thus more suitable for our analysis. The comparison highlights the performance of both the \unfl and \mul, offering insights into their ability to retain or identify influential nodes.  \\

\textbf{Union flattening.}
Common nodes are characterized by a higher degree centrality and eigenvector centrality in all combinations of datasets and approaches, with the exception of \uk with \louv, where lost and gained nodes exhibit higher values for \hash, \men, \rtw, and \urld. Instead, local clustering coefficient distribution is more homogeneous among the different co-actions, with some cases where common communities show higher values.

Therefore, in case of \hash, \men, \rtw, and \urld, in \uk dataset with \louv, \unfl seems to lose the most important nodes, which are the most central and influential in the network, but at the same time it is able to detect new nodes that are not detected by the monomodal approach and that are also central and influential. In all the other cases, \unfl is able to retain the most important nodes in the network, which are the common ones, but at the same time it loses and gains nodes that are less central and influential.  \\

\textbf{Multiplex community detection.}
In all combinations of datasets and approaches, observing \mul, the common nodes have higher values of degree centrality and eigenvector centrality, with respect to the lost and gained nodes. There are some exceptions, such as in \uk with \louv, in which \urld loses nodes that have higher values; or \rpl in \russia with \louv, where gained nodes have higher centrality values. Local clustering coefficient shows more similar average for the lost,  common, and gained nodes, but always with common nodes with higher values. 
Finally, page rank is on average similar for all the co-actions and for the lost, common, and gained nodes, but with a higher variance for the common nodes. 

Hence, degree and eigenvector centrality suggest that \mul is able to detect the most central and influential nodes in the network, which are the same detected by the monomodal approach. Also local clustering coefficient suggests that on average the nodes detected by \mul have similar distributions to the one lost. \\

\textbf{Discussion.}
From the perspective of node centrality, both \mul and \unfl approaches yield results that are generally consistent with those of the individual layers. Previous analyses showed that \mul retains most of the key nodes and communities from each modality, and tends to detect additional ones, that are typically less central. The unique exception is in \uk with \louv, where \unfl exhibits the opposite behavior: it both gains and loses a notable number of nodes and communities, but crucially, it often discards highly central ones while still identifying new, influential actors. This suggests a significant misalignment between the results of \unfl and those of the individual layers.

\subsection{Validation of coordinated behavior}
\label{sec:validation}

\begin{table}[!t]
    \caption{Validation of the detected communities in the \russia coordination network using \louv and \info community detection algorithms. For each layer and approach, we report the size of the largest three communities with size greater or equal to 70, along with the percentage of control and coordinated accounts in each community.}
    \label{tab:iorussia_validation}
	\centering
    \setlength{\tabcolsep}{6pt}
    \scalebox{0.7}{

	\begin{tabular}{llllllll}
	    \toprule
        & & \multicolumn{3}{c}{\textit{\louv}} & \multicolumn{3}{c}{\textit{\info}} \\
        \cmidrule(r{2pt}){3-5} \cmidrule(l{2pt}){6-8}

		& \textbf{approach} & \textbf{size} & \textbf{\%ctrl} & \textbf{\%coord} & \textbf{size} & \textbf{\%ctrl} & \textbf{\%coord} \\
		\midrule
        
\multirow{13}{*}{\rotatebox[origin=c]{90}{\textit{monomodal}}} &\multirow{3}{*}{\rtw} & 292 & 0.000 & 1.000 & 772 & 0.006 & 0.994 \\
        & & 234 & 0.000 & 1.000 & 292 & 1.000 & 0.000 \\
        & & 200  & 0.000 & 1.000 & 205 & 1.000 & 0.000 \\
        \cmidrule{2-8}
& \rpl & 70 & 0.000 & 1.000 & 86 & 0.000 & 1.000 \\
        \cmidrule{2-8}

&  \multirow{3}{*}{\urld} & 208 & 0.000 & 1.000 & 337 & 1.000 & 0.000 \\
        & & 182 & 1.000 & 0.000 & 214 & 0.000 & 1.000 \\
        & & 82  & 1.000 & 0.000 & 115 & 0.000 & 1.000 \\
        \cmidrule{2-8}

&  \multirow{3}{*}{\men} & 324 & 0.012 & 0.988 & \num{1437} & 0.207 & 0.793 \\
        & & 313 & 0.000 & 1.000 & 514 & 1.000 & 0.000 \\
        & & 305  & 0.961 & 0.039 & 162 & 1.000 & 0.000 \\
        \cmidrule{2-8}
& \multirow{3}{*}{\hash} & 318 & 0.000 & 1.000 & \num{1166} & 0.000 & 1.000 \\
        & & 316 & 0.000 & 1.000 & \num{559} & 0.998 & 0.002 \\
        & & 217  & 0.995 & 0.005 & \num{154} & 1.000 & 0.000 \\

\midrule
\multirow{15}{*}{\rotatebox[origin=c]{90}{\textit{multimodal}}} &  \multirow{3}{*}{\unfl (nw)} & 476 & 1.000 & 0.000 & \num{2068} & 0.228 & 0.772 \\
        & & 448 & 0.007 & 0.993 & \num{1761} & 0.999 & 0.001 \\
        & & 442 & 1.000 & 0.000 & \num{89} & 1.000 & 0.000 \\
        \cmidrule{2-8}

        & \multirow{3}{*}{\unfl (ec)} & 485 & 1.000 & 0.000 & \num{1763} & 0.999 & 0.001 \\
        & & 440 & 1.000 & 0.000 &  \num{1588} & 0.002 & 0.998 \\
        & & 421 & 0.971 & 0.029 &  \num{478} & 0.975 & 0.025 \\
        \cmidrule{2-8}

        & \multirow{3}{*}{\unfl (sum)} & 475 & 1.000 & 0.000 & \num{1761} & 0.999 & 0.001 \\
        & & 382 & 0.000 & 1.000 & \num{1588} & 0.002 & 0.998 \\
        & & 370 & 0.968 & 0.032 & \num{478} & 0.975 & 0.025 \\
        \cmidrule{2-8}

        & \multirow{3}{*}{\mul} & 1020 & 0.001 & 0.999 & \num{5012} & 0.194 & 0.806 \\
        & & 922 & 1.000 & 0.000 & \num{3745} & 1.000 & 0.000 \\
        & & 905 & 0.000 & 1.000 & \num{145} & 1.000 & 0.000 \\
	    \bottomrule
	\end{tabular}
    }
\end{table} The previous experiments focused primarily on assessing information that was lost, preserved, or gained across modalities, as well as on identifying signals of potential coordination. This section extends this analysis by directly validating the ability of the proposed methods to detect actual coordinated behavior, thereby strengthening the empirical support for our approach.

Validation in coordinated behavior detection is a crucial, yet challenging task due to the lack of ground truth data~\cite{mannocci2024detection}. However, the \russia dataset provides a unique opportunity for validation, as it includes labels distinguishing between coordinated and control accounts. Specifically, the \russia dataset is composed by the coordinated group of accounts involved in the information operation and the control group of accounts. This labelling enables the validation of the coordinated communities detected by the different approaches. In Table~\ref{tab:iorussia_validation}, we report the validation of the detected communities in the \russia coordination network. For \ind and multimodal approaches implemented with \louv and \info, we report the size of the largest three communities, along with the percentage of control and coordinated accounts in each community. A good separation between these two groups indicates that the approach is effective in identifying coordinated behavior. Indeed, in all the approaches the communities are mainly composed by either the coordinated accounts or the control accounts, with a very small percentage from the other group. This confirms the effectiveness of all the approaches in identifying coordinated communities.

% \begin{figure}[!t]
%     \centering

%     \begin{subfigure}[b]{0.35\textwidth}
%         \centering
%         \includegraphics[width=0.9\textwidth]{./figure_13a.png}
%         \caption{\louv number of coordinated communities.}
%         \label{fig:val_iorussia_louvain}
%     \end{subfigure}
% \hspace{0.5cm}
%     \begin{subfigure}[b]{0.35\textwidth}
%         \centering
%         \includegraphics[width=0.9\textwidth]{./figure_13b.png}
%         \caption{\info number of coordinated communities.}
%         \label{fig:val_iorussia_infomap}
%     \end{subfigure}

%      \caption{Number of coordinated communities lost, common, and gained detected by \unfl and \mul in \russia dataset.}
%     \label{fig:val_iorussia}
% \end{figure}

\begin{figure}[!htbp]
    \centering
    \includegraphics[width=0.7\textwidth]{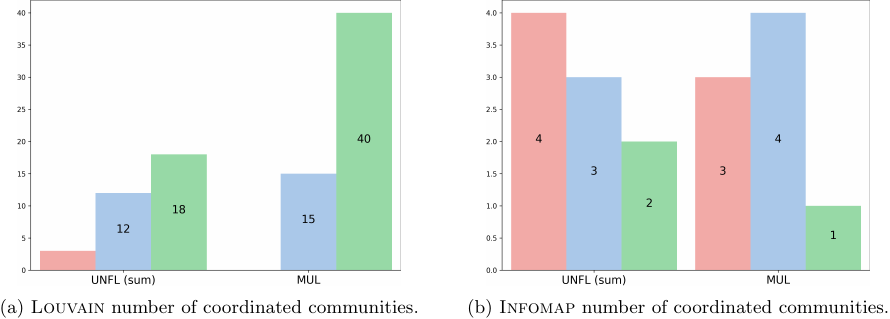}
    \caption{Number of coordinated communities lost, common, and gained detected by \unfl and \mul in \russia dataset.}
    \label{fig:val_iorussia}
\end{figure}
 
Since the distinction between coordinated and non-coordinated communities is well defined, we cross-reference this information with the labels lost, common, and gained. Our goal is to assess how many coordinated communities are lost or gained by the \unfl and \mul approaches.
Figure~\ref{fig:val_iorussia} reports the number of coordinated communities detected by \unfl and \mul (using both \louv and \info). Specifically, Figure~\ref{fig:val_iorussia}a shows the results for \louv, where \mul detects 40 coordinated communities -- substantially more than \unfl. \unfl and \mul detect, respectively, 12 and 15 coordinated common communities, showing similar detection power.
Figures~\ref{fig:val_iorussia}b shows the results for \info. Here, the number of coordinated communities are very similar across lost, common, and gained categories, with the only notable exception being that all three common communities detected by \unfl are coordinated.
Overall, \louv clearly performs better than \info, while \mul is generally more effective at identifying a larger number of coordinated communities. We can conclude that the choice of community detection algorithm has a strong impact on the final outcomes.

% \begin{figure}[!t]
%     \centering

%     \begin{subfigure}[b]{0.45\textwidth}
%         \centering
%         \includegraphics[width=1\textwidth]{./figure_14a.png}
%         \caption{\uk dataset -- \unfl.}
%         \label{fig:coord_uk_louvain_flat_weighted_sum_louvain}
%     \end{subfigure}
% \hspace{0.5cm}
%     \begin{subfigure}[b]{0.45\textwidth}
%         \centering
%         \includegraphics[width=1\textwidth]{./figure_14b.png}
%         \caption{\uk dataset -- \mul.}
%         \label{fig:coord_uk_louvain_multimodal}
%     \end{subfigure}

%      \begin{subfigure}[b]{0.45\textwidth}
%         \centering
%         \includegraphics[width=1\textwidth]{./figure_14c.png}
%         \caption{\russia dataset -- \unfl.}
%         \label{fig:coord_iorussia_louvain_flat_weighted_sum_louvain}
%     \end{subfigure}
% \hspace{0.5cm}
%     \begin{subfigure}[b]{0.45\textwidth}
%         \centering
%         \includegraphics[width=1\textwidth]{./figure_14d.png}
%         \caption{\russia dataset -- \mul.}
%         \label{fig:coord_iorussia_louvain_multimodal}
%     \end{subfigure}

%     \caption{Level of coordination of lost, common, and gained communities detected by \louv approach.}
%     \label{fig:coord_louvain}
% \end{figure}

\begin{figure}[!htbp]
    \centering
    \includegraphics[width=0.7\textwidth]{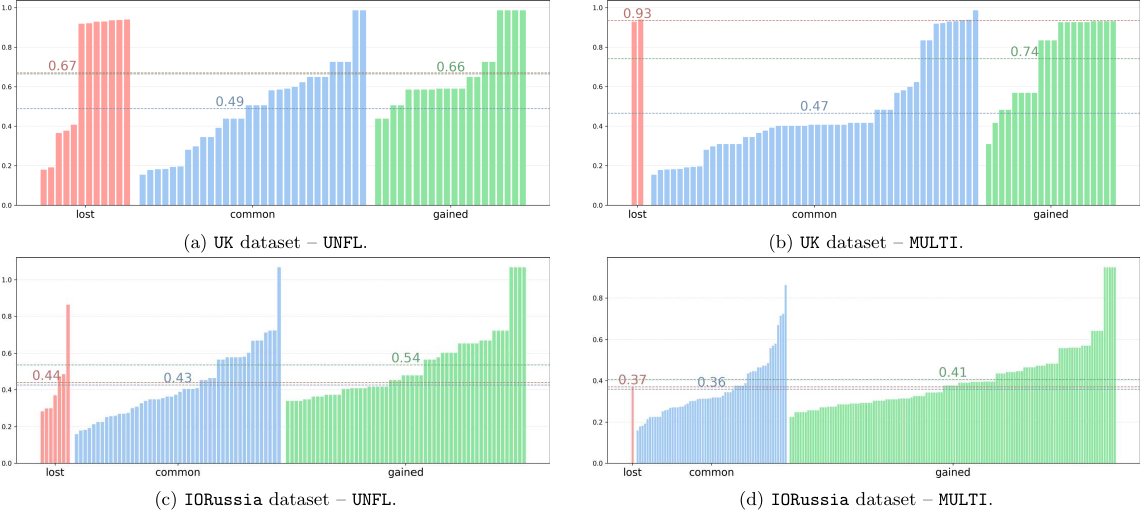}
    \caption{Level of coordination of lost, common, and gained communities detected by \louv approach.}
    \label{fig:coord_louvain}
\end{figure}

Until now, we validated the results only on \russia dataset since the labeling is available only for this dataset. However, we can also evaluate the results of the coordination detection through an internal measure, which is always available. Since the edge weights represent the level of coordination between two nodes, we can compute a coordination score for each community. Figure~\ref{fig:coord_louvain} and Figure~\ref{fig:coord_infomap} in Appendix~\ref{sec:app_infomap} show, respectively for \louv and \info approach, the average weight, i.e. coordination level for each community grouped by the lost, common, gained labeling. We  also report the mean of the averages for each label, indicating it with the horizontal line, colored according to the label.

In \uk, lost and gained communities detected by \mul exhibit higher coordination levels compared to those detected by \unfl, while common communities display similar coordination scores across the two approaches. This implies that, although \mul loses only two communities, they are highly coordinated ones. However, Figure~\ref{fig:coord_infomap}A shows that \unfl loses several communities with high coordination scores as well as others with low scores, resulting in an overall mean that lies in the middle. In \russia, the situation is more balanced across approaches. \unfl yields higher coordination levels for lost, common, and gained communities -- both with \louv and \info\ -- compared to \mul. Thus, in the \russia case, the advantage of \mul is less evident: the coordination levels of lost communities are not particularly low, and the coordination levels of gained communities are not especially high for either approach. 

\section{Discussion and conclusions}
We explored the role of multimodality in the detection and analysis of coordinated behavior, addressing its challenges and opportunities. To tackle this complexity, we introduced a novel methodology that integrates multiple types of user actions into a unified analytical framework for detecting multimodal coordinated behavior. By comparing monomodal approaches with flattened representations and the multiplex community detection, our analysis highlighted the strengths and weaknesses of each approach, offering valuable insights into the detection and characterization of coordination in social media.

\subsection{RQ1: Monomodal contributions}
The first research question investigates how different monomodal approaches contribute to the analysis of coordinated behavior. Our findings suggest that there is no universally dominant modality. As shown in Section~\ref{sec:communities_comparison}, \rtw and \men exhibit a high normalized mutual information score and identify fully overlapping communities, indicating strong agreement between these co-action layers. In contrast, the comparison between \rtw and \hash reveals a partial overlap—some communities are shared, while others are lost—highlighting moderate complementarity. Finally, \rtw shares no communities with \rpl and only one with \urld, demonstrating substantial divergence in the coordinated structures each layer captures.
Beyond metrics such as actor and edge coverage, which offer some anticipatory insights, it remains challenging to determine in advance which co-actions best reveal coordination. These findings underscore the importance of incorporating multiple co-action types to ensure that no significant form of coordination is overlooked.
Additionally, Section~\ref{sec:common_com_char} shows that even when monomodal approaches detect the same communities, these do not necessarily share similar structural characteristics. This is evident for \rtw and \urld, less so for \rtw and \hash, and absent for \rtw and \men. These results further emphasize the value of a multimodal framework, capable of integrating diverse and complementary signals of coordination.

\subsection{RQ2: Multimodal vs. monomodal and different operationalizations}
The second research question examines the contribution of multimodal approaches compared to monomodal ones in analyzing coordinated behavior. \unfl implemented with different flattening strategies produces similar outcomes, suggesting that the specific method of flattening has limited influence on the results (see heatmaps' results in Appendix~\ref{sec:app_heatmap_uk_louvain},Appendix~\ref{sec:app_heatmap_uk_infomap}, Appendix~\ref{sec:app_heatmap_iorussia_louvain}, and Appendix~\ref{sec:app_heatmap_iorussia_infomap} for empirical evidence). 
However, Section~\ref{sec:mul_communities_comparison} demonstrates that the act of network flattening itself significantly impacts the detection of coordinated communities. In particular, the \unfl approach shares only a subset of communities with \rtw, \men, and \hash, while failing to capture nearly all of those identified by \urld. This indicates a substantial loss of information contributed by each monomodal layer. In contrast, the \mul approach shows greater robustness: it retains almost all communities detected by monomodal methods and also uncovers additional coordinated structures. A notable exception emerges in the \uk dataset with \info-based approaches, where \unfl loses only one community and identifies many common and gained ones, performing better than \mul in this specific instance. However, this outcome remains isolated and does not generalize across datasets, but it highlights the importance of selection of the right community detection algorithm. Indeed, in literature \info is never used for coordinated behavior detection, while \louv is the most exploited one~\cite{mannocci2024detection}. These findings raise concerns about the use of network flattening in the literature and advocate for the adoption of more expressive, though complex, multimodal methods.
Section~\ref{sec:mul_com_char} further supports the advantage of \mul, highlighting its ability to preserve influential nodes while also identifying new ones. By comparison, also \unfl tends to detect the central nodes identified by \ind, but differently from \mul it is not able, in general, to detect new communities.
Overall, \mul proves to be a more reliable approach for capturing both the breadth and depth of coordinated behavior.

\subsection{Limitations and future works}
This work presents the first comparative analysis of methods to operationalize the detection of multimodal coordinated behavior. Indeed, we defined multiple operationalizations of multimodality, assessing the contributions of individual monomodal approaches, and evaluating the benefits of a multimodal framework over monomodal ones. We conducted experiments on two datasets, \uk and \russia, comparing methods built on two community detection approaches, \info and \louv. The primary limitation of this study is that the analysis was conducted only on Twitter/X platform, due to data availability. As a result, the generalizability to other social media platforms of our findings remains uncertain, and the conclusions drawn may be specific to the characteristics and dynamics of the observed cases. Future work will need to validate the proposed methodology across multiple platforms and datasets, ideally covering diverse social phenomena and contexts, to assess the robustness and consistency of our conclusions.

Our experiments focused on identifying information lost, preserved, or gained across modalities and on detecting signals of potential coordination. Section \ref{sec:validation} complements this by validating the ability of the proposed methods to identify actual coordinated behavior. Nonetheless, these results are not universally generalizable: this remains a methodological study, and broader claims will require applying the approach in multiple empirical settings and events. Further validation in diverse real-world scenarios is therefore necessary to establish its wider applicability.

Such efforts will be crucial to determine whether the observed patterns hold more broadly and to establish the general applicability of our multimodal framework. Given the lack of prior literature on multimodal coordination, we addressed two foundational research questions, necessarily leaving out additional operationalizations of multimodality. For instance, exploring alternative community detection algorithms, such as multilayer clique percolation and ABACUS, could introduce new levels of multimodality within the intrinsic trade-off of coordination analysis (see Figure~\ref{fig:multimodality_tradeoff})~\cite{magnani2021community}. Furthermore, future work could refine the integration across layers by systematically investigating the $\omega$ and $w_{inter}$ parameters, which control the strength of inter-layer connections between the same user across modalities, respectively in \glouv and \ginfo algorithms.
Our current study focuses on five common co-actions, but the framework could be expanded to incorporate additional modalities, such as multimedia content sharing, which plays a central role on many social media platforms.

Moreover, although our approach identifies coordinated behavior through community structures, we acknowledge that real-world coordination may not always appear as clearly separated groups. By definition, coordination still involves multiple users acting synergistically, but these groups may overlap, with users participating in different coordinated activities at the same time. Overlapping community detection methods could capture such patterns~\cite{magnani2021community}, although they have not yet been explored in coordinated behavior research. Extending our framework to overlapping structures is an interesting future direction, particularly in the multimodal setting, where coordination signals are heterogeneous and the problem becomes even more challenging.

Finally, combining this approach with methods that account for temporal dynamics would yield deeper insights into how coordinated behavior evolves over time~\cite{mannocci2024detection}. This would introduce the additional challenge of integrating temporal and multiplex network representations -- an ambitious yet promising direction for future research.

In conclusion, we provide a solid foundation for understanding multimodal coordinated behavior, advancing both methodological and practical approaches to its detection and analysis. By offering a flexible and expandable framework, this work contributes to the broader effort of uncovering coordinated activities in social media and opens the way for future research in this domain.

\section*{Acknowledgements}
This work was partly supported by SoBigData.it which receives funding from European Union – NextGenerationEU – National Recovery and Resilience Plan (Piano Nazionale di Ripresa e Resilienza, PNRR) – Project: ``SoBigData.it – Strengthening the Italian RI for Social Mining and Big Data Analytics” – Prot. IR0000013 – Avviso n. 3264 del 28/12/2021.; and by project SERICS (PE00000014) under the NRRP MUR program funded by the EU – NGEU; and by the European Union -- Next Generation EU, Mission 4 Component 1, for project PIANO (CUP B53D23013290006); and by the ERC project DEDUCE (Data-driven and User-centered Content Moderation) under grant \#101113826; and by the HORIZON Europe projects TANGO - Grant Agreement n. 101120763; and by the European Commission under the NextGeneration EU programme – National Recovery and Resilience Plan (PNRR), under agreements: PNRR - M4C2 - Investimento 1.3, Partenariato Esteso PE00000013 - "FAIR - Future Artificial Intelligence Research" - Spoke 1 "Human-centered AI"; and by eSSENCE, an e-Science collaboration funded as a strategic research area of Sweden.

\section*{CRediT authorship contribution statement}
\textbf{Lorenzo Mannocci}: Conceptualization, Data curation, Formal analysis, Investigation, Methodology, Software,
Validation, Visualization, Writing – original draft, Writing – review \& editing. \textbf{Stefano Cresci}: Conceptualization, Methodology, Supervision, Writing – original draft, Writing – review \& editing. \textbf{Matteo Magnani}: Conceptualization, Methodology, Supervision, Writing – review \& editing. \textbf{Anna Monreale}: Conceptualization, Methodology, Supervision, Writing – review \& editing . \textbf{Maurizio Tesconi}: Conceptualization, Funding acquisition, Methodology, Project administration, Resources, Supervision, Writing – review \& editing.

\bibliographystyle{elsarticle-num-names} 
\bibliography{mybib}
\clearpage

\newcommand{\authorbio}[3]{\vspace{1em}
\noindent
\begin{tabularx}{\textwidth}{@{}p{0.15\textwidth}@{\hspace{1em}}X@{}}
  \includegraphics[width=\linewidth]{#1} &
  \vspace{-6em}\raggedright\scriptsize\textbf{#2} #3 \\
\end{tabularx}
}

\authorbio{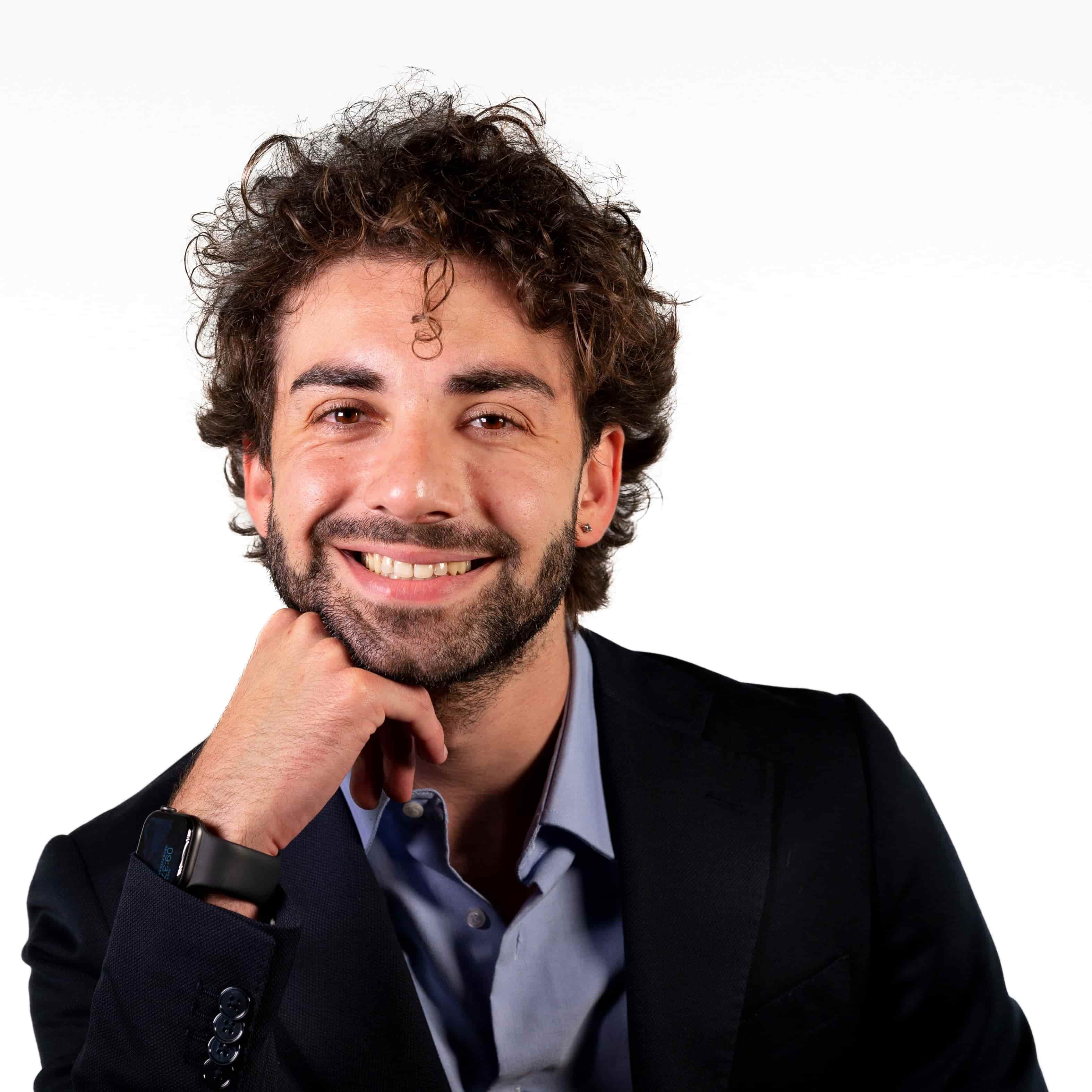}
  {Lorenzo Mannocci}
  {received the Phd.D. degree in the Italian National PhD in Artificial Intelligence at the University of Pisa and IIT-CNR.Currently, he is a Research Fellow on the european TANGO project on Hybrid Artificial Intelligence models for detection of (in)authentic behaviour and content. He is interested in the overlap between web science and data science, particularly in the study of bot detection and coordinated online behavior. He also conducts research in Explainable AI and Human-Computer Interaction, with a focus on using RAG-based systems.}

\authorbio{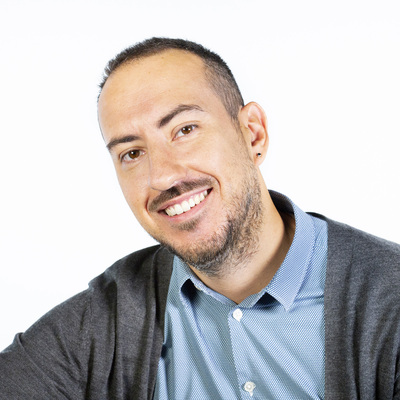}
  {Stefano Cresci}
  {is a Researcher at the Institute for Informatics and Telematics of the National Research Council (IIT-CNR) in Pisa, Italy. His interests lay at the intersection of Web and data science, with a focus on content moderation and human-centered AI. On these topics, he published more than 100 peer reviewed articles in venues such as PNAS, KDD, IJCAI, WebConf, ICWSM, and CSCW. Currently, he leads a prestigious ERC project on data-driven and personalized content moderation. He has received multiple awards including the IEEE Next-Generation Data Scientist Award and the ERCIM Cor Baayen Young Researcher Award.}

\authorbio{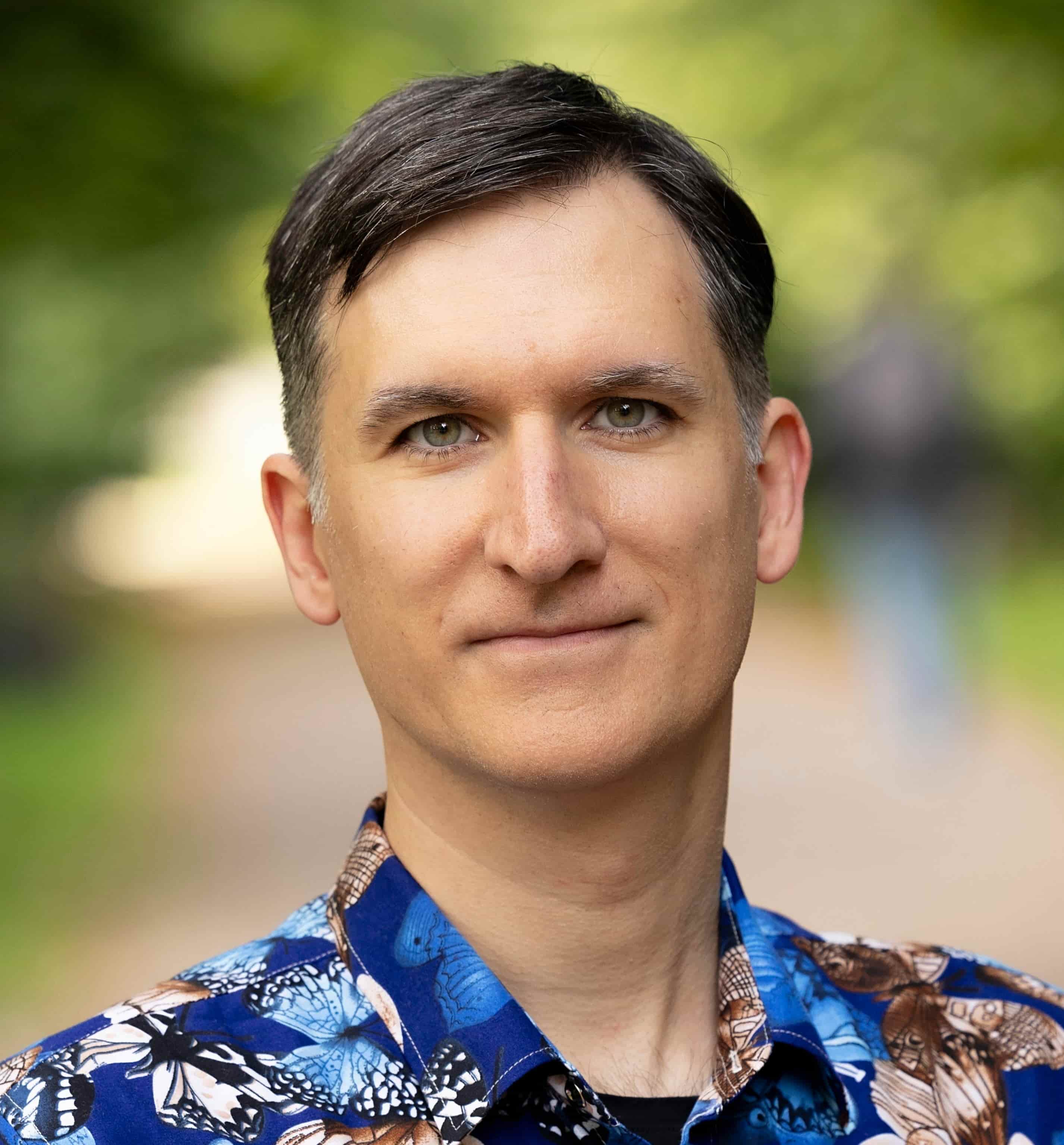}
  {Matteo Magnani}
  {is professor of data science at the Department of Information Technology, Uppsala University, where he leads the Uppsala Information Laboratory (InfoLab). Matteo’s research is on computational methods for the social sciences, in particular social media analysis, (multilayer) social networks, (visual) political communication, and social cybersecurity.}

\authorbio{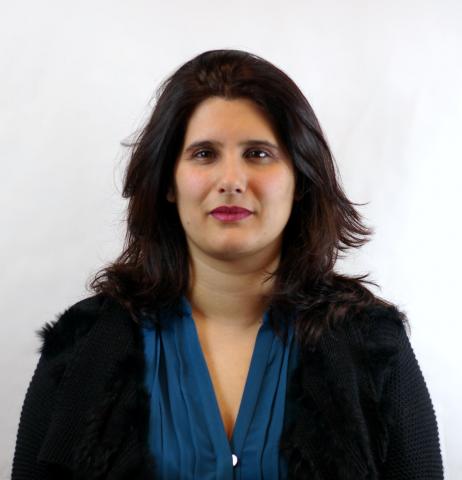}
  {Anna Monreale}
  {is an Assistant Professor at the Computer Science Department of the University of Pisa and a member of the KDD LAB. She has been a visiting student at the Stevens Institute of Technology, USA (2010). Her research interests include big data, social networks analysis, legal and ethical issues rising in mining social and human data. She earned her Ph.D. in Computer Science from the University of Pisa in June 2011 with a dissertation on privacy-by-design in data mining. She is a Privacy-by-Design Ambassador and was a member of the EU Panel of Experts on the Open Science Cloud.}

\authorbio{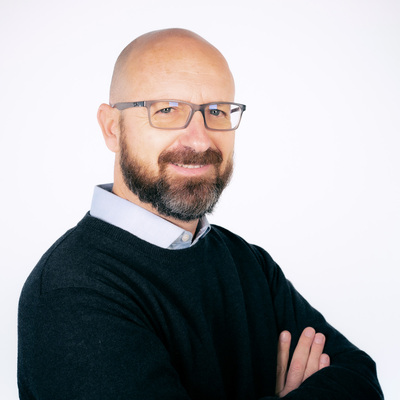}
  {Maurizio Tesconi}
  {(PhD) is a researcher in Computer Science, heading the Cyber Intelligence Lab at the CNR Institute of Informatics and Telematics. His core research areas include Artificial Intelligence, Big Data, Web Mining, and Social Network Analysis for Open Source Intelligence. Maurizio is a prolific author, with numerous publications in prestigious international journals and conferences. He is also a member of the permanent team of the European Laboratory on Big Data Analytics and Social Mining (SoBigData). He further leads impactful projects for public administrations, focusing on AI-driven solutions for secure societies.}
 \clearpage

\appendix

\section{Additional results}
\label{sec:app_infomap}
\renewcommand{\thefigure}{A.\arabic{figure}}
In this section, we present all results for the \info approaches. We also report the characterization of common communities, including the starplots (Figure~\ref{fig:metrics_starplot_iorussia}) and the PCA visualization (Figure~\ref{fig:pca_monomodal_comparison_iorussia}) obtained with \louv on \russia.

\setcounter{figure}{0}

% \begin{figure}[!ht]
%     \centering
%     \begin{subfigure}[b]{0.3\textwidth}
%         \centering
%         \includegraphics[width=1\textwidth]{./figure_A1a.png}
%         \caption{\textit{NMI} \uk dataset.}
%         \label{fig:nmi_infomap_uk}
%     \end{subfigure}
% \hspace{0.5cm}
%     \begin{subfigure}[b]{0.3\textwidth}
%         \centering
%         \includegraphics[width=1\textwidth]{./figure_A1b.png}
%         \caption{\textit{NMI} \russia dataset.}
%         \label{fig:nmi_infomap_iorussia}
%     \end{subfigure}

%     \caption{Normalized Mutual Information (NMI) heatmap comparing the communities detected with the \info algorithm, considering only those with size greater than 200 (for \uk) and 70 (for \russia), across all pairs of co-actions.}
%     \label{fig:nmi_infomap}
% \end{figure}

\begin{figure}[!htbp]
    \centering
    \includegraphics[width=0.7\textwidth]{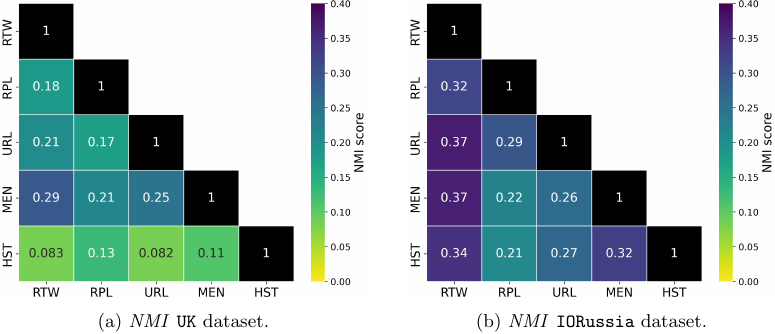}
    \caption{Normalized Mutual Information (NMI) heatmap comparing the communities detected with the \info algorithm, considering only those with size greater than 200 (for \uk) and 70 (for \russia), across all pairs of co-actions.}
    \label{fig:nmi_infomap}
\end{figure}

%  \begin{figure}[!ht]
%     \centering
%     \begin{minipage}{0.8\textwidth}
%         \centering
%         \includegraphics[width=0.5\linewidth]{./figure_6_legend_horizontal.png}
%         \label{fig:legend_starplot1_infomap_co_retweet}
%     \end{minipage}
    
% \begin{subfigure}[b]{0.33\textwidth}
%         \centering
%         \includegraphics[width=1\textwidth]{./figure_A2a.png}
%         \caption{\rtw vs. \ind \textit{communities} (\uk).}
%         \label{fig:communities_flux_co_retweet_infomap_uk}
%     \end{subfigure}
% \hspace{0.5cm}
%     \begin{subfigure}[b]{0.33\textwidth}
%         \centering
%         \includegraphics[width=1\textwidth]{./figure_A2b.png}
%         \caption{\rtw vs. \ind \textit{nodes} (\uk).}
%         \label{fig:users_flux_co_retweet_infomap_uk}
%     \end{subfigure}

% \begin{subfigure}[b]{0.33\textwidth}
%         \centering
%         \includegraphics[width=1\textwidth]{./figure_A2c.png}
%         \caption{\rtw vs. \ind \textit{communities} (\russia).}
%         \label{fig:communities_flux_co_retweet_infomap_iorussia}
%     \end{subfigure}
% \hspace{0.5cm}
%     \begin{subfigure}[b]{0.33\textwidth}
%         \centering
%         \includegraphics[width=1\textwidth]{./figure_A2d.png}
%         \caption{\rtw vs. \ind \textit{nodes} (\russia).}
%         \label{fig:users_flux_co_retweet_infomap_iorussia}
%     \end{subfigure}

%     \caption{Bar charts for \info approach showing the number of the lost, common, and gained communities and nodes switching from \ind to \rtw in \russia dataset.}
%     \label{fig:flux_co_retweet_infomap}
% \end{figure} 

\begin{figure}[!htbp]
    \centering
    \includegraphics[width=0.6\textwidth]{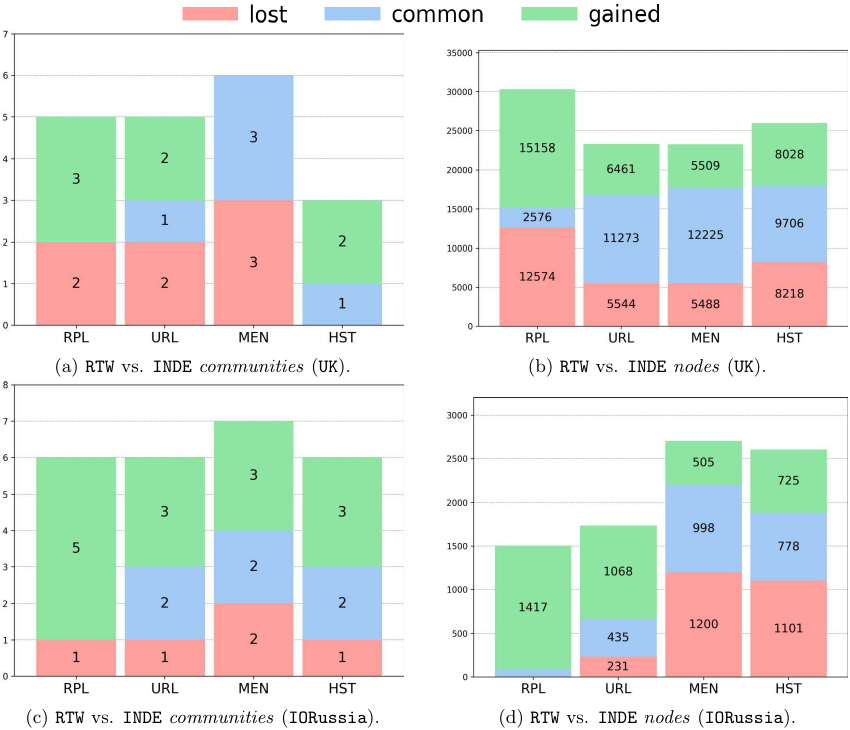}
    \caption{Bar charts for \info approach showing the number of the lost, common, and gained communities and nodes switching from \ind to \rtw in \russia dataset.}
    \label{fig:flux_co_retweet_infomap}
\end{figure}

% \begin{figure}[!t]
%     \centering
%     \begin{minipage}{0.8\textwidth}
%         \centering
%         \includegraphics[width=0.4\linewidth]{./figure_7_cosine_similarity_legend.png}
%         \label{fig:legend_cosine_similarity_infomap}
%     \end{minipage}

%        \begin{subfigure}[b]{0.45\textwidth}
%         \centering
%         \includegraphics[width=1\textwidth]{./figure_A3a.png}
%         \caption{\uk dataset -- \info.}
%         \label{fig:cos_sim_uk_infomap}
%     \end{subfigure}
% \hspace{0.5cm}
%     \begin{subfigure}[b]{0.45\textwidth}
%         \centering
%         \includegraphics[width=1\textwidth]{./figure_A3b.png}
%         \caption{\russia dataset -- \info.}
%         \label{fig:cos_sim_iorussia_infomap}
%     \end{subfigure}
%     \caption{Cosine similarity between the vectors of the communities detected with \info in common with \rtw communities. The horizontal lines report the average cosine similarity for each layer's communities.}
%     \label{fig:cosine_similarity_infomap}
% \end{figure} 

\begin{figure}[!ht]
    \centering
    \includegraphics[width=0.7\textwidth]{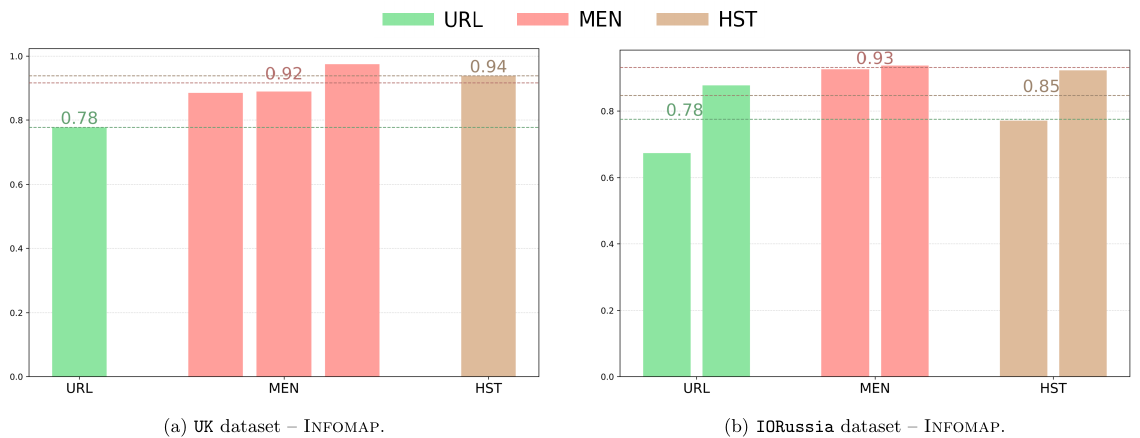}
    \caption{Cosine similarity between the vectors of the communities detected with \info in common with \rtw communities. The horizontal lines report the average cosine similarity for each layer's communities.}
    \label{fig:cosine_similarity_infomap}
\end{figure}

% \begin{figure}[!t]
%     \centering
% \begin{minipage}{0.8\textwidth}
%         \centering
%         \includegraphics[width=0.5\linewidth]{./figure_8_RTW_starplot_legend.png}
%         \label{fig:legend_starplot_uk_infomap}
%     \end{minipage}
%     \vspace{0.3cm}

% \begin{subfigure}[b]{0.3\textwidth}
%         \centering
%         \includegraphics[width=0.9\linewidth]{./figure_A4a.png}
%         \subcaption{\rtw vs. \men (\info).}
%     \end{subfigure}\hfill
%     \begin{subfigure}[b]{0.3\textwidth}
%         \centering
%         \includegraphics[width=0.9\linewidth]{./figure_A4b.png}
%         \subcaption{\rtw vs. \hash (\info).}
%     \end{subfigure}\hfill
%     \begin{subfigure}[b]{0.3\textwidth}
%         \centering
%          \includegraphics[width=0.9\linewidth]{./figure_A4c.png}
%          \subcaption{\rtw vs. \urld (\info).}
%     \end{subfigure}
    
%     \caption{Example of metrics comparison of three couple of common communities between \rtw and \men, \hash, and \urld{} in \uk dataset with \info.}
%     \label{fig:metrics_starplot_uk_infomap}
% \end{figure}

\begin{figure}[!htbp]
    \centering
    \includegraphics[width=0.8\textwidth]{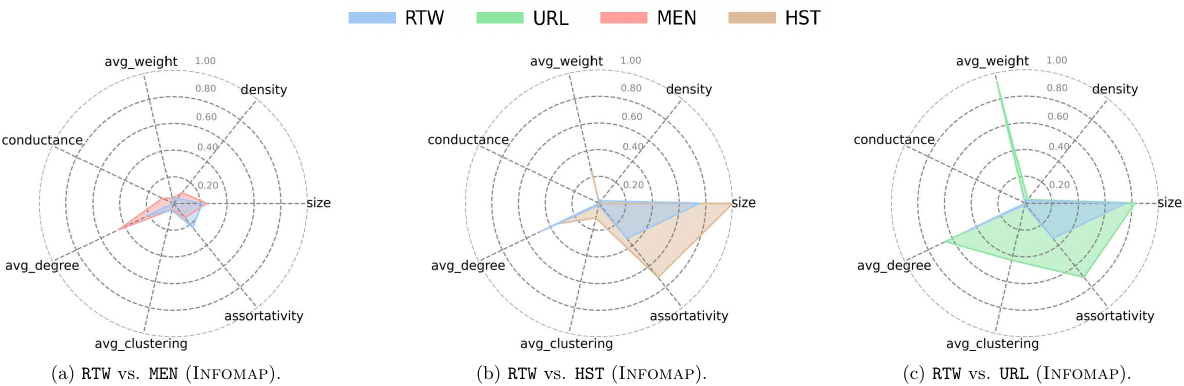}
    \caption{Example of metrics comparison of three couple of common communities between \rtw and \men, \hash, and \urld{} in \uk dataset with \info.}
    \label{fig:metrics_starplot_uk_infomap}
\end{figure}

\begin{figure}[!htbp]
    \centering
    \includegraphics[width=0.8\textwidth]{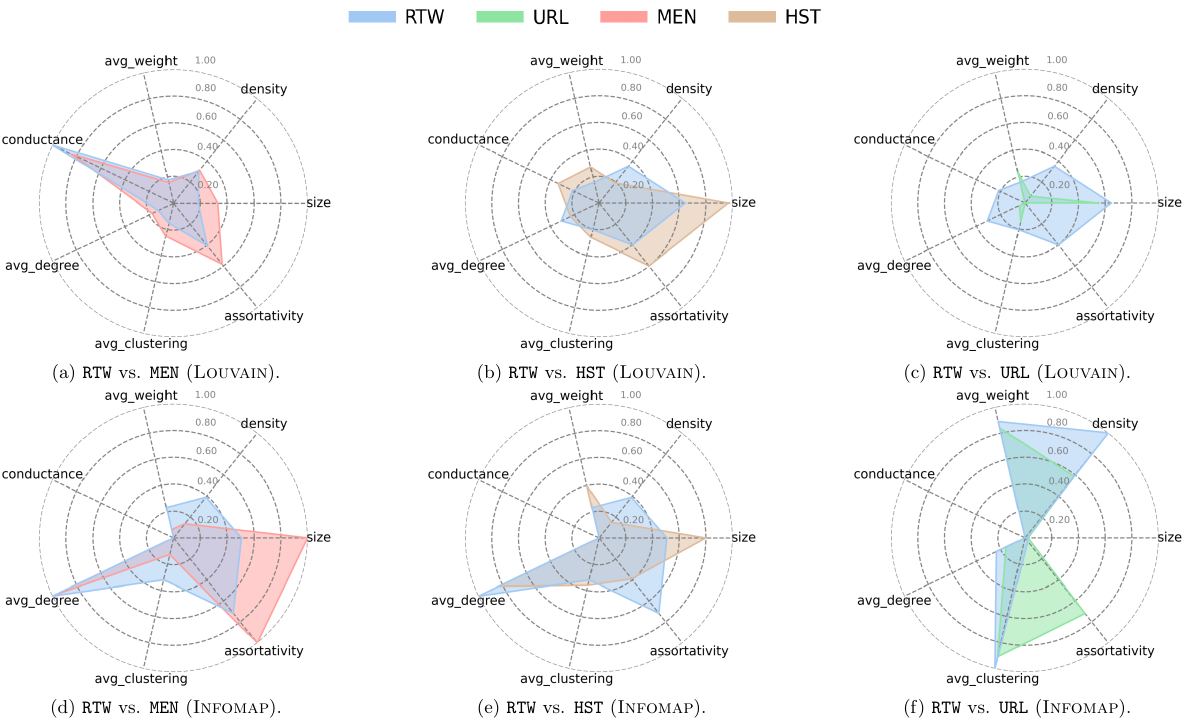}
    \caption{Example of metrics comparison of three couple of common communities between \rtw and \men, \hash, and \urld in \russia dataset with \louv and \info.}
    \label{fig:metrics_starplot_iorussia}
\end{figure}

 \begin{figure}[!htbp]
    \centering
        \includegraphics[width=0.35\textwidth]{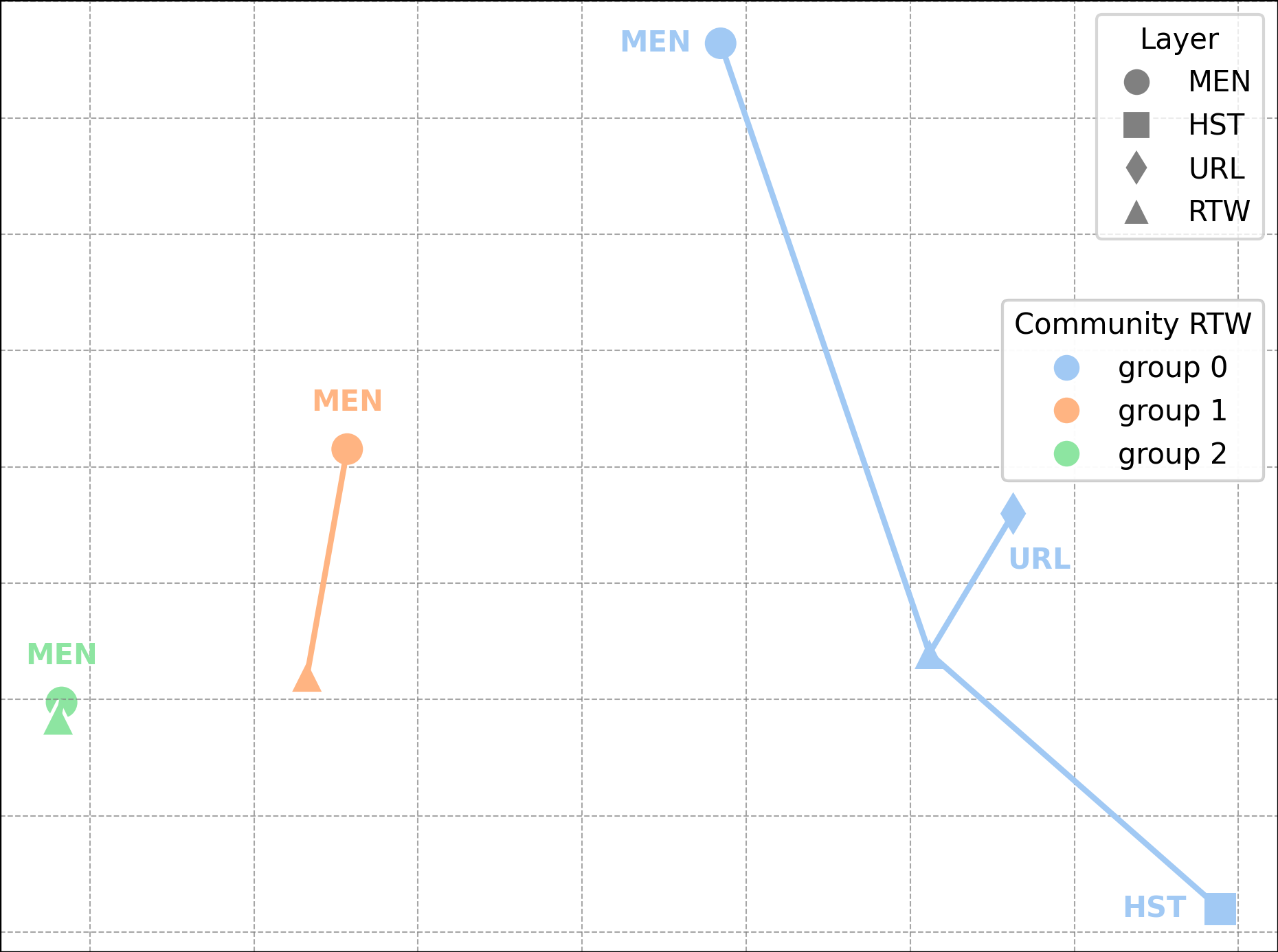}
        \caption{Visualization, for \uk with \info approach of common communities between \rtw and other co-actions in a two-dimensional space using PCA. Each community is represented by a vector of metrics, i.e., size, density, average degree, average weight, average clustering coefficient, conductance, and assortativity. These high dimensional vectors are projected into a two-dimensional space using PCA. Communities are represented with distinct point shapes based on the co-action and are linked to their corresponding \rtw community. Each \rtw community may be associated with multiple communities from different co-actions, with all connected communities sharing the same color for clarity.}
        \label{fig:pca_infomap_uk}
    \label{fig:pca_monomodal_comparison_uk_infomap}
\end{figure}

% \begin{figure}[!t]
%     \centering
%     \begin{subfigure}[b]{0.35\textwidth}
%         \centering
%         \includegraphics[width=1\textwidth]{./figure_A7a.png}
%         \caption{\russia dataset -- \louv.}
%         \label{fig:pca_louvain_iorussia}
%     \end{subfigure}
% \hspace{0.5cm}
%     \begin{subfigure}[b]{0.35\textwidth}
%         \centering
%         \includegraphics[width=1\textwidth]{./figure_A7b.png}
%         \caption{\russia dataset - \info.}
%         \label{fig:pca_infomap_iorussia}
%     \end{subfigure}

%     \caption{Visualization for \russia with \info and \louv, of common communities between \rtw and other co-actions in a two-dimensional space using PCA. Each community is represented by a vector of metrics, i.e., size, density, average degree, average weight, average clustering coefficient, conductance, and assortativity. These high dimensional vectors are projected into a two-dimensional space using PCA. Communities are represented with distinct point shapes based on the co-action and are linked to their corresponding \rtw community. Each \rtw community may be associated with multiple communities from different co-actions, with all connected communities sharing the same color for clarity.}
%     \label{fig:pca_monomodal_comparison_iorussia}
% \end{figure}

\begin{figure}[!htbp]
    \centering
    \includegraphics[width=0.7\textwidth]{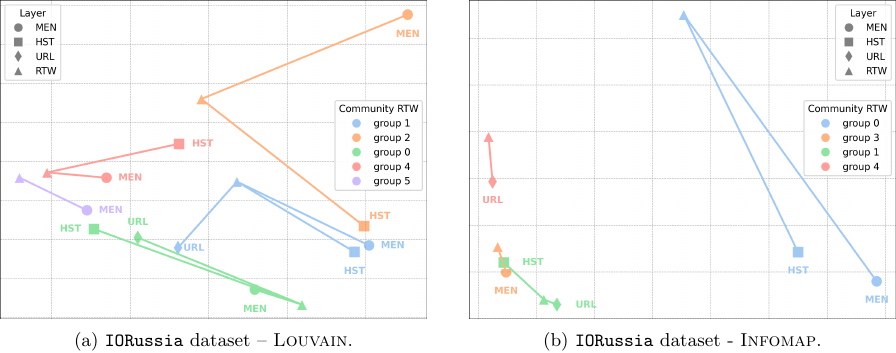}
    \caption{Visualization for \russia with \info and \louv, of common communities between \rtw and other co-actions in a two-dimensional space using PCA. Each community is represented by a vector of metrics, i.e., size, density, average degree, average weight, average clustering coefficient, conductance, and assortativity. These high dimensional vectors are projected into a two-dimensional space using PCA. Communities are represented with distinct point shapes based on the co-action and are linked to their corresponding \rtw community. Each \rtw community may be associated with multiple communities from different co-actions, with all connected communities sharing the same color for clarity.}
    \label{fig:pca_monomodal_comparison_iorussia}
\end{figure}

%  \begin{figure}[!ht]
%     \centering
%     \begin{minipage}{0.8\textwidth}
%         \centering
%         \includegraphics[width=0.5\linewidth]{./figure_6_legend_horizontal.png}
%         \label{fig:legend_starplot_uk_multmodal_barchart}
%     \end{minipage}
%     \vspace{0.3cm}
    
% \begin{subfigure}[b]{0.35\textwidth}
%         \centering
%         \includegraphics[width=\textwidth]{./figure_A8a.png}
%         \caption{\unfl (sum) vs. \ind \rev(\uk).}
%         \label{fig:communities_flux_flat_sum_uk_infomap}
%     \end{subfigure}
% \hspace{0.5cm}
%     \begin{subfigure}[b]{0.3\textwidth}
%         \centering
%         \includegraphics[width=\textwidth]{./figure_A8b.png}
%         \caption{\mul vs. \ind \rev(\uk).}
%         \label{fig:communities_flux_ginfomap_uk}
%     \end{subfigure}

% \begin{subfigure}[b]{0.3\textwidth}
%         \centering
%         \includegraphics[width=\textwidth]{./figure_A8c.png}
%         \caption{\unfl (sum) vs. \ind \rev(\russia).}
%         \label{fig:communities_flux_flat_sum_iorussia_infomap}
%     \end{subfigure}
% \hspace{0.5cm}
%     \begin{subfigure}[b]{0.35\textwidth}
%         \centering
%         \includegraphics[width=\textwidth]{./figure_A8d.png}
%         \caption{\mul vs. \ind \rev(\russia).}
%         \label{fig:communities_flux_ginfomap_iorussia}
%     \end{subfigure}

%     \caption{Bar charts showing the number of lost, common, and gained \textit{communities} switching from \ind to \unfl (sum) and \mul (in \uk and \russia dataset).}
%     \label{fig:communities_flux_multimodal_infomap}
% \end{figure}

\begin{figure}[!htbp]
    \centering
    \includegraphics[width=0.6\textwidth]{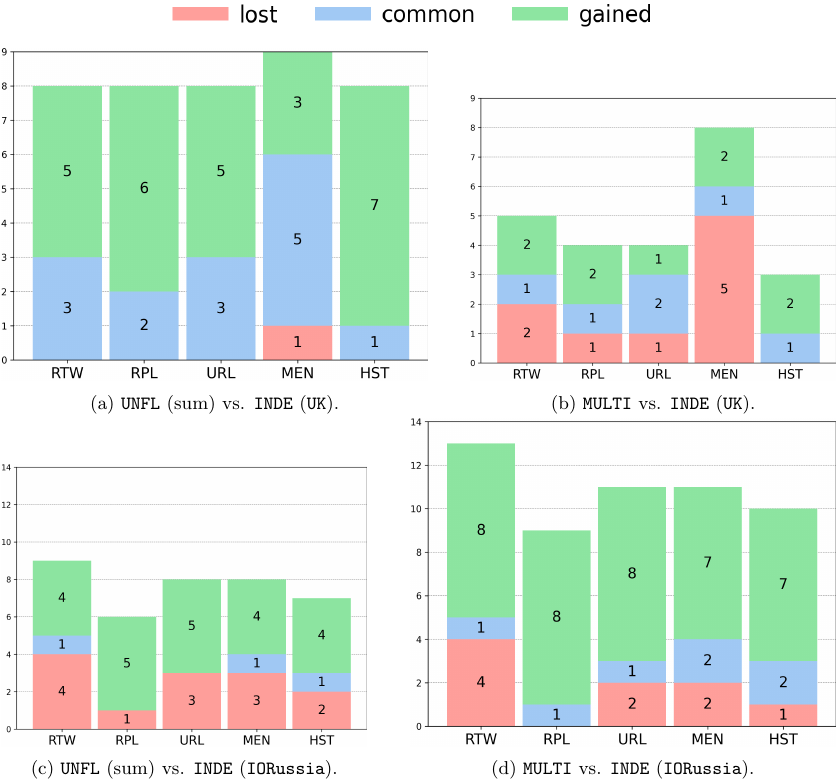}
    \caption{Bar charts showing the number of lost, common, and gained \textit{communities} switching from \ind to \unfl (sum) and \mul (in \uk and \russia dataset).}
    \label{fig:communities_flux_multimodal_infomap}
\end{figure}

% \begin{figure}[!ht]
%     \centering
%     \begin{minipage}{0.8\textwidth}
%         \centering
%         \includegraphics[width=0.5\linewidth]{./figure_6_legend_horizontal.png}
%         \label{fig:legend_starplot_uk_multimodal_barchart_louvain}
%     \end{minipage}

%     \vspace{0.5cm}

%     \begin{subfigure}[b]{0.35\textwidth}
%         \centering
%         \includegraphics[width=\textwidth]{./figure_A9a.png}
%         \caption{\unfl (sum) vs. \ind (\info).}
%         \label{fig:users_flux_flat_sum_uk_infomap}
%     \end{subfigure}
% \hspace{0.5cm}
%     \begin{subfigure}[b]{0.35\textwidth}
%         \centering
%         \includegraphics[width=\textwidth]{./figure_A9b.png}
%         \caption{\mul vs. \ind (\info).}
%         \label{fig:users_flux_ginfomap_uk}
%     \end{subfigure}

% \begin{subfigure}[b]{0.35\textwidth}
%         \centering
%         \includegraphics[width=\textwidth]{./figure_A9c.png}
%         \caption{\unfl (sum) vs. \ind (\info).}
%         \label{fig:users_flux_flat_sum_iorussia_infomap}
%     \end{subfigure}
% \hspace{0.5cm}
%     \begin{subfigure}[b]{0.35\textwidth}
%         \centering
%         \includegraphics[width=\textwidth]{./figure_A9d.png}
%         \caption{\mul vs. \ind (\info).}
%         \label{fig:users_flux_ginfomap_iorussia}
%     \end{subfigure}
    
%     \caption{Bar charts showing the number of lost, common, and gained \textit{nodes} switching from \ind to \unfl (sum) and \mul (in \uk and \russia dataset).}
%     \label{fig:users_flux_multimodal_infomap}
% \end{figure}

\begin{figure}[!htbp]
    \centering
    \includegraphics[width=0.6\textwidth]{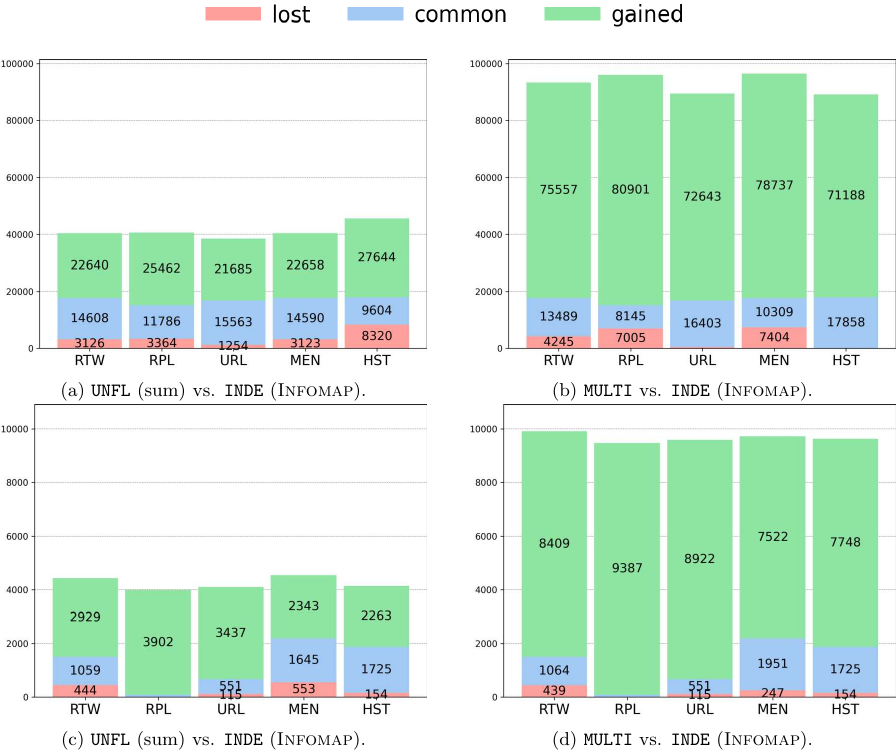}
    \caption{Bar charts showing the number of lost, common, and gained \textit{nodes} switching from \ind to \unfl (sum) and \mul (in \uk and \russia dataset).}
    \label{fig:users_flux_multimodal_infomap}
\end{figure}

\begin{figure}[!htbp]
    \centering
    \includegraphics[width=\textwidth]{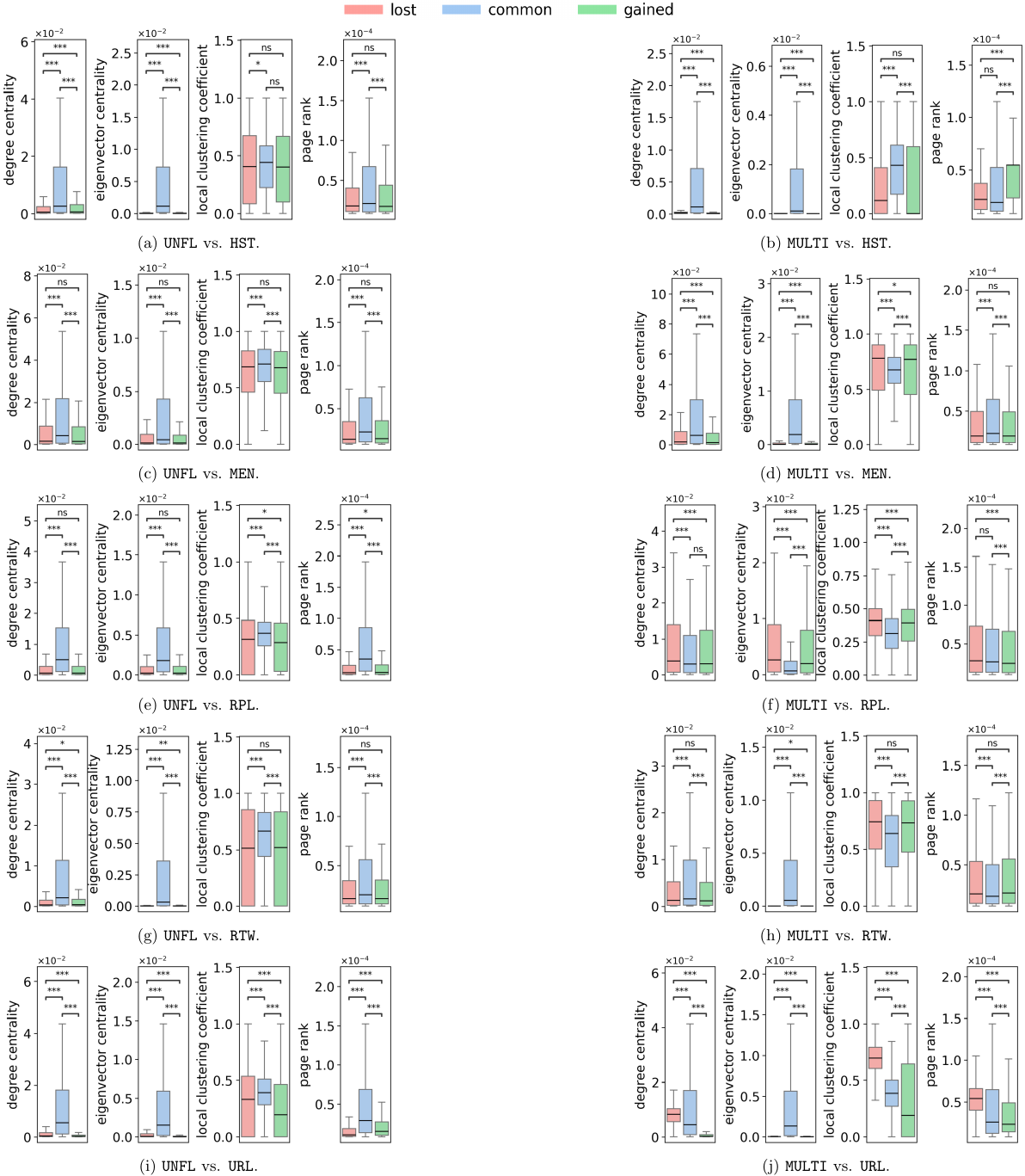}
    \caption{Lost, common, gained nodes characterization for \unfl and \mul (\info) in \uk dataset. We use the notation \textit{ns} and * to indicate statistical significance levels. Specifically, \textit{ns} denotes that there is no significant difference between the groups, with a p-value in the range $p \geq 0.05$. For statistically significant results, we used the following annotations based on p-value ranges: *: $p \leq 0.05$; **: $p \leq 0.01$; ***: $p \leq 0.001$.}
    \label{fig:node_metrics_boxplot_uk_infomap}
\end{figure}

\begin{figure}[!htbp]
    \centering
    \includegraphics[width=\textwidth]{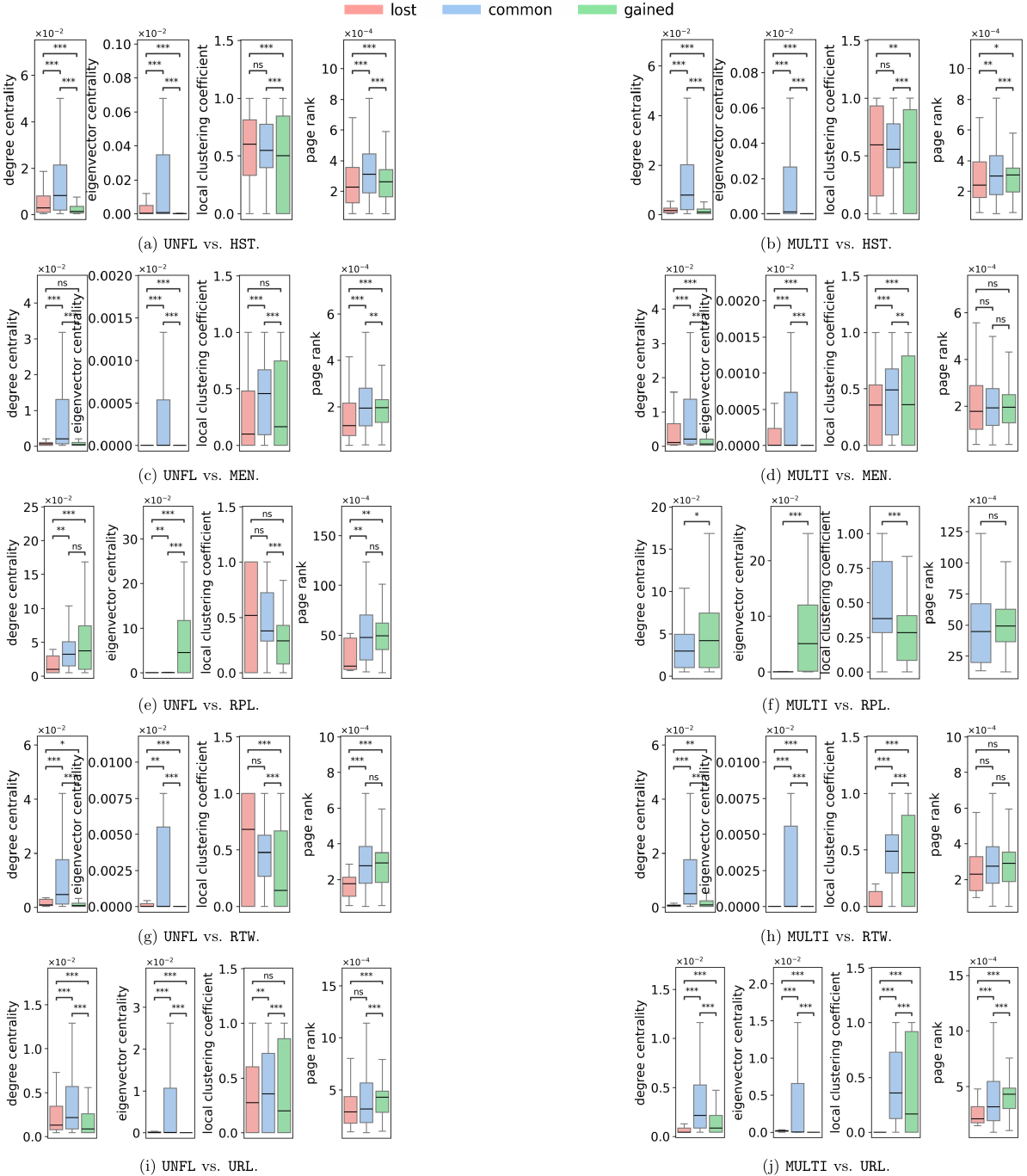}
    \caption{Lost, common, gained nodes characterization for \unfl and \mul (\louv) in \russia dataset. We use the notation \textit{ns} and * to indicate statistical significance levels. Specifically, \textit{ns} denotes that there is no significant difference between the groups, with a p-value in the range $p \geq 0.05$. For statistically significant results, we used the following annotations based on p-value ranges: *: $p \leq 0.05$; **: $p \leq 0.01$; ***: $p \leq 0.001$.}
    \label{fig:node_metrics_boxplot_iorussia_louvain}
\end{figure}

\begin{figure}[!htbp]
    \centering
    \includegraphics[width=\textwidth]{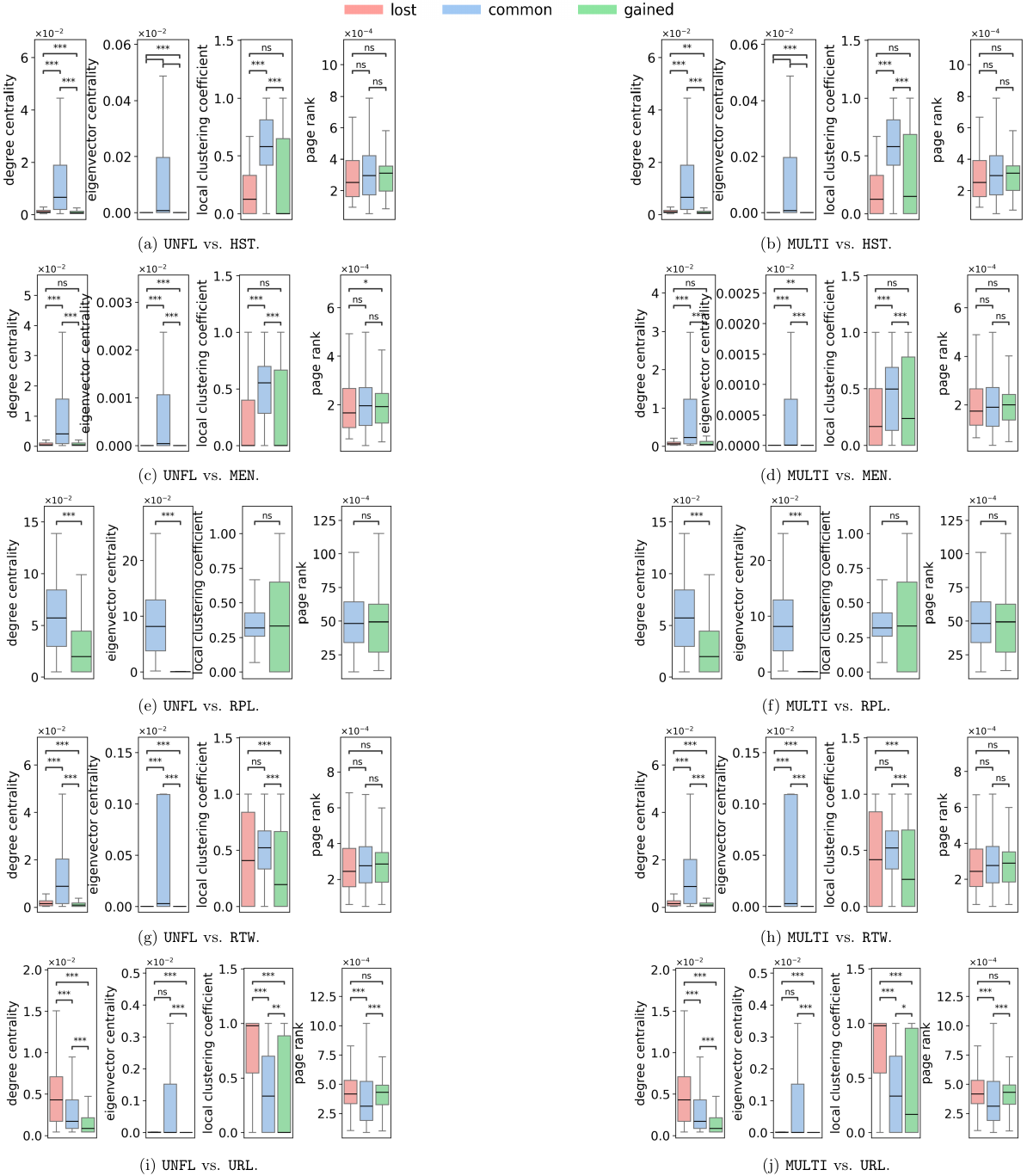}
    \caption{Lost, common, gained nodes characterization for \unfl and \mul (\info) in \russia dataset. We use the notation \textit{ns} and * to indicate statistical significance levels. Specifically, \textit{ns} denotes that there is no significant difference between the groups, with a p-value in the range $p \geq 0.05$. For statistically significant results, we used the following annotations based on p-value ranges: *: $p \leq 0.05$; **: $p \leq 0.01$; ***: $p \leq 0.001$.}
    \label{fig:node_metrics_boxplot_iorussia_infomap}
\end{figure}

% \begin{figure}[!t]
%     \centering

%      \begin{subfigure}[b]{0.45\textwidth}
%         \centering
%         \includegraphics[width=1\textwidth]{./figure_A13a.png}
%         \caption{\uk dataset -- \unfl.}
%         \label{fig:coord_uk_infomap_flat_weighted_sum}
%     \end{subfigure}
% \hspace{0.5cm}
%     \begin{subfigure}[b]{0.45\textwidth}
%         \centering
%         \includegraphics[width=1\textwidth]{./figure_A13b.png}
%         \caption{\uk dataset -- \mul.}
%         \label{fig:coord_uk_infomap_multimodal}
%     \end{subfigure}

%      \begin{subfigure}[b]{0.45\textwidth}
%         \centering
%         \includegraphics[width=1\textwidth]{./figure_A13c.png}
%         \caption{\russia dataset -- \unfl.}
%         \label{fig:coord_iorussia_infomap_flat_weighted_sum}
%     \end{subfigure}
% \hspace{0.5cm}
%     \begin{subfigure}[b]{0.45\textwidth}
%         \centering
%         \includegraphics[width=1\textwidth]{./figure_A13d.png}
%         \caption{\russia dataset -- \mul.}
%         \label{fig:coord_iorussia_infomap_multimodal}
%     \end{subfigure}    
%     \caption{Level of coordination of lost, common, and gained communities detected by \info approach.}
%     \label{fig:coord_infomap}
% \end{figure}

\begin{figure}[!htbp]
    \centering
    \includegraphics[width=0.7\textwidth]{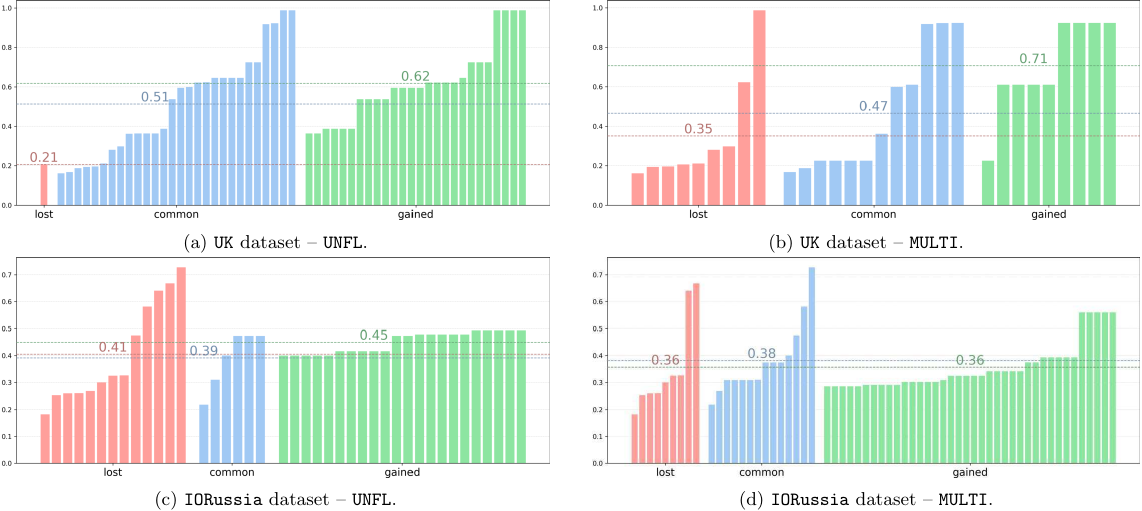}
    \caption{Level of coordination of lost, common, and gained communities detected by \info approach.}
    \label{fig:coord_infomap}
\end{figure}

 \clearpage
\newpage
\section{\uk heatmap overlapping matrix of \louv communities}
We report the heatmaps of the overlapping matrix between \louv communities for \uk. Specifically between \rtw and \ind, \mul and \ind, \unfl (all 3 flattening strategies) and \ind.

\label{sec:app_heatmap_uk_louvain}
\renewcommand{\thefigure}{B.\arabic{figure}}
\setcounter{figure}{0}

% \begin{figure}[!ht]
% \centering

% \begin{minipage}{0.48\textwidth}
%     \centering
%     \includegraphics[width=0.53\textwidth, height=17cm]{./figure_A14a.png}
%     \caption{\rtw vs. \ind.}
%     \label{fig:heatmap_retweet_uk_louvain}
% \end{minipage}
% \hfill
% \begin{minipage}{0.48\textwidth}
%     \centering
%     \includegraphics[width=0.9\textwidth, height=17cm]{./figure_A14b.png}
%     \caption{\mul vs. \ind.}
%     \label{fig:heatmap_multimodal_basic_and_uk_louvain}
% \end{minipage}

% \end{figure}

% \begin{figure}[!ht]
% \centering
% \begin{minipage}{0.33\textwidth}
%         \centering
%         \includegraphics[width=\textwidth, height=19cm]{./figure_A15a.png}
%         \caption{\unfl (nw) vs. \ind.}
%         \label{fig:heatmap_flat_nw_uk_louvain}
%     \end{minipage}
%     \hfill
% \begin{minipage}{0.29\textwidth}
%         \centering
%         \includegraphics[width=\textwidth, height=19cm]{./figure_A15b.png}
%         \caption{\unfl (ec) vs. \ind.}
%         \label{fig:heatmap_flat_ec_uk_louvain}
%     \end{minipage}
%     \hfill
% \begin{minipage}{0.35\textwidth}
%         \centering
%         \includegraphics[width=\textwidth, height=19cm]{./figure_A15c.png}
%         \caption{\unfl (sum) vs. \ind.}
%         \label{fig:heatmap_flat_sum_uk_louvain}
%     \end{minipage}
    
% \end{figure}

% --- first figure ---
\begin{figure}[!htbp]
\centering
\includegraphics[width=0.24\textwidth]{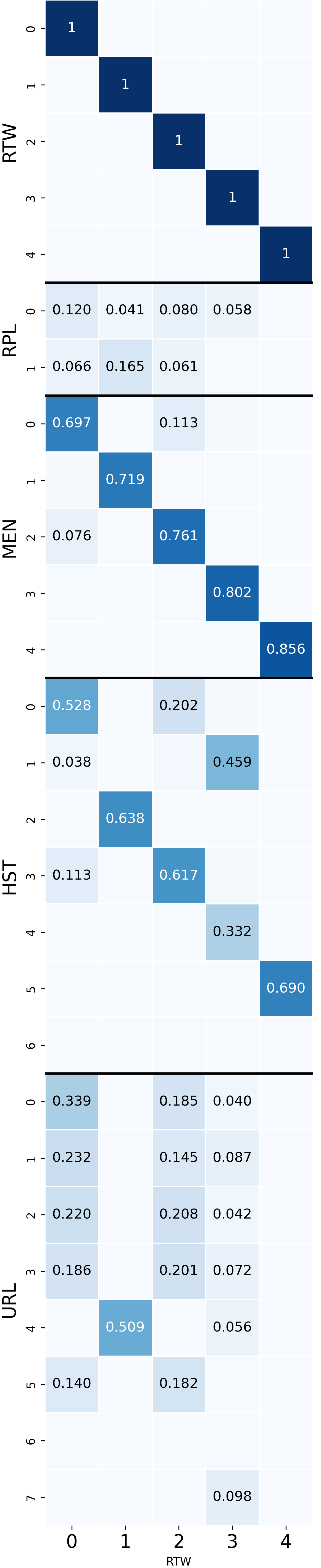}
\caption{\rtw vs. \ind.}
\label{fig:heatmap_retweet_uk_louvain}
\end{figure}

% --- second figure ---
\begin{figure}[!htbp]
\centering
\includegraphics[width=0.6\textwidth]{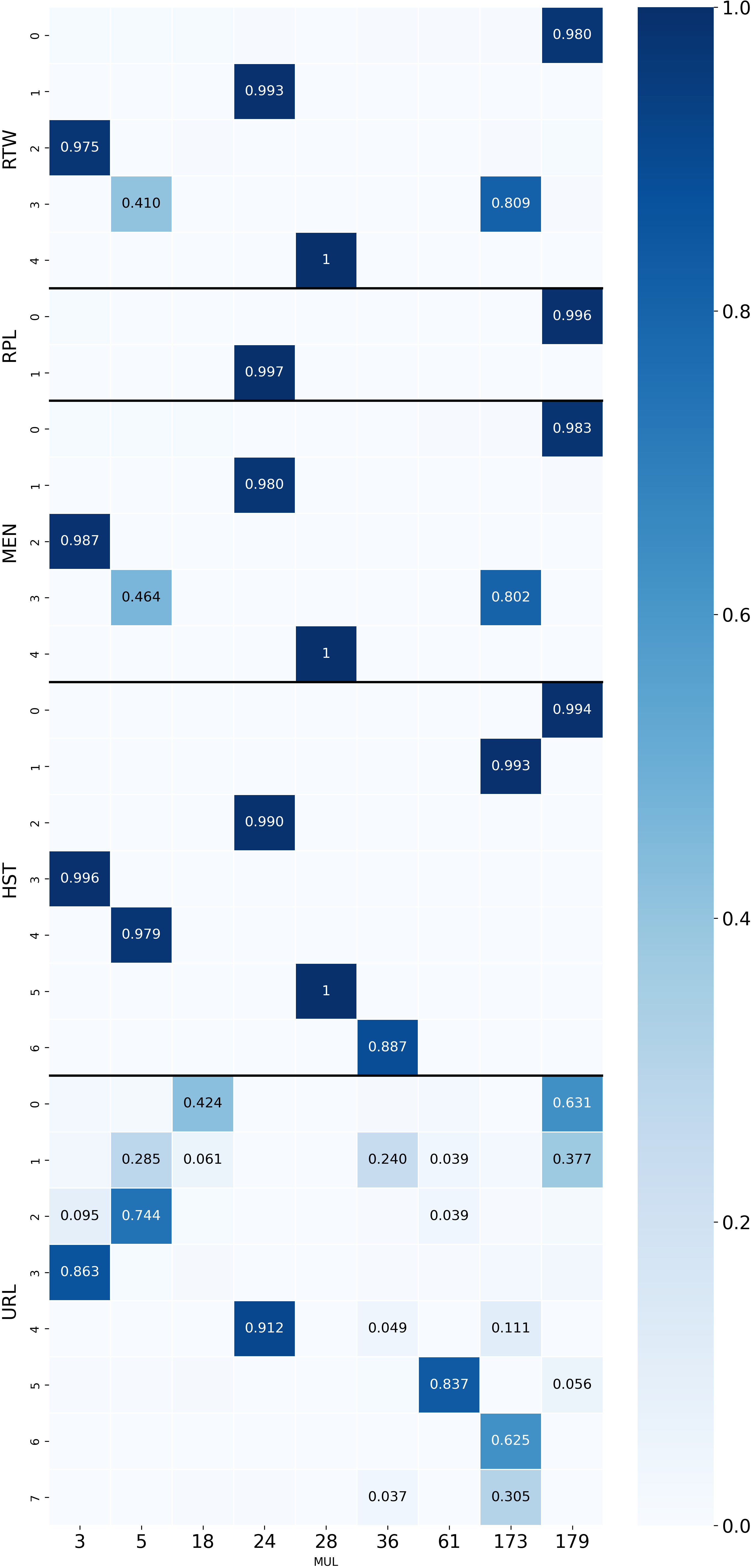}
\caption{\mul vs. \ind.}
\label{fig:heatmap_multimodal_basic_and_uk_louvain}
\end{figure}

% ---------------------

% --- first figure ---
\begin{figure}[!htbp]
\centering
\includegraphics[width=0.33\textwidth,height=19cm]{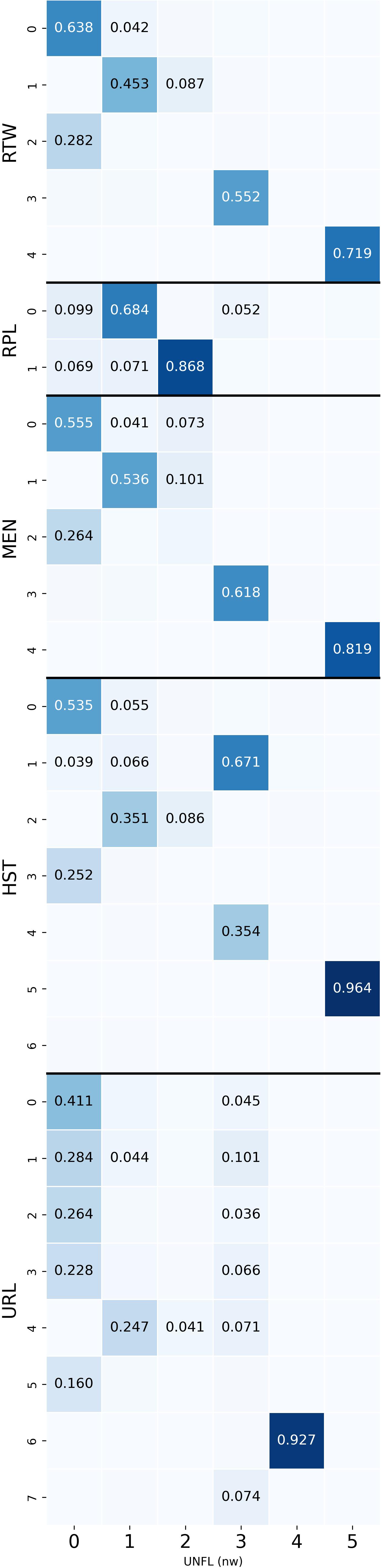}
\caption{\unfl (nw) vs. \ind.}
\label{fig:heatmap_flat_nw_uk_louvain}
\end{figure}

% --- second figure ---
\begin{figure}[!htbp]
\centering
\includegraphics[width=0.29\textwidth,height=19cm]{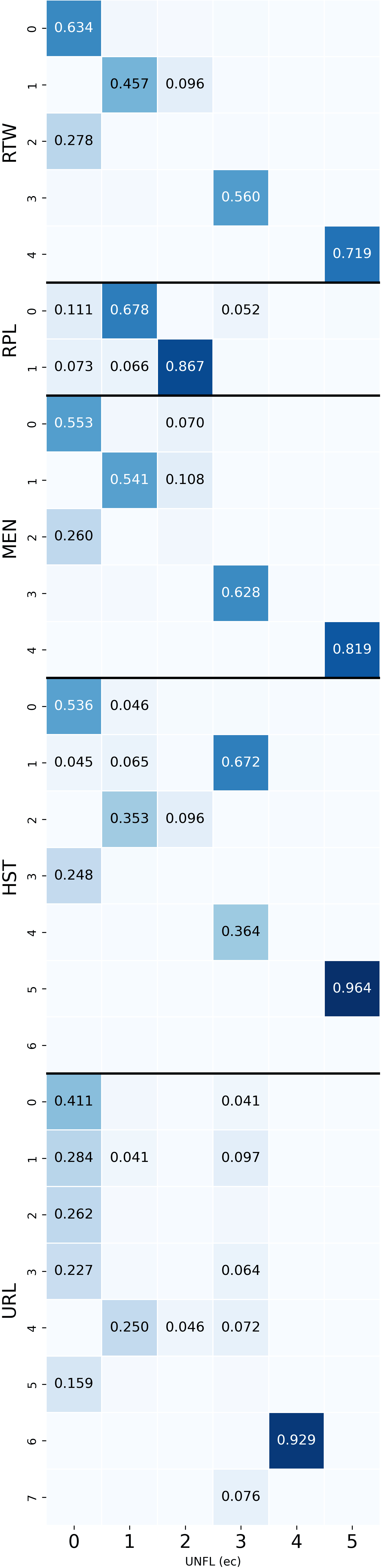}
\caption{\unfl (ec) vs. \ind.}
\label{fig:heatmap_flat_ec_uk_louvain}
\end{figure}

% --- third figure ---
\begin{figure}[!htbp]
\centering
\includegraphics[width=0.35\textwidth,height=19cm]{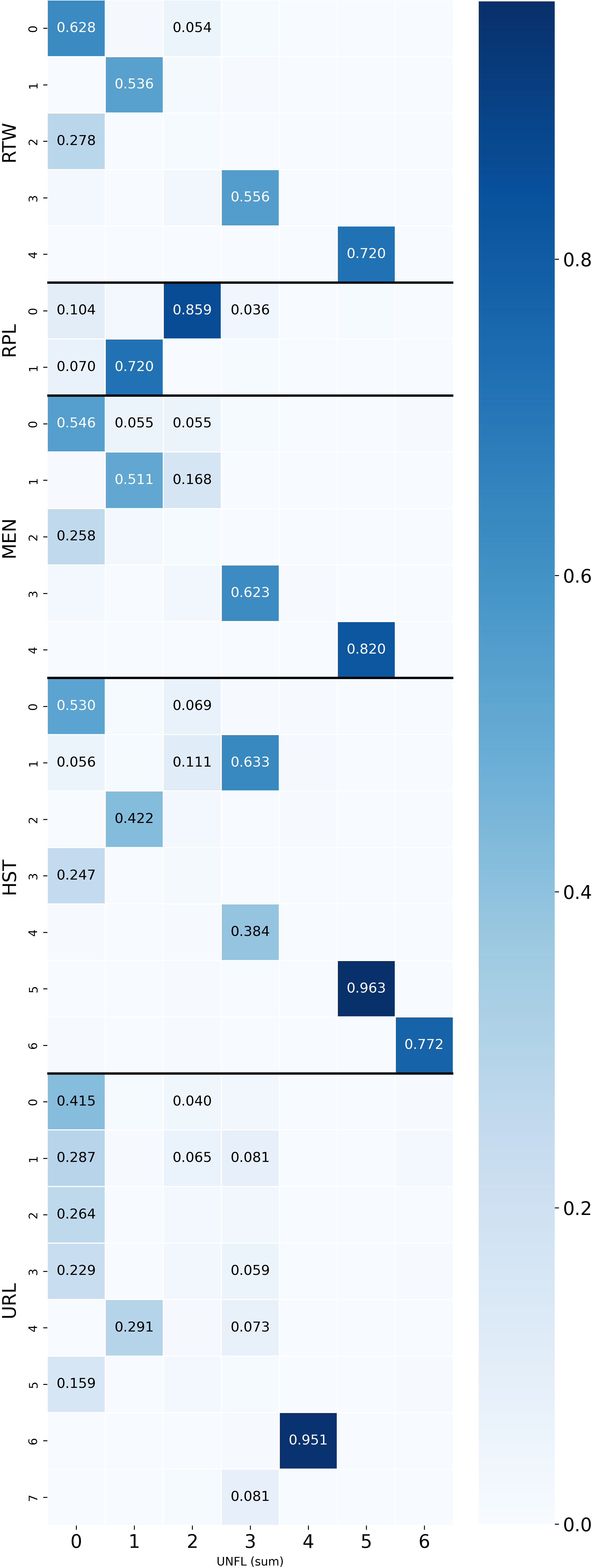}
\caption{\unfl (sum) vs. \ind.}
\label{fig:heatmap_flat_sum_uk_louvain}
\end{figure}

 \clearpage
\newpage

\section{\uk heatmap overlapping matrix of \info communities}
We report the heatmaps of the overlapping matrix between \info communities for \uk. Specifically between \rtw and \ind, \mul and \ind, \unfl (all 3 flattening strategies) and \ind.

\label{sec:app_heatmap_uk_infomap}
\renewcommand{\thefigure}{C.\arabic{figure}}
\setcounter{figure}{0}

% \begin{figure}[!ht]
% \centering
    
% \begin{minipage}{0.48\textwidth}
%         \centering
%         \includegraphics[width=0.6\textwidth, height=18cm]{./figure_A16a.png}
%         \caption{\rtw vs. \ind.}
%         \label{fig:heatmap_retweet_uk_infomap}
%     \end{minipage}
%     \hfill
% \begin{minipage}{0.48\textwidth}
%         \centering
%         \includegraphics[width=0.8\textwidth, height=18cm]{./figure_A16b.png}
%         \caption{\mul vs. \ind.}
%         \label{fig:heatmap_multimodal_basic_and_uk_infomap}
%     \end{minipage}
    
% \end{figure}

% --- first figure ---
\begin{figure}[!htbp]
\centering
\includegraphics[width=0.24\textwidth]{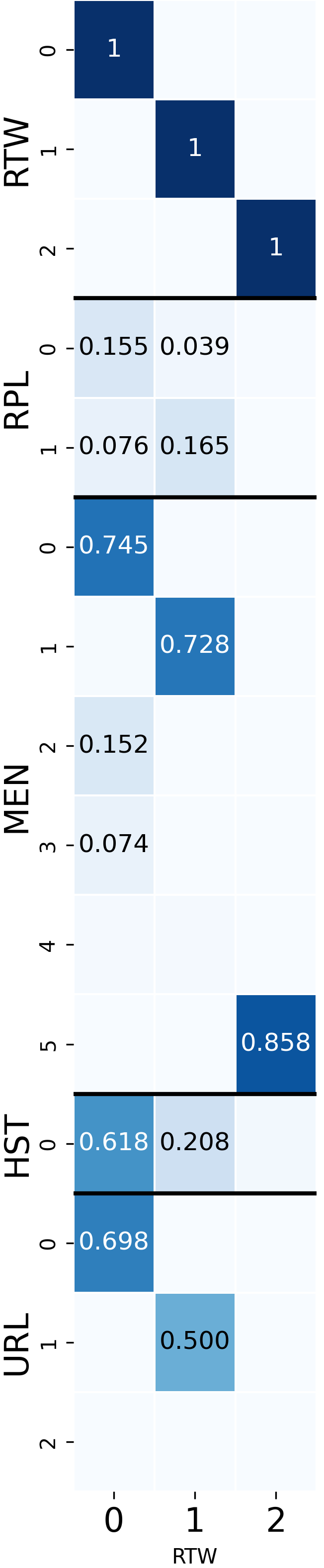}
\caption{\rtw vs. \ind.}
\label{fig:heatmap_retweet_uk_infomap}
\end{figure}

% --- second figure ---
\begin{figure}[!htbp]
\centering
\includegraphics[width=0.33\textwidth]{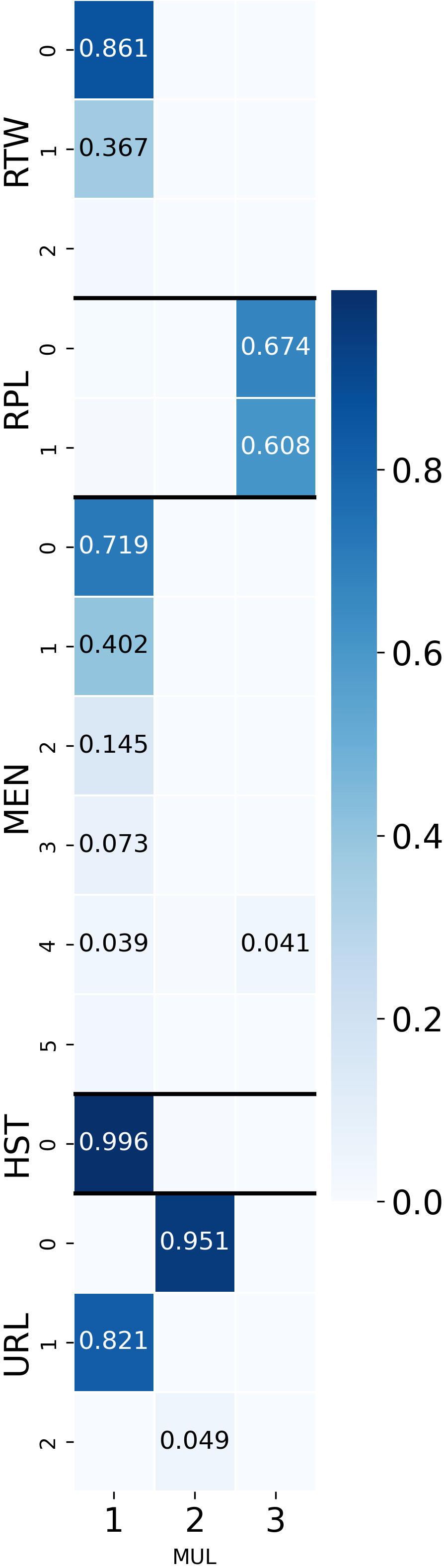}
\caption{\mul vs. \ind.}
\label{fig:heatmap_multimodal_basic_and_uk_infomap}
\end{figure}

\begin{figure}[!htbp]
    \centering
    \includegraphics[width=0.9\textwidth, height=20cm]{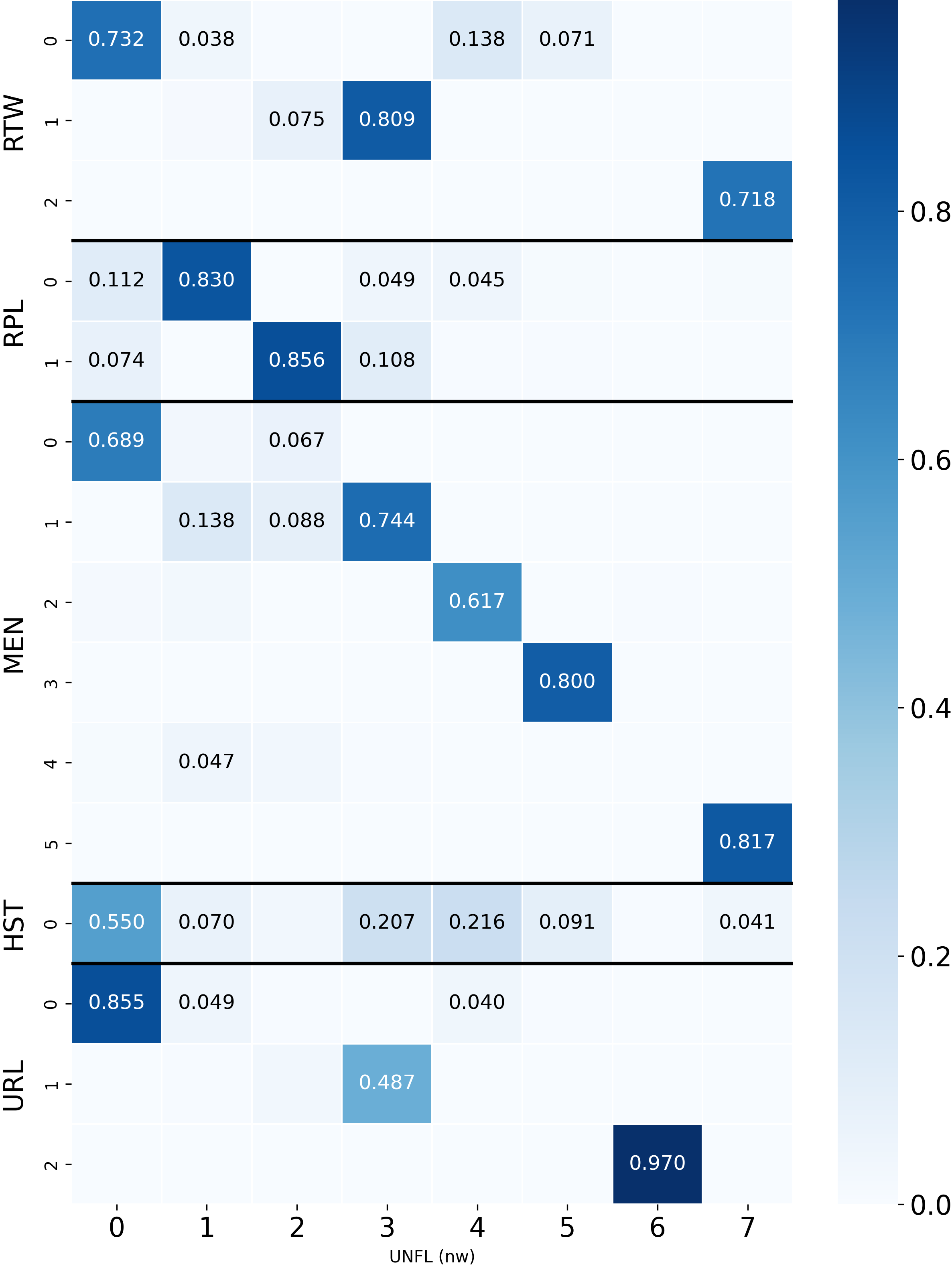}
    \caption{\unfl (nw) vs. \ind.}
    \label{fig:heatmap_flat_nw_uk_infomap}
\end{figure}

\begin{figure}[!htbp]
    \centering
    \includegraphics[width=0.9\textwidth, height=20cm]{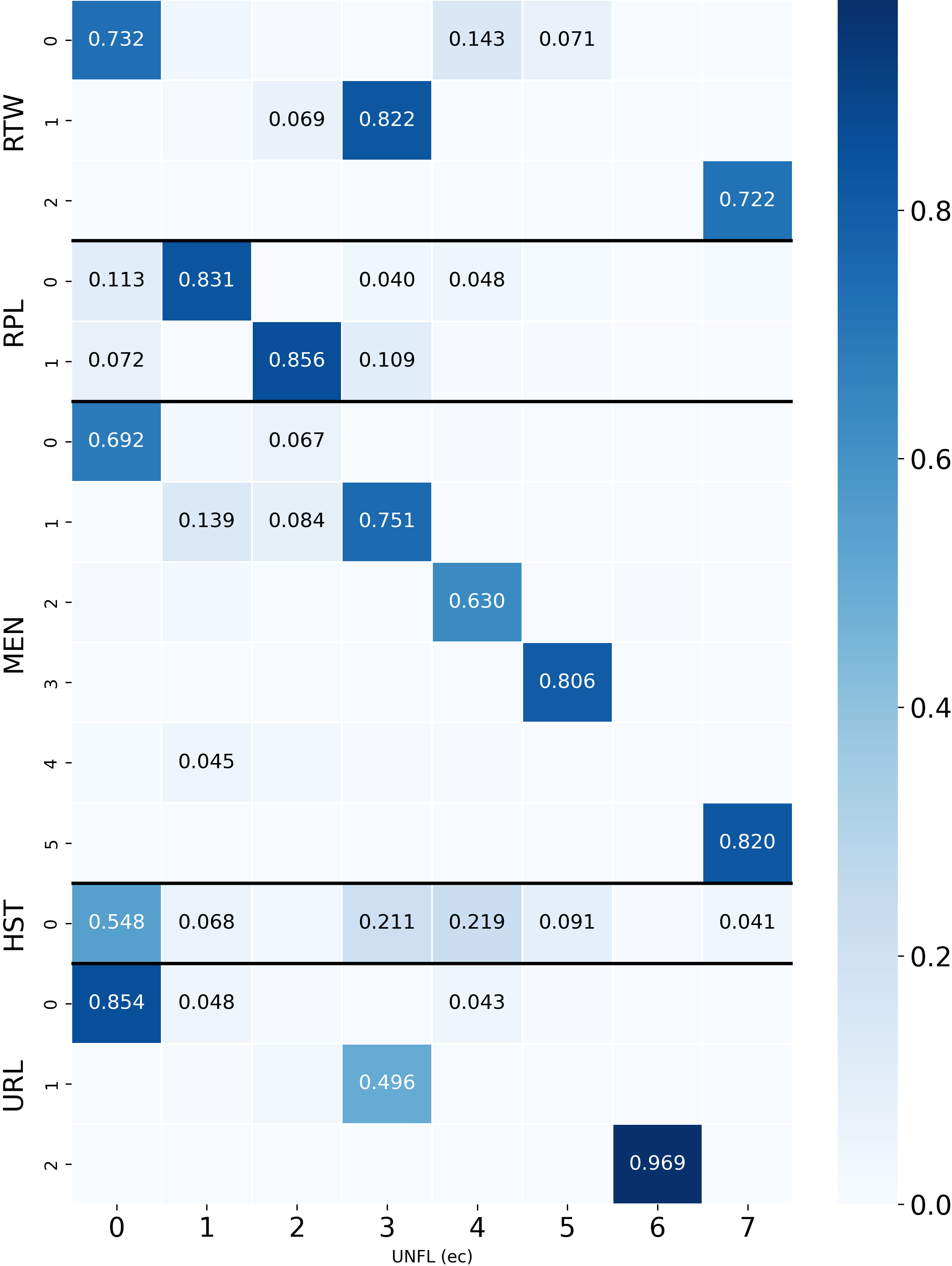}
    \caption{\unfl (ec) vs. \ind.}
    \label{fig:heatmap_flat_ec_uk_infomap}
\end{figure}

\begin{figure}[!htbp]
    \centering
    \includegraphics[width=0.9\textwidth, height=20cm]{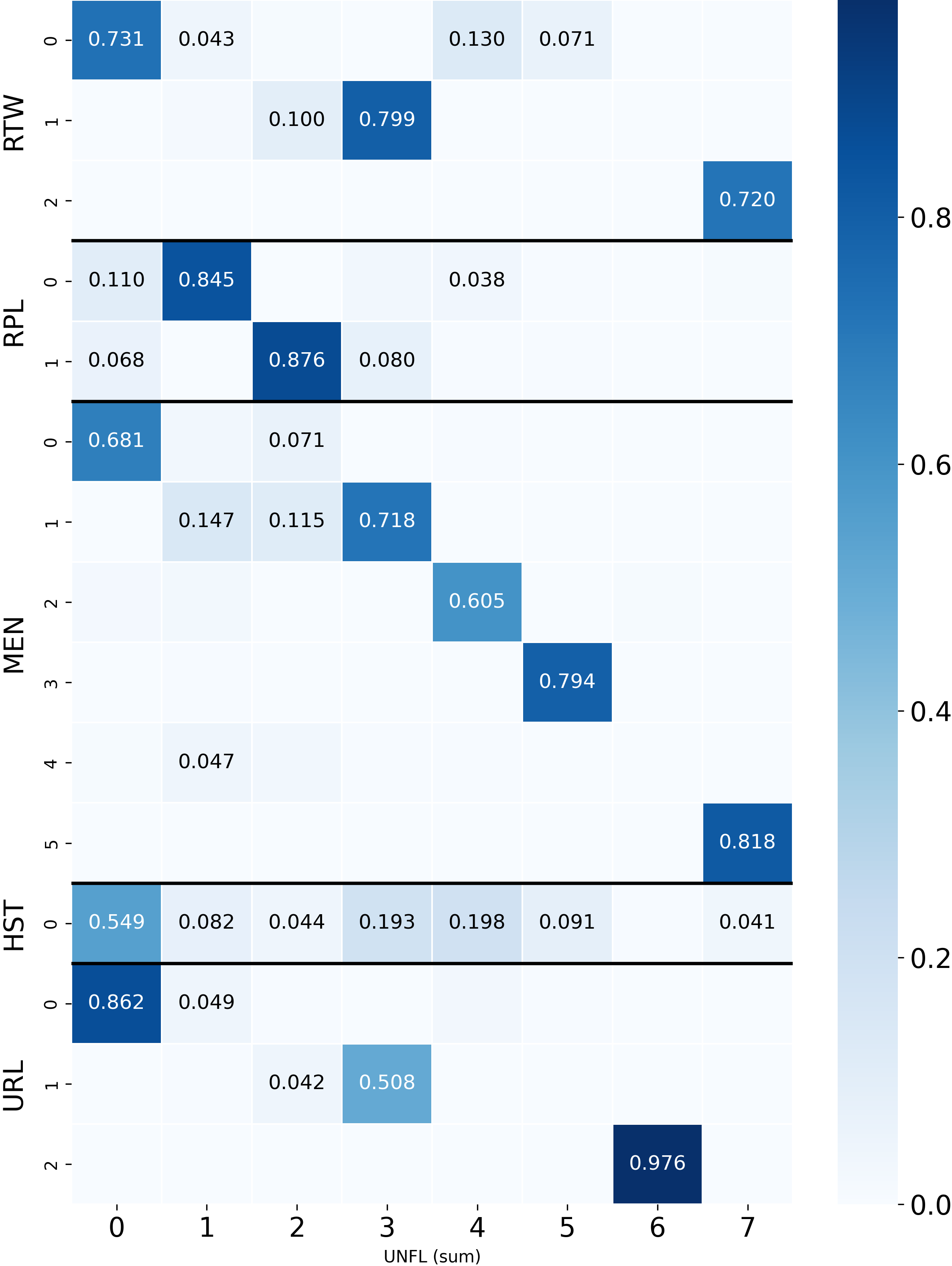}
    \caption{\unfl (sum) vs. \ind.}
    \label{fig:heatmap_flat_sum_uk_infomap}
\end{figure}

\clearpage
\newpage
\section{\russia heatmap overlapping matrix of \louv communities}
We report the heatmaps of the overlapping matrix between \louv communities for \russia. Specifically between \rtw and \ind, \mul and \ind, \unfl (all 3 flattening strategies) and \ind.
\label{sec:app_heatmap_iorussia_louvain}
\renewcommand{\thefigure}{D.\arabic{figure}}
\setcounter{figure}{0}

\begin{figure}[!htbp]
    \centering
    \includegraphics[width=0.4\textwidth, height=18cm]{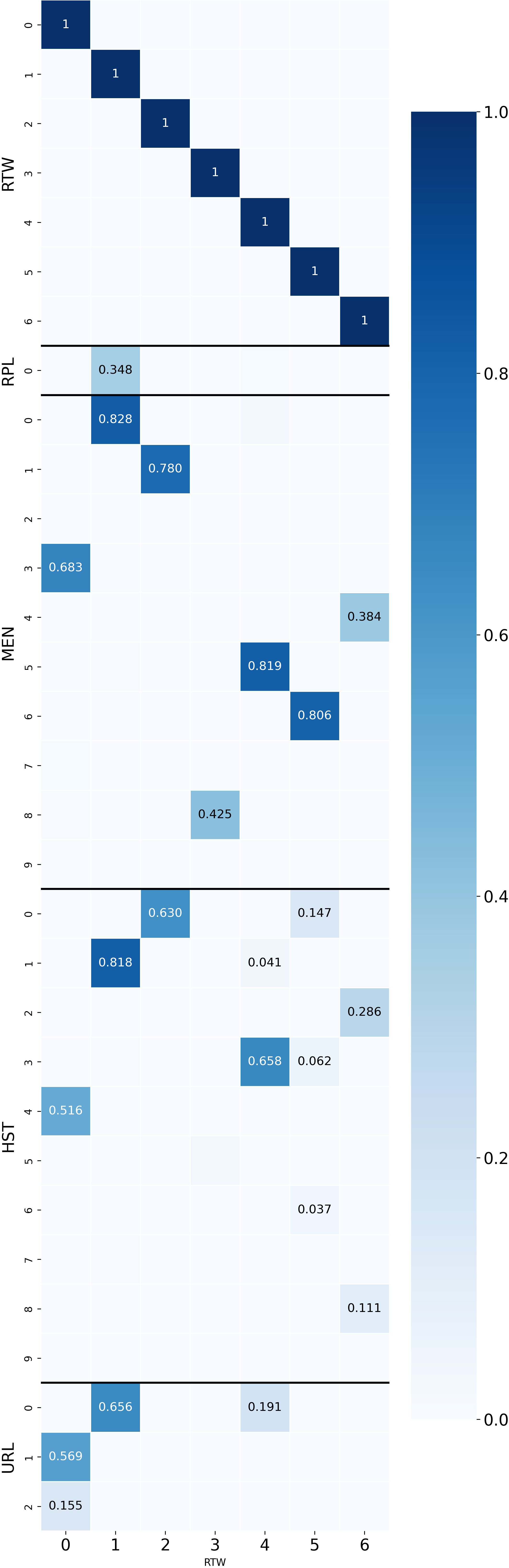}
    \caption{\rtw vs. \ind.}
    \label{fig:heatmap_retweet_iorussia_louvain}
\end{figure}

\begin{figure}[!htbp]
    \centering
    \includegraphics[width=1\textwidth, height=18cm]{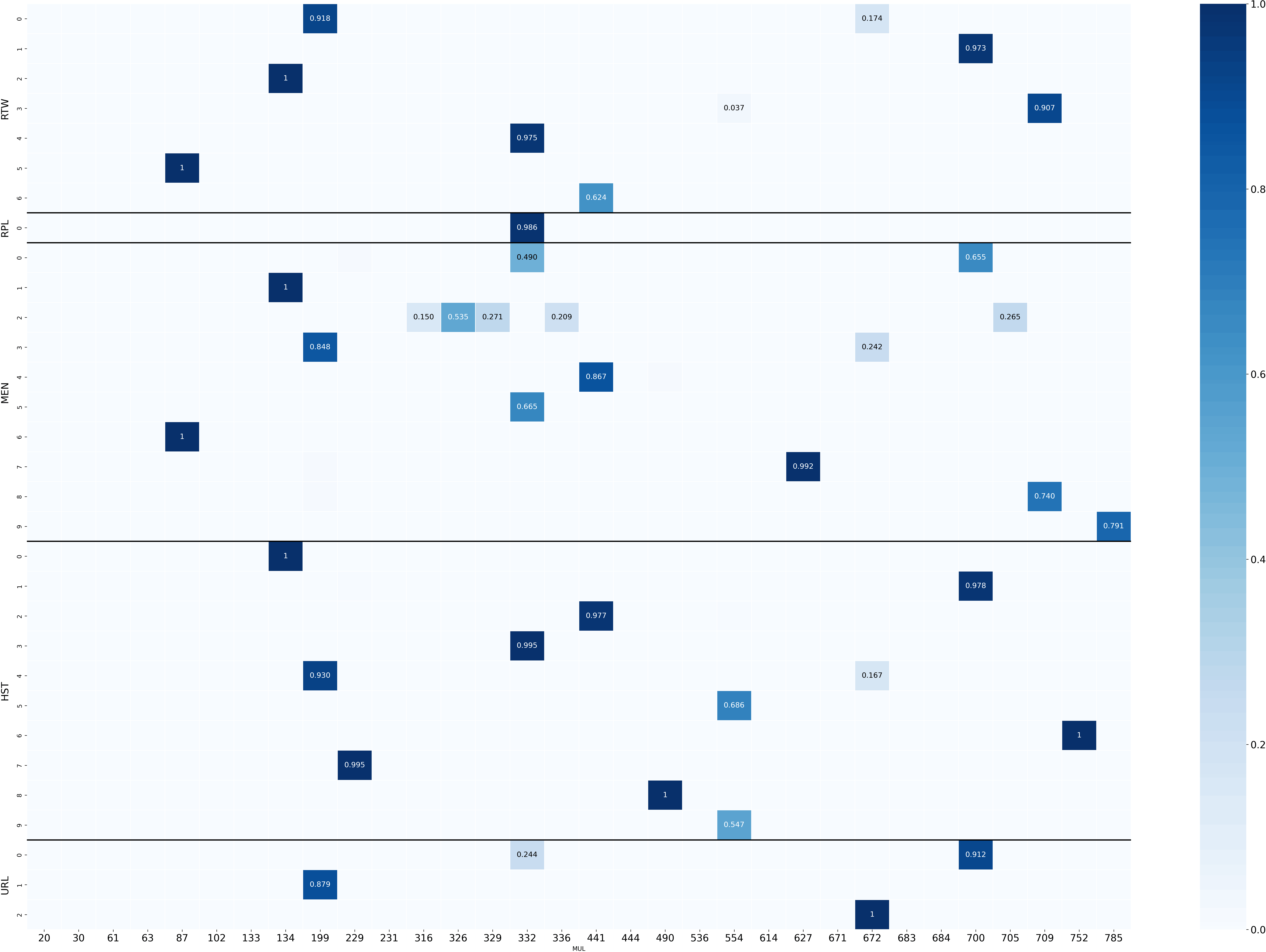}
    \caption{\mul vs. \ind.}
    \label{fig:heatmap_multimodal_basic_and_iorussia_louvain}
\end{figure}

\begin{figure}[!htbp]
    \centering
    \includegraphics[width=0.8\textwidth, height=19cm]{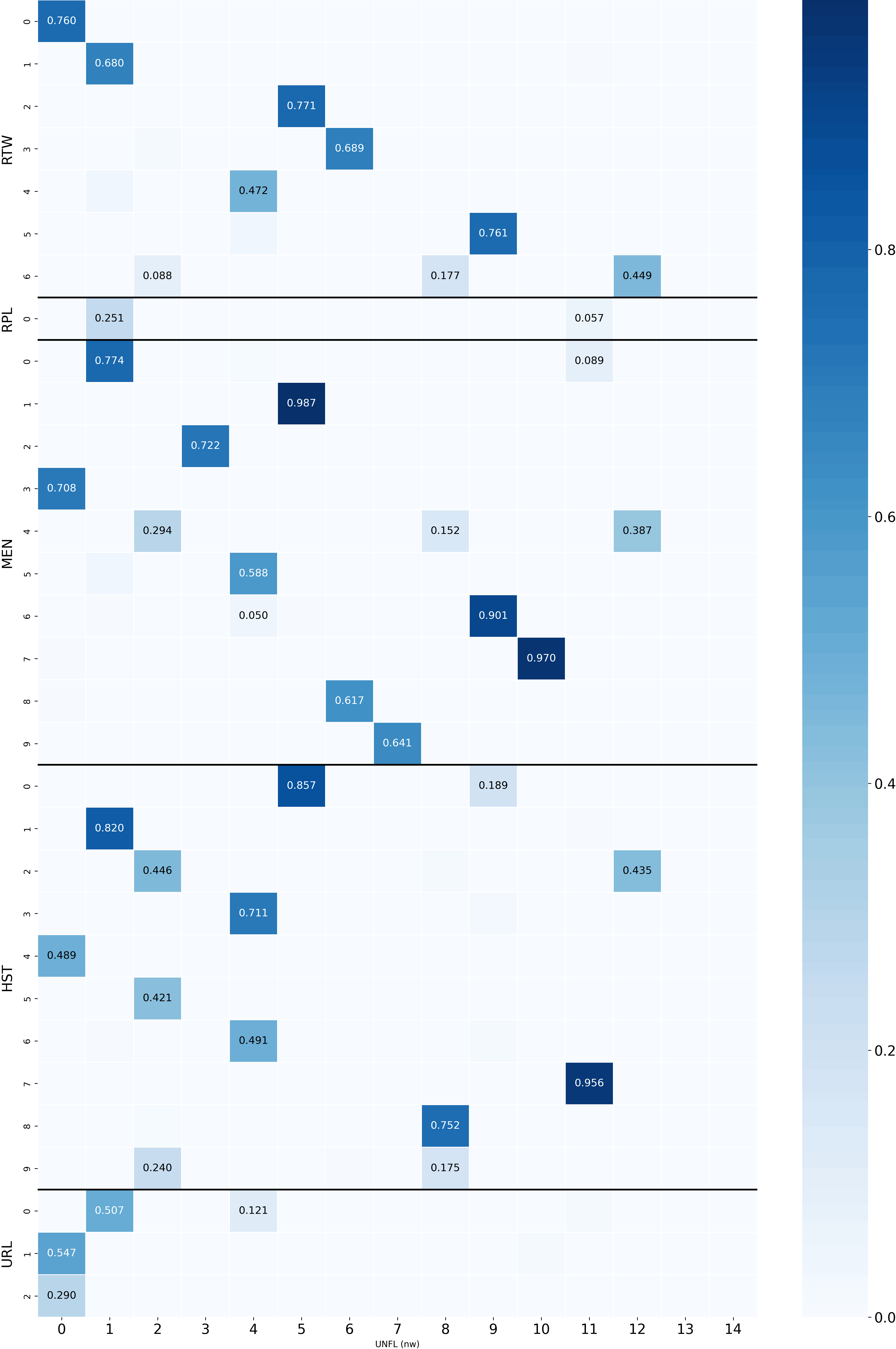}
    \caption{\unfl (nw) vs. \ind.}
    \label{fig:heatmap_flat_nw_iorussia_louvain}
\end{figure}

\begin{figure}[!htbp]
    \centering
    \includegraphics[width=0.8\textwidth, height=19cm]{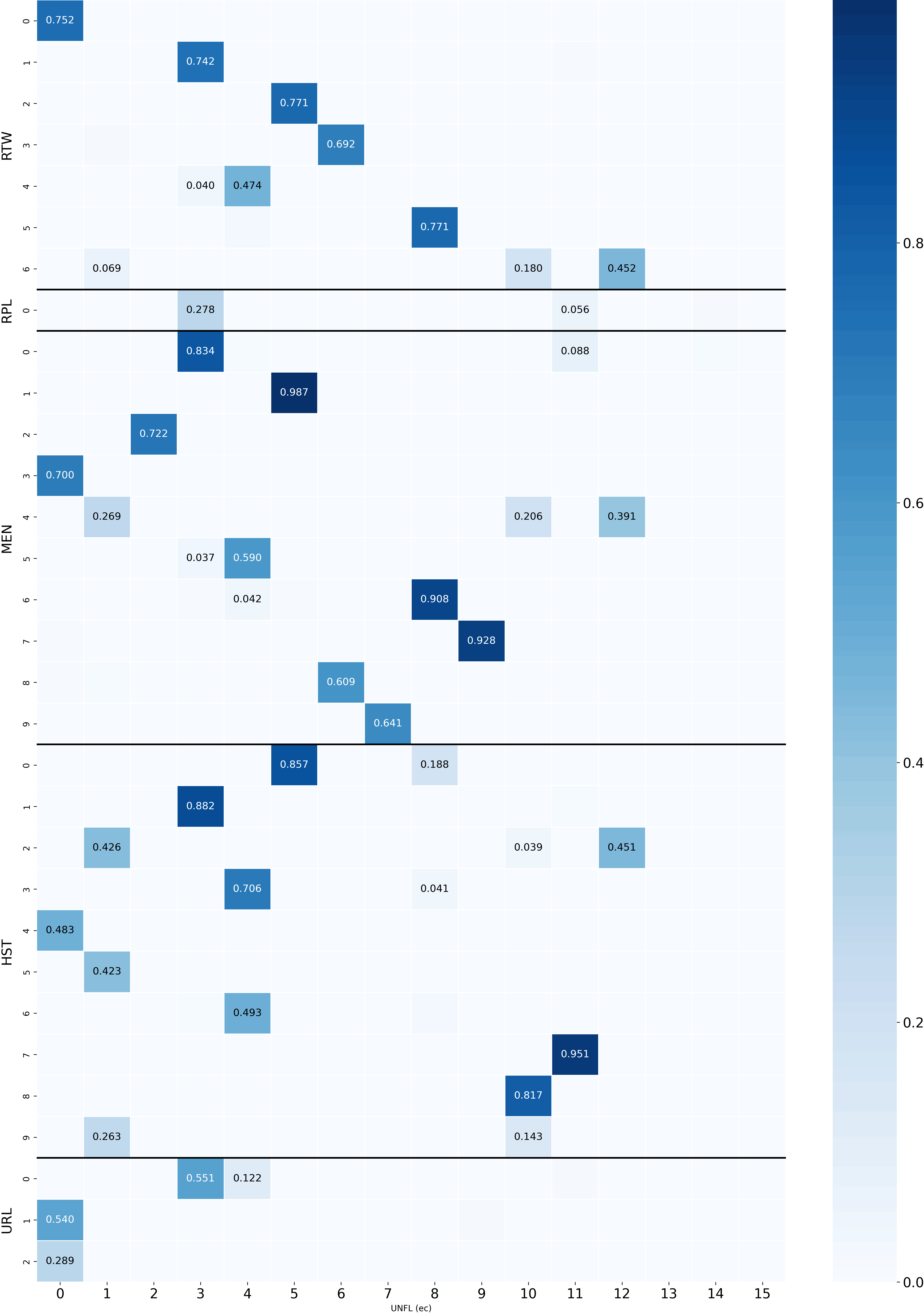}
    \caption{\unfl (ec) vs. \ind.}
    \label{fig:heatmap_flat_ec_iorussia_louvain}
\end{figure}

\begin{figure}[!htbp]
    \centering
    \includegraphics[width=0.8\textwidth, height=19cm]{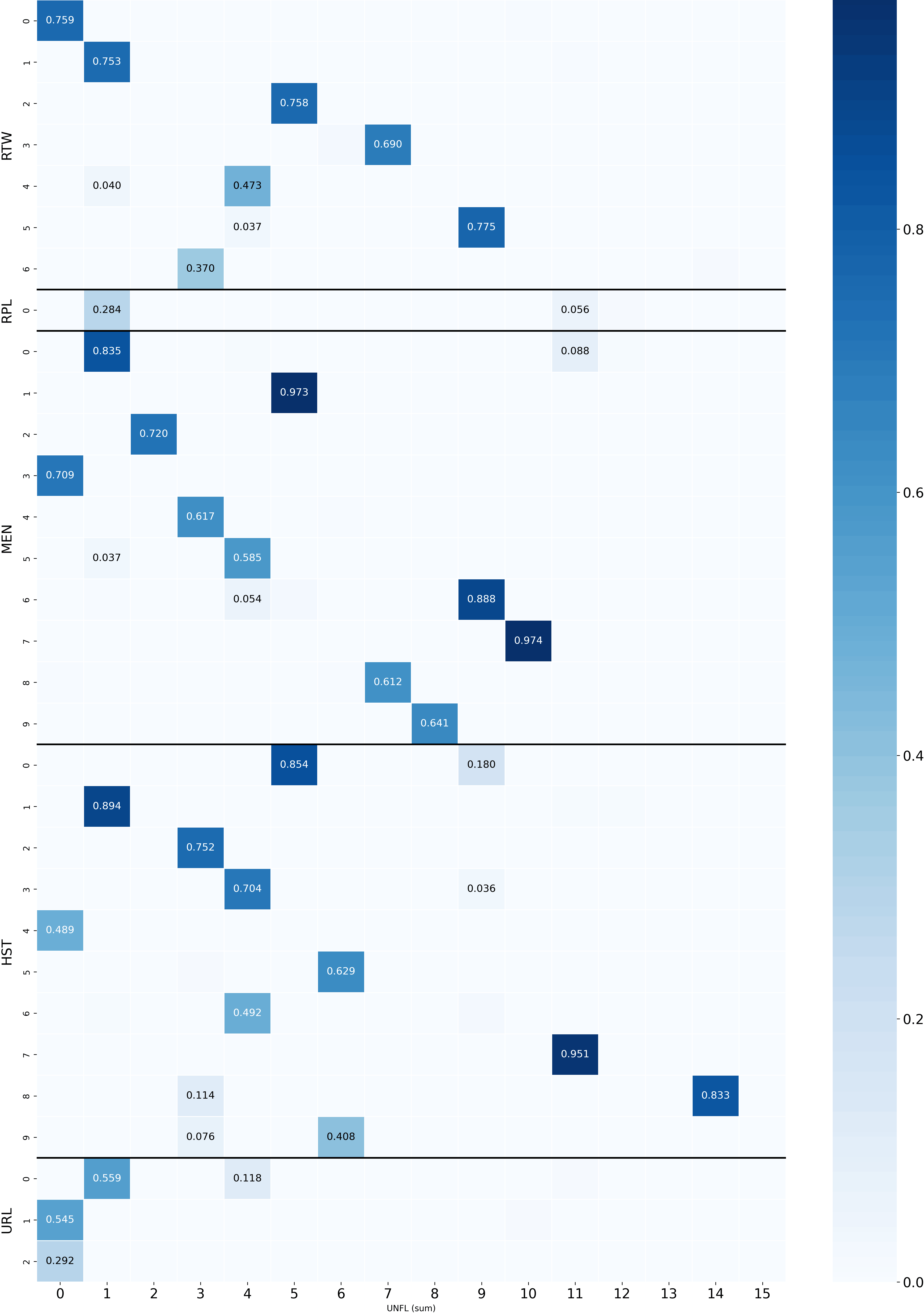}
    \caption{\unfl (sum) vs. \ind.}
    \label{fig:heatmap_flat_sum_iorussia_louvain}
\end{figure}
 \clearpage
\newpage

\section{\russia heatmap overlapping matrix of \info communities}
\label{sec:app_heatmap_iorussia_infomap}
\renewcommand{\thefigure}{E.\arabic{figure}}
We report the heatmaps of the overlapping matrix between \info communities for \russia. Specifically between \rtw and \ind, \mul and \ind, \unfl (all 3 flattening strategies) and \ind.
\setcounter{figure}{0}

\begin{figure}[!htbp]
    \centering
    \includegraphics[width=0.55\textwidth, height=18cm]{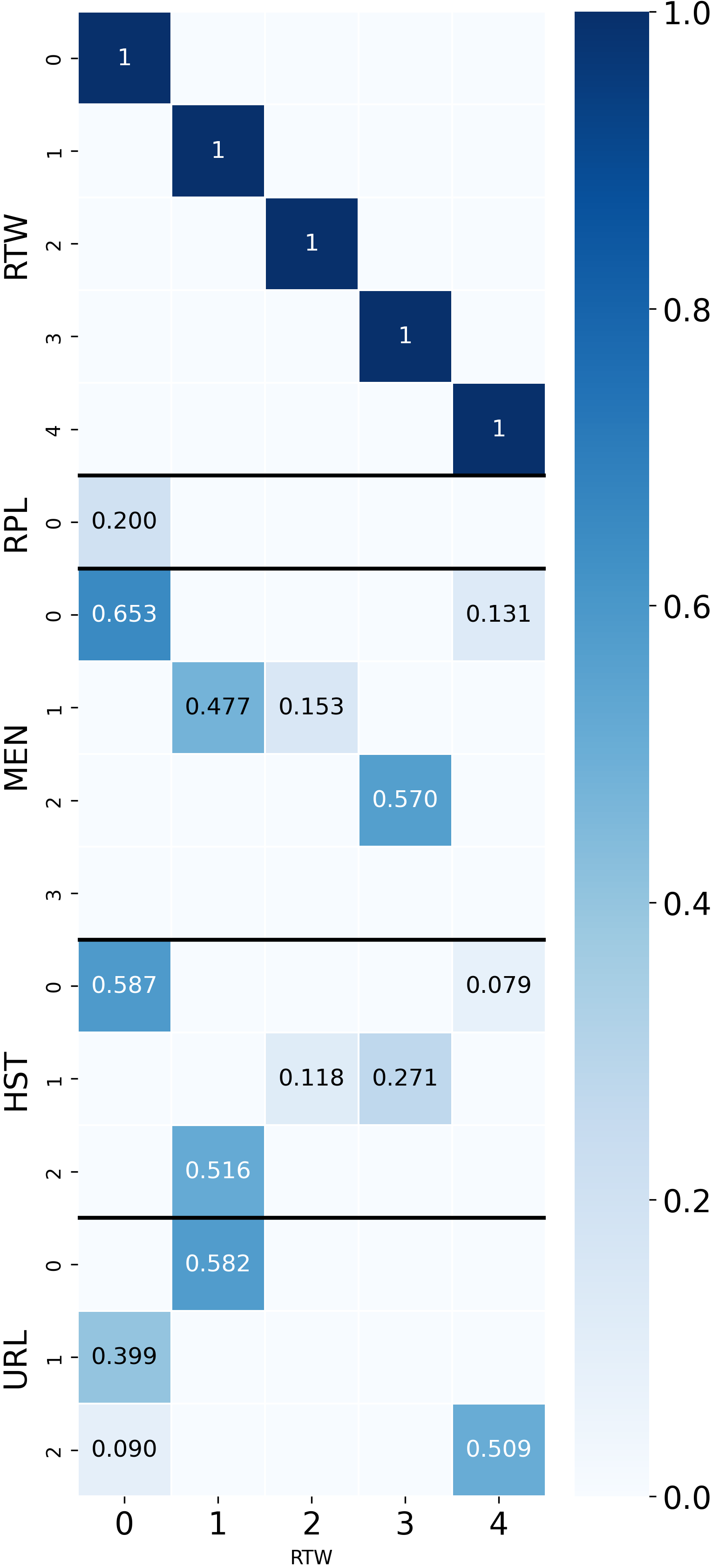}
    \caption{\rtw vs. \ind.}
    \label{fig:heatmap_retweet_iorussia_infomap}
\end{figure}

\begin{figure}[!htbp]
    \centering
    \includegraphics[width=0.9\textwidth, height=19cm]{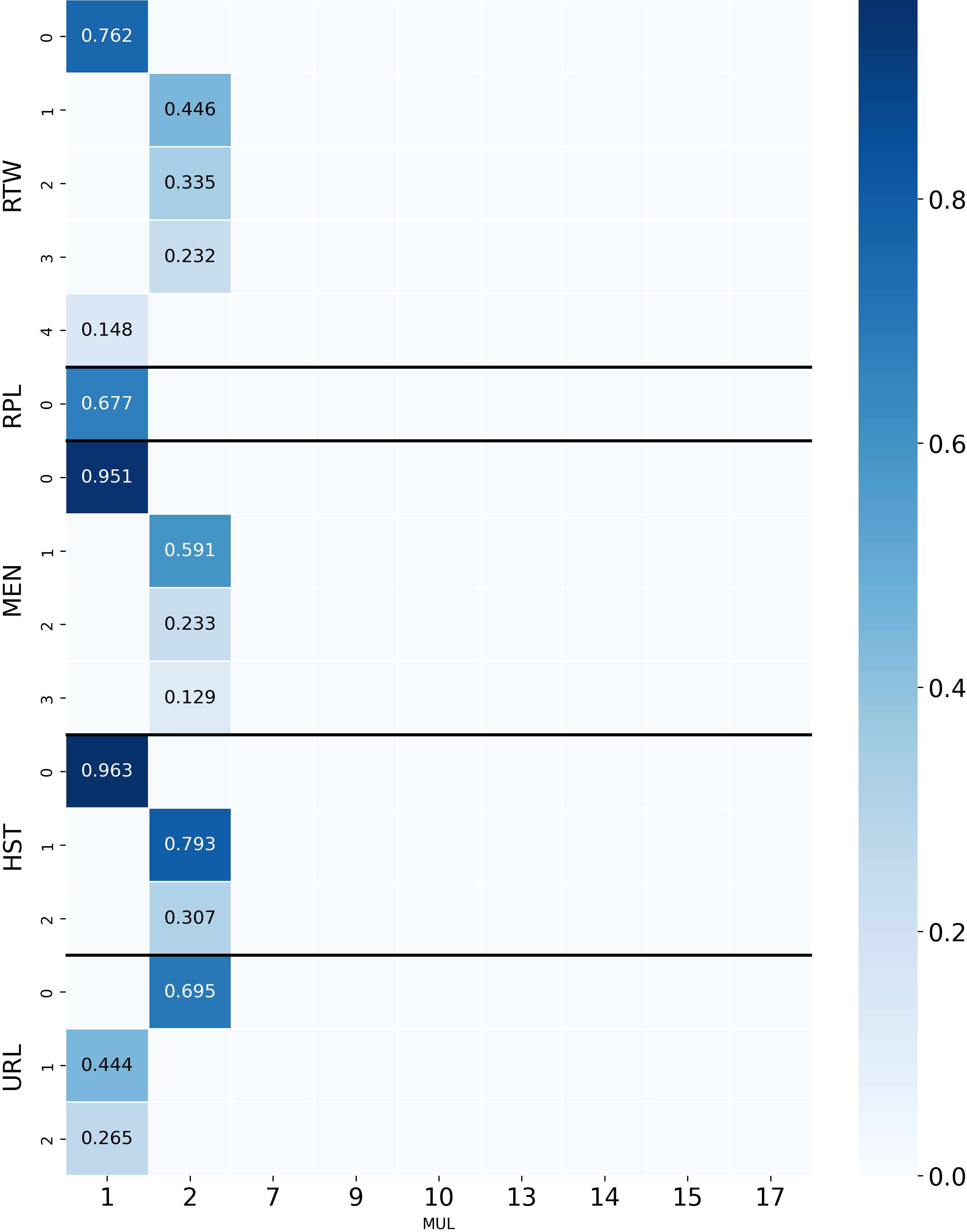}
    \caption{\mul vs. \ind.}
    \label{fig:heatmap_multimodal_basic_and_iorussia_infomap}
\end{figure}

% \begin{figure}[!ht]
%     \centering
%     \begin{minipage}{0.3\textwidth}
%         \centering
%         \includegraphics[width=\textwidth, height=16cm]{./figure_A27a.png}
%         \caption{\unfl (nw) vs. \ind.}
%         \label{fig:heatmap_flat_nw_iorussia_infomap}
%     \end{minipage}
%     \hfill
%     \begin{minipage}{0.33\textwidth}
%         \centering
%         \includegraphics[width=\textwidth, height=16cm]{./figure_A27b.png}
%         \caption{\unfl (ec) vs. \ind.}
%         \label{fig:heatmap_flat_ec_iorussia_infomap}
%     \end{minipage}
%     \hfill
%     \begin{minipage}{0.34\textwidth}
%         \centering
%         \includegraphics[width=\textwidth, height=16cm]{./figure_A27c.png}
%         \caption{\unfl (sum) vs. \ind.}
%         \label{fig:heatmap_flat_sum_iorussia_infomap}
%     \end{minipage}
% \end{figure}

% --- figure A27a ---
\begin{figure}[!htbp]
\centering
\includegraphics[width=0.3\textwidth,height=19cm]{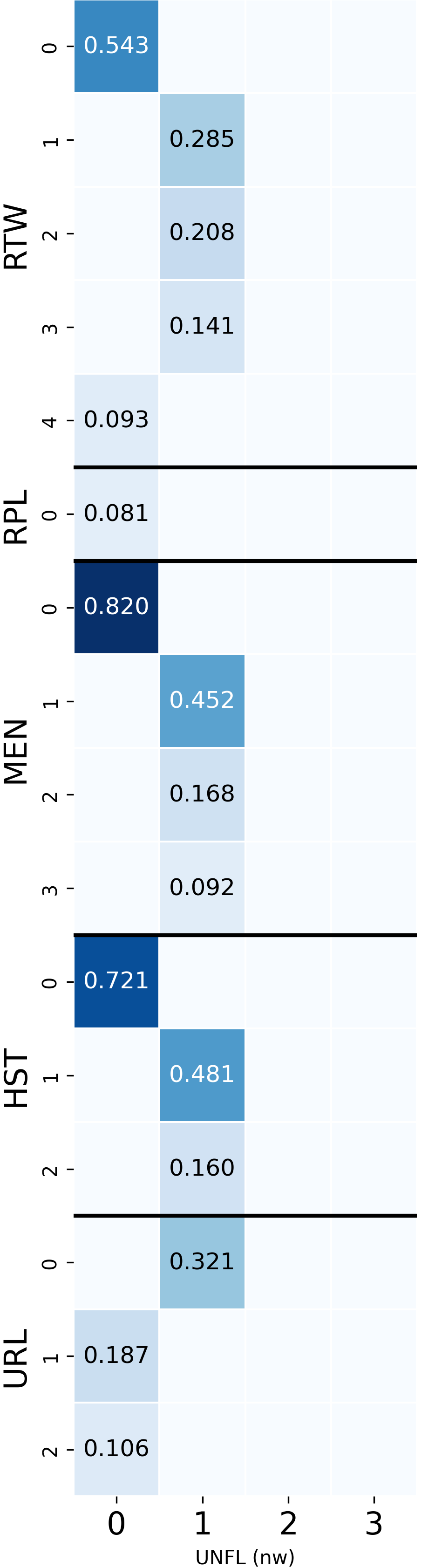}
\caption{\unfl (nw) vs. \ind.}
\label{fig:heatmap_flat_nw_iorussia_infomap}
\end{figure}

% --- figure A27b ---
\begin{figure}[!htbp]
\centering
\includegraphics[width=0.35\textwidth,height=19cm]{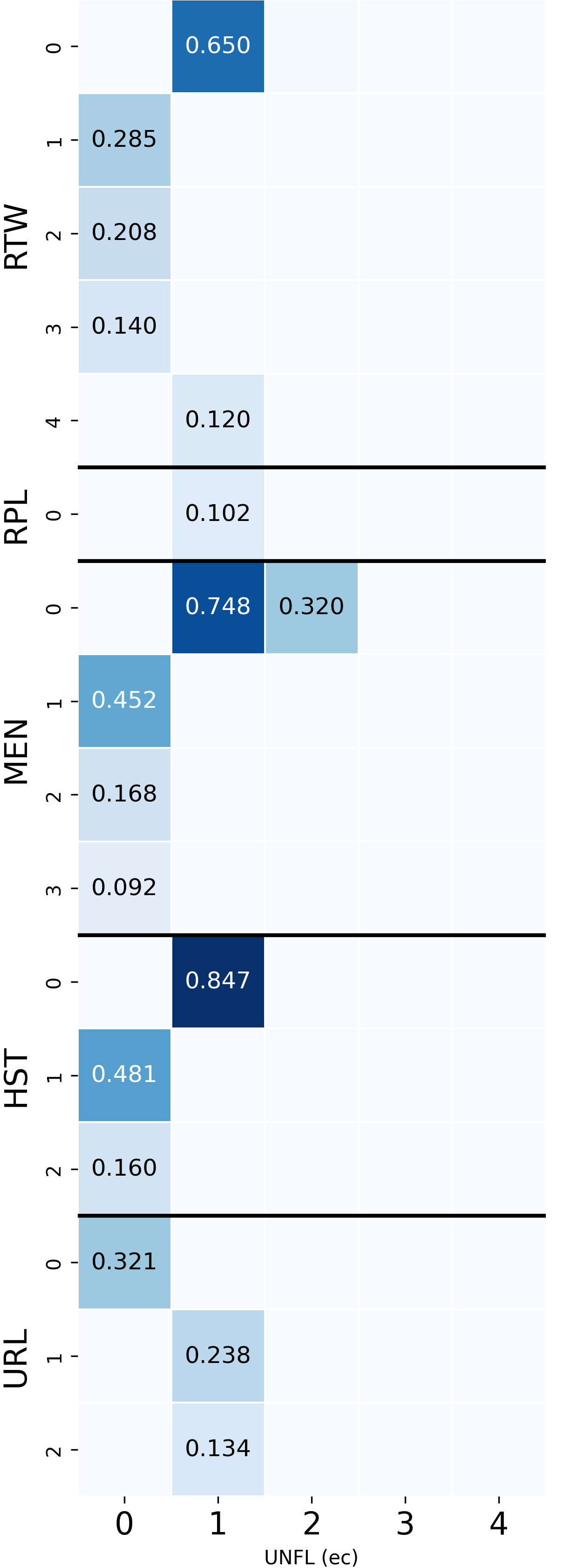}
\caption{\unfl (ec) vs. \ind.}
\label{fig:heatmap_flat_ec_iorussia_infomap}
\end{figure}

% --- figure A27c ---
\begin{figure}[!htbp]
\centering
\includegraphics[width=0.44\textwidth,height=19cm]{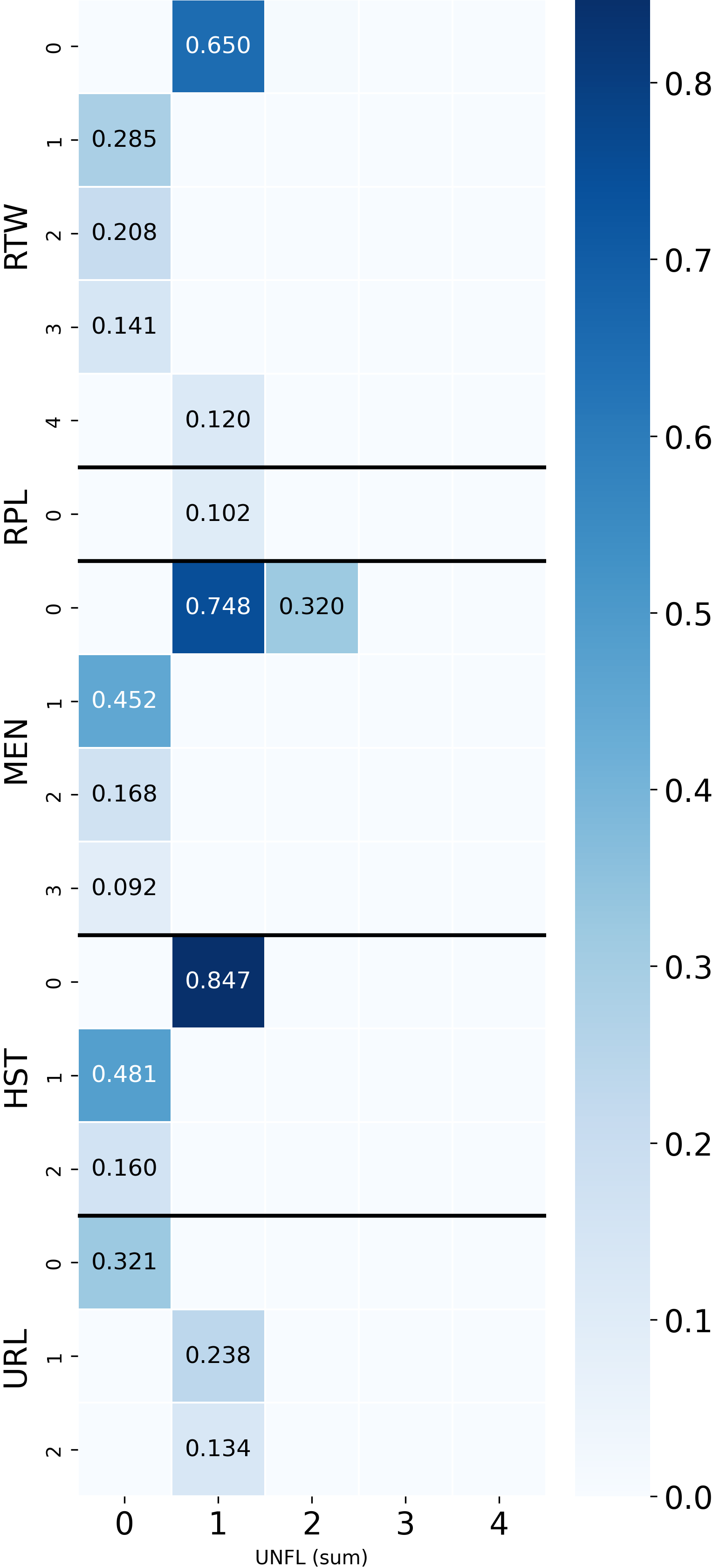}
\caption{\unfl (sum) vs. \ind.}
\label{fig:heatmap_flat_sum_iorussia_infomap}
\end{figure}

 \clearpage
\newpage

 \clearpage
\end{document}